\documentclass[letterpaper,twocolumn,10pt]{article}
\usepackage{usenix}
\usepackage{authblk} 
\usepackage{abstract} 

\usepackage{tocloft}
\usepackage[toc]{appendix}

\newcommand{\AppendixTOC}{
  \begingroup
  \setcounter{tocdepth}{-1} 
  \addtocontents{toc}{\protect\setcounter{tocdepth}{2}} 
  \tableofcontents
  \addtocontents{toc}{\protect\setcounter{tocdepth}{3}} 
  \endgroup
}

\usepackage[english]{babel}
\usepackage{tikz}
\usetikzlibrary{quantikz}

\usepackage{subcaption}

\usepackage{amsmath}
\usepackage{braket}

\usepackage[inline]{enumitem}
\usepackage{caption}

\usepackage{array}

\usepackage{booktabs}
\usepackage{tabularx}

\usepackage{siunitx}

\usepackage[nolist,printonlyused]{acronym}

\usepackage[all]{hypcap}

\usepackage[capitalise,noabbrev]{cleveref}

\definecolor{codegreen}{rgb}{0,0.6,0}
\definecolor{codegray}{rgb}{0.5,0.5,0.5}
\definecolor{codepurple}{rgb}{0.58,0,0.82}
\definecolor{backcolour}{rgb}{0.95,0.95,0.92}

\usepackage{listings}
\usepackage[
    giveninits=true,  
    maxnames=50,      
    maxcitenames=2,   
    urldate=long,     
    backend=biber,    
    sorting=none,
]{biblatex}

\acrodef{AE}{Atomic Ensembles}
\acrodef{API}{Application Programming Interface}
\acrodef{ASIC}{Application-Specific Integrated Circuit}
\acrodef{CNPU}{Classical Network Processing Unit}
\acrodef{CPLD}{Complex Programmable Logic Device}
\acrodef{CPU}{Central Processing Unit}
\acrodef{CR}{Charge-Resonance}
\acrodef{DD}{Dynamical Decoupling}
\acrodef{DQC}{Delegated Quantum Computation}
\acrodef{DQP}{Distributed Queue Protocol}
\acrodef{EGP}{Entanglement Generation Protocol}
\acrodef{EMU}{Entanglement Management Unit}
\acrodef{EPR}{Entanglement Pair Request}
\acrodef{ER}{Entanglement Request}
\acrodef{FPGA}{Field Programmable Gate Array}
\acrodef{HAL}{Hardware Abstraction Layer}
\acrodef{IPC}{Inter-Process Communication}
\acrodef{IT}{Ion Traps}
\acrodef{LGT}{Local Gate Tomography}
\acrodef{MW}{Microwave}
\acrodef{NetQASM}{Quantum Network Assembly Language}
\acrodef{NV}{Nitrogen Vacancy} 
\acrodef{OS}{Operating System}
\acrodef{OSI}{Open Systems Interconnect}
\acrodef{QASM}{Quantum Assembly Language}
\acrodef{QDevice}{Quantum Device}
\acrodef{QDriver}{QDevice Driver}
\acrodef{QEGP}{Quantum Entanglement Generation Protocol}
\acrodef{QMMU}{Quantum Memory Management Unit}
\acrodef{QNetStack}{Quantum Network Stack}
\acrodef{QNodeOS}{Quantum Network Operating System}
\acrodef{QNP}{Quantum Network Protocol}
\acrodef{QNPU}{Quantum Network Processing Unit}
\acrodef{QPU}{Quantum Processing Unit}
\acrodef{PID}{Proportional–Integral–Derivative}
\acrodef{PSB}{Phonon-Side Band}
\acrodef{PMT}{Photomultiplier Tube}
\acrodef{RO}{Readout}
\acrodef{SDK}{Software Development Kit}
\acrodef{SoC}{System on a Chip}
\acrodef{SP}{Spinpump}
\acrodef{SPI}{Serial Peripheral Interface}
\acrodef{SSRO}{Single-Shot Readout}
\acrodef{TTL}{Transistor-Transistor Logic}
\acrodef{ZPL}{Zero-Phonon Line}

\newcommand{\CaPlus}{$^{40}$Ca$^+$}

\setlist[enumerate,itemize]{
    topsep=0.75ex,
    itemsep=0.75ex,
    parsep=0ex,
    partopsep=0ex,
    leftmargin=*
}

\newlist{inlinelist}{enumerate*}{1}
\setlist[inlinelist]{label=(\arabic*)}

\makeatletter
\renewcommand\paragraph{\@startsection{paragraph}{4}{\z@}%
    {1ex \@plus 1ex \@minus 0.2ex}%
    {-0.5em}%
    {\normalfont\normalsize\bfseries\maybe@addperiod}%
}
\newcommand{\maybe@addperiod}[1]{%
    #1\@addpunct{.}%
}
\makeatother
\title{Design and demonstration of an operating system for executing applications on quantum network nodes}

\author[1,2,$\dag$]{Carlo Delle Donne}
\author[1,$\dag$]{Mariagrazia Iuliano}
\author[1,2,$\dag$]{Bart van der Vecht}
\author[1]{Guilherme M. Ferreira}
\author[1]{Hana Jirovská}
\author[1]{Thom J. W. van der Steenhoven}
\author[1,2]{Axel Dahlberg}
\author[1,2]{Matt Skrzypczyk}
\author[4]{Dario Fioretto}
\author[4]{Markus Teller}
\author[4]{Pavel Filippov}
\author[1]{Alejandro R.-P. Montblanch}
\author[1]{Julius Fischer}
\author[1]{H. Benjamin van Ommen}
\author[1]{Nicolas Demetriou}
\author[6]{Dominik Leichtle}
\author[6]{Luka Music}
\author[5]{Harold Ollivier}
\author[1]{Ingmar te Raa}
\author[1]{Wojciech Kozlowski}
\author[1]{Tim H. Taminiau}
\author[3]{Przemysław Pawełczak}
\author[4]{Tracy E. Northup}
\author[1]{Ronald Hanson}
\author[1,2,*]{Stephanie Wehner}

\affil[1]{\normalsize \textit{QuTech and Kavli Institute of Nanoscience, Delft University of Technology, 2628 CJ Delft, The Netherlands}}
\affil[2]{\normalsize \textit{Quantum Computer Science, Department of Software Technology, Faculty of Electrical Engineering, Mathematics and Computer Science, Delft University of Technology, 2628 XE Delft, The Netherlands}}
\affil[3]{\normalsize \textit{Embedded Systems, Department of Software Technology, Faculty of Electrical Engineering, Mathematics and Computer Science, Delft University of Technology, 2628 XE Delft, The Netherlands}}
\affil[4]{\normalsize \textit{Institut für Experimentalphysik, Universität Innsbruck, Technikerstraße 25, 6020 Innsbruck, Austria}}
\affil[5]{\normalsize \textit{QAT, DIENS, Ecole Normale Supérieure -- PSL University, CNRS, INRIA. 45 rue d'Ulm 75005 Paris, France}}
\affil[6]{\normalsize \textit{LIP6, CNRS, Sorbonne Université, 4 Place Jussieu, 75005, Paris, France}}
\affil[$\dag$]{These authors contributed equally to this work}
\affil[*]{To whom correspondence should be addressed; E-mail: s.d.c.wehner@tudelft.nl}

\begin{document}
\twocolumn[
  \begin{@twocolumnfalse}
    \maketitle
    \vspace{10em}
    \begin{abstract}
       The goal of future quantum networks is to enable new internet applications that are impossible to achieve using solely classical communication\cite{kimble_2008_quantum,wehner_2018_stages,vanMeter_book}. Up to now, demonstrations of quantum network applications\cite{barz_2012_demonstration,drmota_verifiable_2024,nadlinger_device-independent_2022} and functionalities\cite{hermans2022qubit,iuliano2024qubit,matsukevich_quantum_2009,langenfeld_quantum_2021,pfaff_unconditional_2014,chou_deterministic_2018} on quantum processors have been performed in ad-hoc software that was specific to the experimental setup, programmed to perform one single task (the application experiment) directly into low-level control devices using expertise in experimental physics. Here, we report on the design and implementation of the first architecture capable of executing quantum network applications on quantum processors in platform-independent high-level software. We demonstrate the architecture’s capability to execute applications in high-level software, by implementing it as a quantum network operating system – QNodeOS – and executing test programs including a delegated computation from a client to a server\cite{broadbent_2009_ubqc} on two quantum network nodes based on nitrogen-vacancy (NV) centers in diamond\cite{doherty_2013,childress2013diamond}. We show how our architecture allows us to maximize the use of quantum network hardware, by multitasking different applications on a quantum network for the first time. Our architecture can be used to execute programs on any quantum processor platform corresponding to our system model, which we illustrate by demonstrating an additional driver for QNodeOS for a trapped-ion quantum network node based on a single \CaPlus atom\cite{fioretto_towards_2020}. Our architecture lays the groundwork for computer science research in the domain of quantum network programming, and paves the way for the development of software that can bring quantum network technology to society.
    \end{abstract}
  \end{@twocolumnfalse}
]
\section{Introduction}

The first quantum networks linking multiple quantum processors as end nodes have recently been realized as physics experiments in laboratories~\cite{moehring_2007_ion_traps,ritter_2012_elementary,hofmann_2012_heralded,stockill_2017_phasetuned,jing2019entanglement,stephenson_2020_highrate,pompili_2021_multinode,krutyanskiy_entanglement_2023} and fiber networks~\cite{liu2024creation,stolk2024metropolitan,knaut2024entanglement}, opening the tantalizing possibility of realizing advanced quantum network applications~\cite{wehner_2018_stages} such as data consistency in the cloud~\cite{benor_2005_byzantine}, privacy-enhancing proofs of deletion~\cite{poremba_quantum_2022}, exponential savings in communication~\cite{guerin_exponential_2016}, or secure quantum computing in the cloud~\cite{broadbent_2009_ubqc,childs_2005_secure_qc}. Demonstrations relied either on ad-hoc software, or chose to establish that hardware parameters were in principle good enough to support a given quantum network application, although the application itself was not realized~\cite{nadlinger_device-independent_2022,liu_2022_photonic_diqkd,zhang_2022_diqkd}.

\begin{figure}
\centering
\includegraphics[width=1\linewidth]{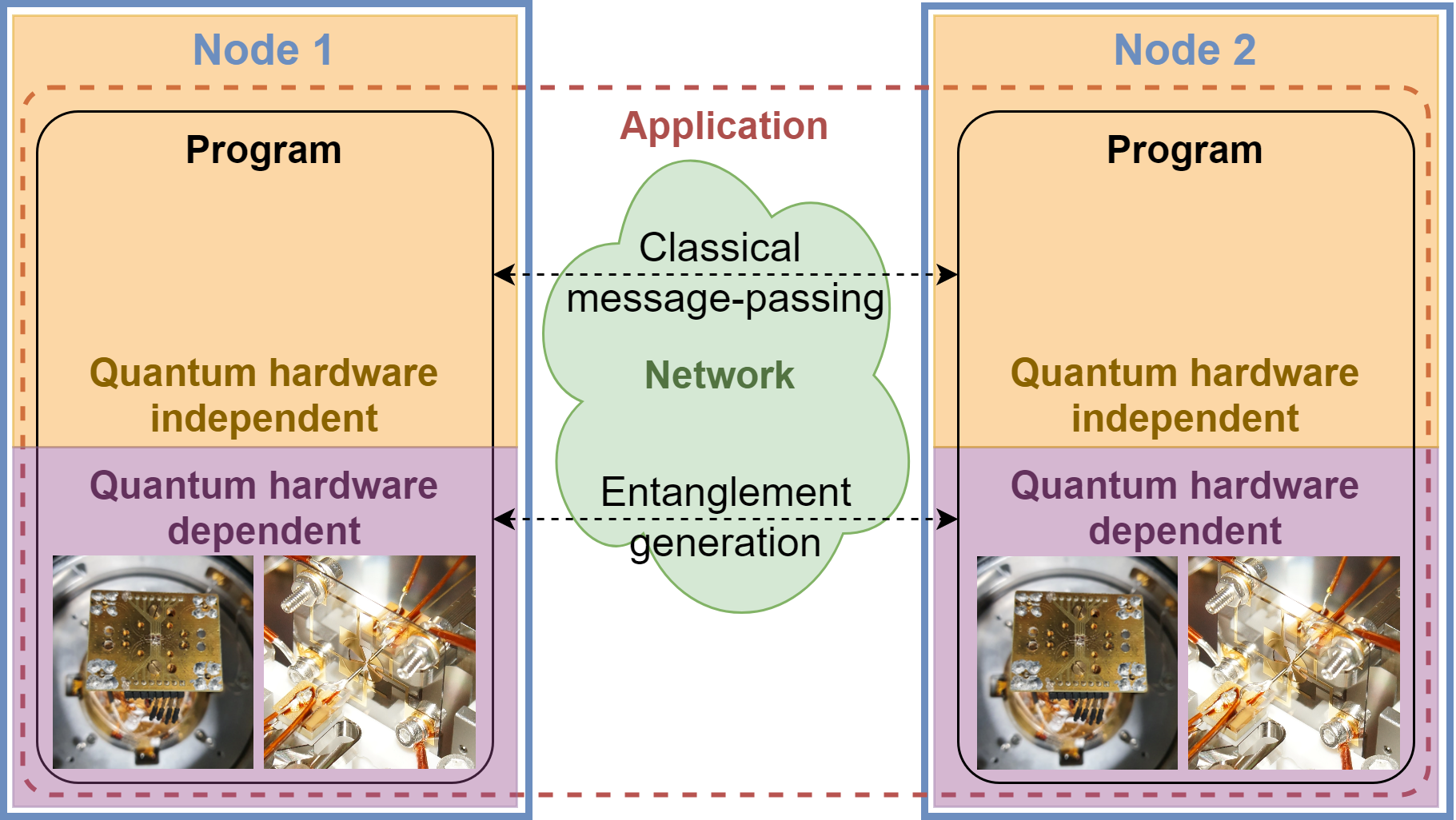}
\caption{\textbf{Application Paradigm.} A quantum networking application consists of multiple programs, each running on one of the end nodes~\cite{dahlberg_2022_netqasm} The distinct programs can only interact via (1) quantum communication (e.g. entanglement generation) and (2) classical communication. This allows a programmer to realize security-sensitive applications, but prohibits a global orchestration of the quantum execution, like one might do in (distributed) quantum computing~\cite{caleffi_distributed_2022} in which a single quantum program is executed on multiple nodes. Programs are written in high-level quantum hardware independent software, and executed on a quantum hardware independent system (our architecture) that controls a hardware dependent system (QDevice, \cref{fig:fig2}) such as a nitrogen-vacancy (NV) center node with a diamond chip (photo taken by authors, left images) or a trapped-ion quantum node~\cite{teller2023integrating} (right images). These platforms constitute physically very different QDevice systems, but can both be programmed by our architecture.}
\label{fig:fig1}
\end{figure}

It is a major challenge to design and implement an architecture that can enable the execution of arbitrary quantum network applications on quantum processors (\cref{fig:fig1}), while enabling programming in high-level software that neither depends on the underlying quantum hardware, nor requires the programmer to understand the physics of the underlying devices.  In the domain of the conventional internet, the possibility of programming arbitrary internet applications in high-level software has led to the realization of radically new communication applications by diverse communities, which had a transformative impact on our society~\cite{castells_impact_2013}. What's more, the advent of programmable hardware and new application areas sparked novel fields of computer science research and guided further hardware development.  A similar development is underway in quantum computing, where the availability of high-level programming tools allows a broad participation in developing applications~\cite{noauthor_quantum_2024}.

In realizing the first such architecture we overcome both fundamental challenges that are inherent to quantum network applications, as well as technological challenges that arise from the current state of the art of quantum network hardware.

\section{Design Considerations and Challenges}
\paragraph{Interactive Classical-Quantum Execution}
The execution of quantum network applications requires a continuing interaction between the quantum and classical parts of the execution, including interactions between different programs (\cref{fig:fig1}). For example, during secure quantum computing in the cloud~\cite{broadbent_2009_ubqc,ma_qenclave-practical_2022}, the program on the server is waiting for classical messages from a remote client before continuing the quantum execution at the server. This is in sharp contrast to quantum computing applications, where one quantum program can be executed in a single batch, without the need to keep quantum states live while waiting for input from other programs. In quantum computing, only relatively low-level interactions between classical and quantum processing are realized, such as in quantum error correction~\cite{lidar2013quantum}, or mid-circuit measurements~\cite{botelho_error_2022}. Higher-level classical-quantum interactions in quantum computing~\cite{bharti_noisy_2022} do not keep qubits live in memory.

We assume that the programs are divided into classical and quantum blocks of instructions (by a programmer or a compiler). Classical blocks consist of local classical operations executed on a conventional classical processor, as well as networked classical operations (i.e. sending messages to remote nodes) executed using network devices. Quantum blocks consist of local quantum operations (gates, measurements, classical control logic), as well as networked quantum operations (entanglement generation) executed on quantum hardware. A single quantum block, in essence, corresponds to a program in quantum computing, and may contain simple classical control logic, such as for the purpose of mid-circuit measurements~\cite{botelho_error_2022}.

\paragraph{Different Hardware Platforms}
Interfacing with different hardware platforms presents technological challenges: currently, a clear line between software and hardware has not been defined, and the low-level control of present-day quantum processor hardware has been built to conduct physics experiments. Early microarchitectures~\cite{bertels_quantum_2020,fu_2019_eqasm} and operating systems~\cite{giortamis_qos_2024,kong_2021_origin} for quantum computing do not address the execution of quantum network applications. We thus have to define a hardware abstraction layer (HAL), capable of interfacing with present-day quantum network setups. 

\paragraph{Timescales}
It is a fundamental challenge that different parts of such a system operate at vastly different timescales. For nodes separated by hundreds of kilometers, the duration of network operations is in the millisecond (ms) regime, and some applications~\cite{wehner_2018_stages} need  significant local classical processing (ms). In contrast, the time to execute quantum operations on processing nodes is in the regime of microseconds ($\mu$s), and the low-level control (including timing synchronization between neighboring nodes to generate entanglement~\cite{humphreys_2018_delivery}) requires nanosecond (ns) precision.

\paragraph{Memory Lifetimes}
Present-day quantum network nodes have short coherence times, posing a technological challenge to ensure operations are executed within the timeframe allowed by the quantum memory.

\paragraph{Scheduling Local and Network Operations}
In contrast to classical networking, entanglement is a form of stateful connection already at the physical layer where both nodes hold one qubit. Heralded entanglement generation requires agreement between neighboring network nodes to trigger entanglement generation in precise time-bins~\cite{dahlberg_2019_egp}, organized into a network schedule~\cite{skrzypczyk_2021_arch} that dictates when nodes make entanglement. It is a technological challenge to manage the interdependencies between the schedule of local operations, and the networked operations, since in all current processing node implementations~\cite{pompili_2021_multinode,drmota_robust_2023}, entanglement generation cannot be performed simultaneously with local operations~\cite{pompili_2021_multinode,krutyanskiy_light-matter_2019}. While interdependencies may be mitigated in the future~\cite{vardoyan_2022_netarch}, this implies that we cannot schedule (i.e. decide when to execute) the execution of local quantum operations independently of the network schedule.

\paragraph{Multitasking}
When executing quantum network applications, one node is typically idle while waiting for the other node before it can continue execution. It is hence a fundamental challenge how we can increase the utility of the system by performing multitasking~\cite{mccullough_design_1965-1,dennis_segmentation_1965}, that is, allowing the concurrent execution of several programs at once to make use of idle times. Consequently, there is a need for managing state and resources for multiple independent programs, including processes, quantum memory management, and entanglement requests. 

\begin{figure*}[htb]
\centering
\includegraphics[width=1\linewidth]{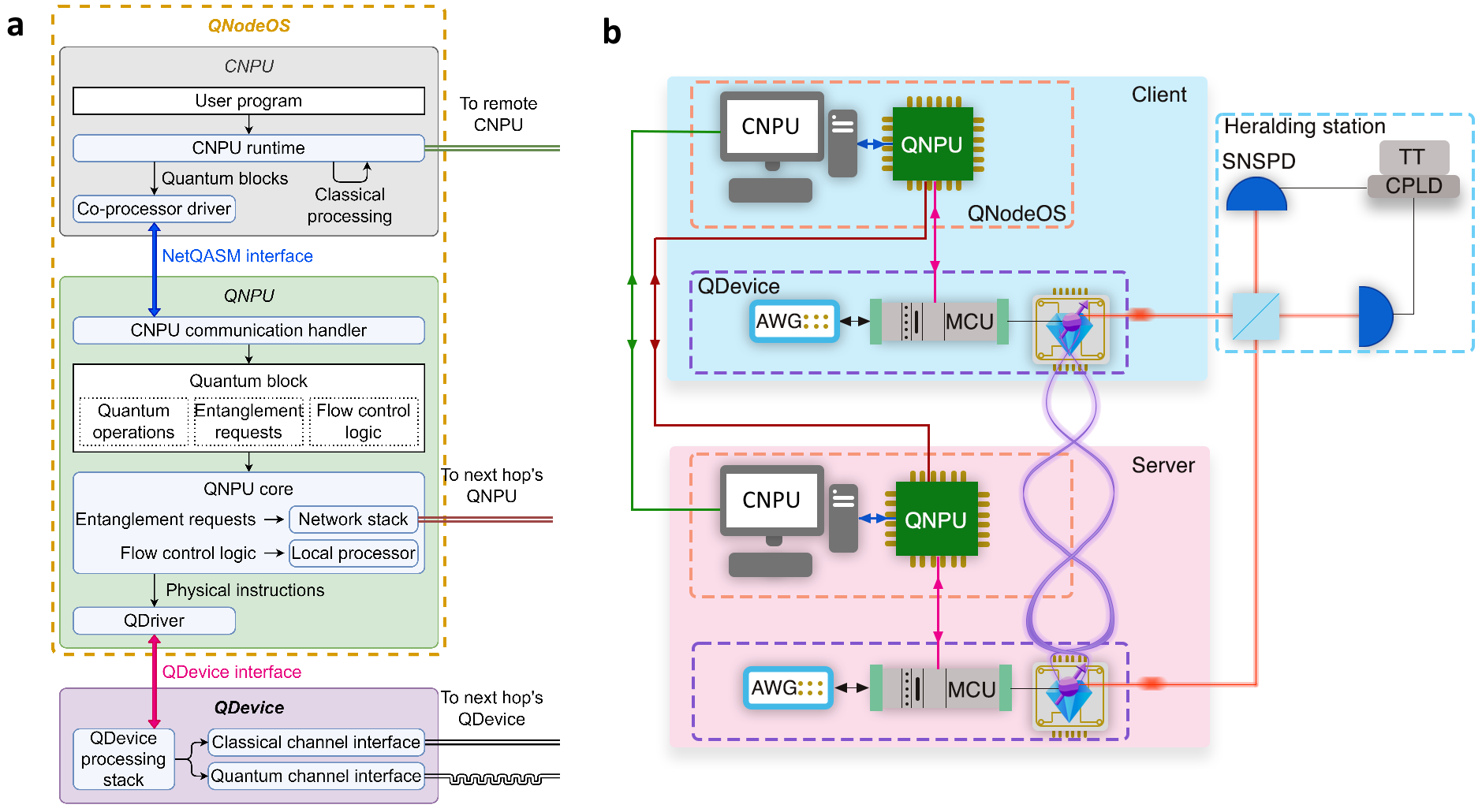}
\caption{\textbf{QNodeOS architecture.} \textbf{(a)} QNodeOS consists of a Classical Network Processing Unit (CNPU) and a Quantum Network Processing Unit (QNPU, classical system). QNodeOS controls a QDevice (quantum hardware and low-level classical control).
\textbf{(b)} Schematic of our implementation of QNodeOS on a two-node setup where both QDevices control a single qubit in a diamond nitrogen-vacancy (NV) center. The CNPU is implemented on a general-purpose PC, and the QNPU on an embedded system, connected via Gigabit Ethernet (blue). The QNPU connects to its QDevice via a serial peripheral interface (SPI, pink). The two QNPUs (brown), and the two CNPUS (green) connect to each other via Gigabit Ethernet. The setup is based on~\cite{pompili_2022_experimental} with two QDevices (including arbitrary waveform generators (AWG) and microcontroller units (MCU); QDevices communicating over a classical DIO interface) and a heralding station composed by a balanced 50:50 beam-splitter (whose output ports are connected to superconducting nanowire single-photon detectors (SNSPD) via optical fibers (red)),  a  TimeTagger (TT), and a \ac{CPLD} that heralds the entanglement generation between QDevices and sends a classical message to the MCU.}
\label{fig:fig2}
\end{figure*}

\section{Architecture}
\label{sec:architecture}
We divide the architecture logically into three main components (\cref{fig:fig2}, \cref{sec:methods}): The Classical Network Processing Unit (CNPU) is responsible for starting the execution of programs, and the execution of classical code blocks; the Quantum Network Processing Unit (QNPU) is responsible for governing the execution of the quantum code blocks; The CNPU and QNPU together form QNodeOS and control the QDevice, which is responsible for executing any quantum operations (gates, measurements, entanglement generation at the physical layer~\cite{dahlberg_2019_egp}) on the quantum hardware. Upon starting the execution the CNPU creates a corresponding CNPU process (a well-known concept in classical operating systems~\cite{dennis_programming_1966-1,tanenbaum_operating_2005}), registers the program on the QNPU (via the QNPU's end node application programming interface (API), \cref{sec:design:qnpu_stack}), which, in turn, creates its own associated QNPU process (including context such as process owner, ID, process state and priority). QNodeOS also defines kernel processes on the QNPU, which are similar to user processes, but are created when the system starts (on boot). The CNPU sends quantum blocks to the QNPU in the form of NetQASM subroutines~\cite{dahlberg_2022_netqasm}. Classical control logic in quantum blocks is executed by the QNPU processor. Quantum gates and measurements (from any QNPU process) and entanglement instructions (from the network process) are delegated to the QDevice by submitting physical instructions (\cref{sec:methods}), after which the QDevice responds back with the result of the instruction. 

To enable different hardware platforms, we introduce a QDriver realizing the HAL for any hardware corresponding to our minimal QDevice system model (\cref{sec:methods}). The QDriver is responsible for translating quantum operations, expressed in NetQASM~\cite{dahlberg_2022_netqasm}, into platform dependent (streams of) physical instructions to the underlying QDevice. We realize a QDriver for the trapped-ion system of~\cite{teller2023integrating,teller2021heating}, and one for NV centers in diamond based on the system of~\cite{hermans2022qubit,pompili_2021_multinode,pompili_2022_experimental}. We validate the trapped-ion QDriver (\cref{fig:fig5}) by implementing and verifying a set of single-qubit gate operations (\cref{sec:methods}), and the QDriver on the NV system as part of the full stack system evaluation (see below). 
To allow for different timescales, we logically divide the architecture into CNPU, QNPU and QDevice which can thus be realized at different timing scale granularities. In our proof-of-concept implementation, we realize the CNPU and QNPU on different devices, reflecting the ms timescales of communication between distant nodes (\cref{sec:methods}).

Ensuring the necessary interactivity requires architectural as well as implementation choices: as programs may depend on messages from remote nodes, the architecture needs to be able to dynamically handle both classical and quantum blocks, even if not known at runtime. Consequently, it is not possible to preload all quantum blocks of the program into the low-level controller of the QDevice ahead of time as done in previous physics experiments. Instead, in our system model the QDevice is capable of executing individual physical instructions similar to a classical CPU. Consequently, the QNPU is continuously ready to receive new NetQASM subroutines from the CNPU, and the QDevice can continuously receive and respond to physical instructions from the QNPU (\cref{sec:methods}).

In our NV QDevice implementation, we address the challenge of interactivity by interleaving specific preloaded pulse sequences (realizing physical instructions sent from QNodeOS) and dynamical decoupling (DD) sequences (protecting quantum memory from decoherence) in an arbitrary waveform generator (AWG)~\cite{zurich_instruments_hdawg_2019}. The DD sequences extend qubit coherence times up to $T_{\text{coh}} = 13(2)$ ms, while arbitrary physical instructions can be handled by triggering the corresponding pulse sequence, without knowing them in advance (\cref{sec:methods}).

To integrate local operations with the network schedule, our architecture first introduces a QNPU scheduler that can choose which of the ready processes is assigned to the local processor (\cref{fig:fig2}) and QDevice. This allows interleaving the execution of different processes directly on the QNPU without incurring delays on the timescale of the CNPU (ms), addressing the challenge of short coherence times. In our implementation, we choose to schedule QNPU processes using a priority based non-preemptive scheduler~\cite{liu_1973_scheduling}, due to limited quantum memory lifetimes, which make it undesirable to pre-empt and temporarily store quantum states while halting the execution. Second, we realize a network process as a kernel process, which manages entanglement generation using the network stack~\cite{dahlberg_2019_egp,kozlowski_2020_qnp} (implemented in~\cite{pompili_2022_experimental} without the ability to execute network applications), including a network schedule that can be determined by a time-division multiple access (TDMA) controller~\cite{skrzypczyk_2021_arch}. The network process handles entanglement requests submitted by user processes, coordinates entanglement generation with the rest of the network via the TDMA controller, interacts with the QDevice and eventually returns entangled qubits to user processes. User processes enter the waiting state when they need entanglement, and become ready again once entanglement was delivered. The network process has the highest scheduling priority, and is consequently given precedence over the execution of any local quantum operations. We remark local operations may still be executed during time-bins already occupied by the network schedule, if a running non-preemptable user process prevents the network process from running, as we indeed observe in our evaluation.

To increase utility, QNodeOS allows multiple programs to be run concurrently, using the QNPU scheduler from above to enable multitasking~\cite{mccullough_design_1965-1,dennis_segmentation_1965} user processes on the QNPU itself. The QNPU hence needs to keep context for each process, including a virtual quantum memory space (as in classical operating systems~\cite{daley_virtual_1968-1}). Similar to classical memory management systems~\cite{peterson_operating_1985}, a quantum memory management unit (QMMU) on the QNPU manages qubit allocations from processes, and translates virtual qubit addresses in NetQASM subroutines to physical addresses in the QDevice. This allows flexibility in translating a virtual qubit address to: (1) a different physical qubit address over time, allowing qubits to be rearranged transparently in the physical memory in the future, or (2) a logical qubit address, when QNodeOS is executed on top of a processor employing quantum error correction~\cite{lidar2013quantum}. Entanglement generation between different pairs of processes at remote nodes are distinguished by Entanglement Request (ER) sockets, inspired by classical sockets~\cite{chesson_network_1975-1,leach_architecture_1983}, which are established once a user process requests entanglement from the network process. In our implementation, processes of the same priority are scheduled first-come-first-served~\cite{peterson_operating_1985}, where the total schedule of the program in our implementation is dependent both on the schedule on the CNPU as well as the QNPU (\cref{sec:methods}).

\begin{figure*}[htbp]
\centering
\includegraphics[width=0.89\linewidth]{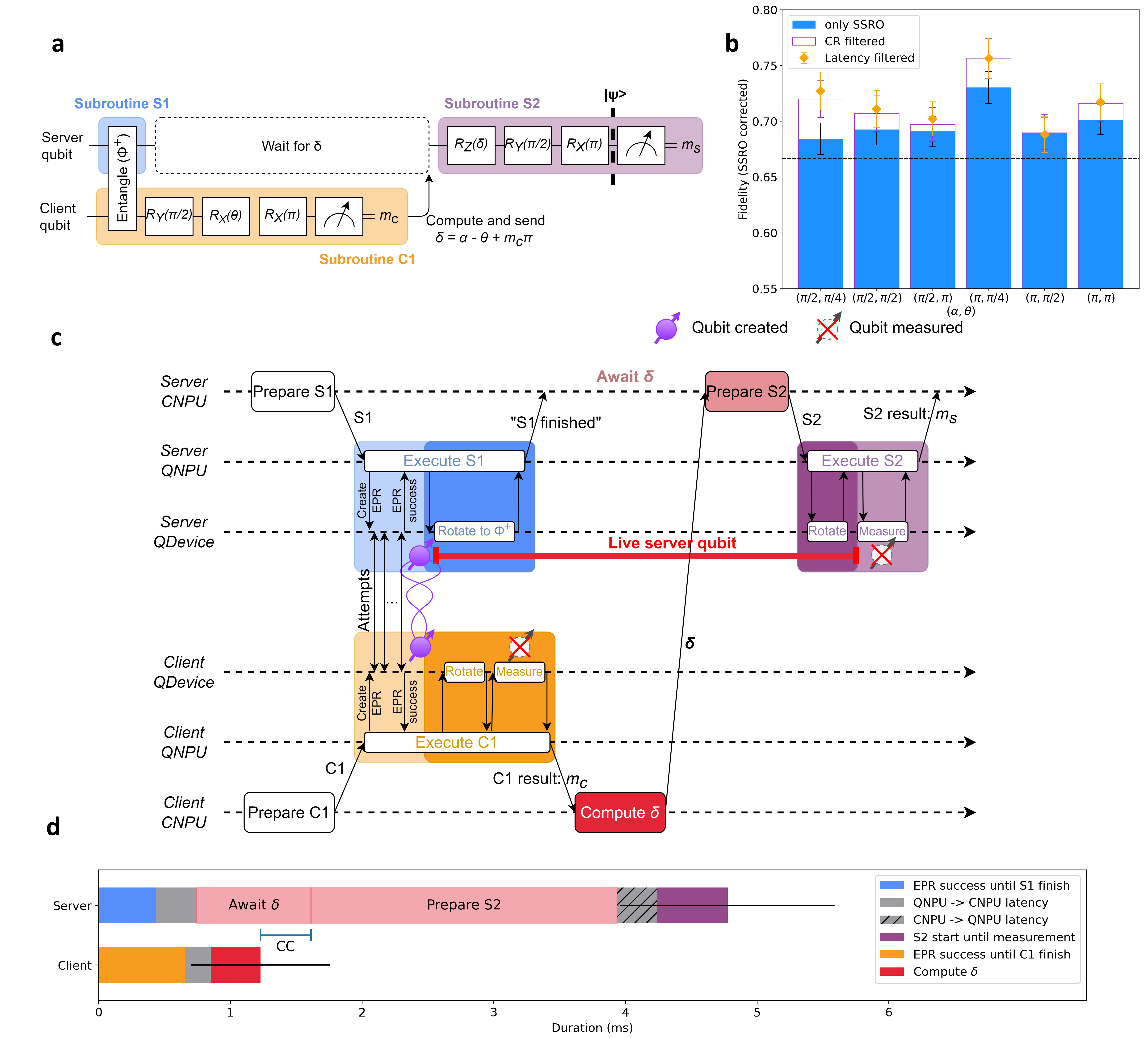}
\caption{\textbf{Delegated computation between two NV center nodes using QNodeOS.} 
\textbf{(a)} Delegated Quantum Computation (DQC) circuit (effective computation: single-qubit rotation $R_Z (\alpha)$, \cref{sec:methods}). The DQC application consists of $k$ repetitions of this circuit (varying measurement bases for tomography on $\ket{\psi}$) realized by two programs: the DQC-client program (client node, repeating the sequence ``quantum block (C1, orange) – classical block (computing $\delta$)'' $k$ times), and the DQC-server program (server node, repeating ``quantum block (S1, blue) - classical block (receiving $\delta$) – quantum block (S2, purple)'' $k$ times). Client and server produce an entangled pair $\ket{\Phi^+} = (\ket{00} + \ket{11}) / \sqrt{2}$ (S1 and first part of C1). The client performs local gates and a measurement (``destroying'' qubit), resulting in outcome bit $m_c$ (rest of C1). Client computes $\delta$ as function of $m_c$ and DQC parameters $\alpha \in [0,2\pi)$ and $\theta \in [0, 2\pi)$, and sends $\delta$ to server (classical message). Meanwhile the server keeps its qubit coherent (alive). Upon receiving $\delta$, the server applies gates depending on $\delta$, resulting in single-qubit state $\ket{\psi}$ (S2) depending only on $\alpha$ and $\theta$.
\textbf{(b)} Experimental results of executing DQC for 6 different sets of $(\alpha, \theta)$ parameters ($k=1200$, i.e. 7200 executions of the circuit of~\ref{fig:fig3}a). The fidelity of the resulting server state to the target state $\ket{\psi}$ is estimated using single-qubit tomography (1200 measurement results per data point), and corrected for known tomography errors (SSRO, blue), and post-selected for Charge-Resonance (CR) check validation (purple), and post-selected for latencies (orange) (\cref{sec:methods}).
\textbf{(c)} Sequence diagram including the interaction CNPU-QNPU-QDevice for one execution of the DQC circuit of \ref{fig:fig3}a on QNodeOS (repeated $k=1200$ times in each experiment) (time flows to the right; not to scale). CNPUs prepare NetQASM subroutines (C1, S1, S2), and send them to their respective QNPUs. CNPUs also do classical computation (computing $\delta$) and communication (message containing $\delta$). QNPUs execute subroutines, sending physical instructions to their QDevices. Entanglement is generated by QDevices doing a batch of attempts, resulting in the heralding of a two-qubit entangled state (Bell pair) rotated to $\ket{\Phi^+}$ by the server.
\textbf{(d)} Processing times and latencies while server qubit is live (time frame red line 3c, averaged over all 7200 circuit executions except executions with latency spikes, see~\cref{sec:methods}), including CNPU-QNPU communication latencies, CNPU processing on both nodes and client-server communication latency (CC) (average total of $\sim 4.8 (\pm 0.8)$ ms, error bars for the sum of individual segments (variance per segment in~\cref{sec:processing_time_latencies}).}
\label{fig:fig3}
\end{figure*}

\section{Demonstrations}
\paragraph{Delegated Computation}
We first validate our architecture and implementation by the first successful execution of an arbitrary – i.e. not preloaded – execution of a quantum network application in high-level software on quantum processors. We implement QNodeOS on a two-node setup of NV centers using one qubit per node (\cref{fig:fig2}, \cref{sec:methods}). We choose to execute an elementary form of delegated quantum computation (DQC)~\cite{broadbent_2009_ubqc} from a client to a server, because the client and server programs jointly realize repetitions of a circuit (\cref{fig:fig3}a) that triggers all parts of our system (\cref{fig:fig3}c). 
We first verify that the quantum result (fidelity) was found to be above the classical bound~\cite{massar_optimal_1995} $> 2/3$, which verifies that QNodeOS can successfully handle interactive applications consisting of entanglement generation, millisecond-scale memory lifetimes, and classical message passing. The non-perfect fidelity (\cref{fig:fig3}b) comes mainly from two sources: a noisy entangled state with fidelity 0.72(2) (quantum hardware limitation), and decoherence in the server qubit (depending on $T_{\text{coh}}$) due to waiting for several milliseconds (classical software latencies, \cref{fig:fig3}d).
We proceed to characterize latencies. As expected, we find that the duration that the server qubit must remain alive is dominated (> 50\%) by processing in the CNPU, which could be improved by caching the preparation of S2, and implementing the CNPU and QNPU on one board (Outlook). We observe that CNPU processing time varies significantly (standard deviation 30\%, \cref{sec:processing_time_latencies}), due to limited scheduling control over CNPU processes (\cref{sec:methods}).  Using an a priori estimate of what delays lead to too low a quality of execution (i.e. delays that are too long for the server qubit to be stored with sufficiently high quality), we discard application iterations in which the CNPU latencies spiked by more than 8.95 ms. This lead the discarding of 2\% of iterations in post-processing (\cref{sec:methods}).

\begin{figure*}[htbp]
\centering
\includegraphics[width=0.95\linewidth]{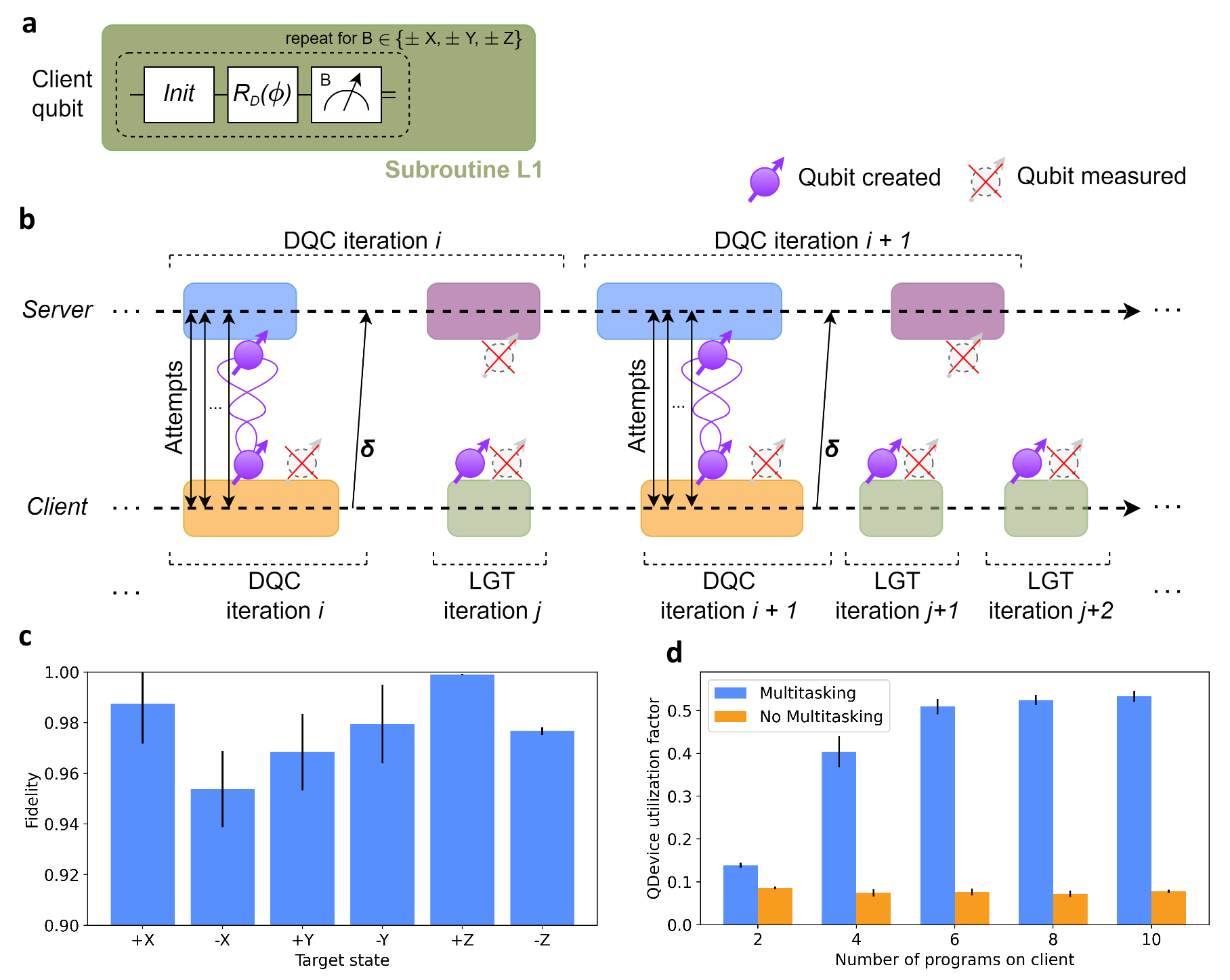}
\caption{\textbf{Multitasking experiment on two NV centers with QNodeOS}.
\textbf{(a)} Local Gate Tomography (LGT) Circuit. A single NetQASM subroutine (L1) executes the following 6 times for different bases $B \in \{\pm X, \pm Y, \pm Z\}$: initialize qubit to $\ket{0}$, rotate around fixed axis $D \in \{X,Y\}$ by angle $\ket{\phi}$, measure in $B$. The LGT application consists of a single LGT program, which submits subroutine L1 for execution to the QNPU (fixed $D$ and $\phi$) $k$ times in succession.
\textbf{(b)} Example sequence diagram illustrating concurrent execution (multitasking) of the DQC application (\cref{fig:fig3}) and the LGT program on the client: time slice in which two DQC circuit repetitions (\cref{fig:fig3}a) are realized (2 subroutines on the client (orange), 4 on the server (blue and purple)), and three LGT circuit repetitions (3 subroutines, green). The client QNPU receives subroutines for both the DQC program and the LGT program, which the QNPU scheduler can interleave: While the server executes S2 (purple), the client cannot yet execute the next S1 (orange) since it involves joint entanglement generation. In this idle time, the client can execute a number of LGT subroutines (number can vary).
\textbf{(c)} Results of multitasking LGT (client) and DQC (on both server and client). For each input pair $(D, \phi) \in \{ (X,0), (X,\pi), (Y,pi/2), (Y,-\pi/2), (X,-\pi/2), (X,\pi/2) \}$ (6 cardinal states $\{\pm X, \pm Y, \pm Z\}$), the following experiment was performed: simultaneously (1) a single LGT program was initiated on the client ($k=1000$), (2) a single DQC-client program was initiated on the client ($k=200$ successive subroutines), and (3) a single DQC-server program was initiated at the server ($k=200$, i.e. 400 successive subroutines). This resulted in a total of 6000 LGT subroutine executions and 36000 LGT measurement results, yielding plotted fidelity estimates for the LGT quantum state before measurement. Results are the same as running LGT on its own (no multitasking with DQC), as expected (\cref{sec:multitasking-tomography}).
\textbf{(d)} Scaling number of programs on the client. For $N \in \{1,2,3,4,5\}$, we initiate at the same time: (1) $N$ LGT programs (each using $k=100$) on the client, (2) N DQC-client programs on the client (each using $k=60$), and (3) $N$ DQC-server programs on the server (each using $k=60$). This results in $2N$ programs active at the same time on the client, each continuously submitting subroutines from the CNPU to the QNPU, where the QNPU scheduler chooses which process to execute when. Each experiment was repeated but with multitasking disabled on the client. Plot shows the utilization factor of the QDevice (fraction of time spent executing instructions), corrected for variable entanglement generation duration (\cref{sec:methods}), with (blue) and without (orange) multitasking, showing that multitasking can increase device utilization.}
\label{fig:fig4}
\end{figure*}

\begin{figure}[htb]
\centering
\includegraphics[width=1\linewidth]{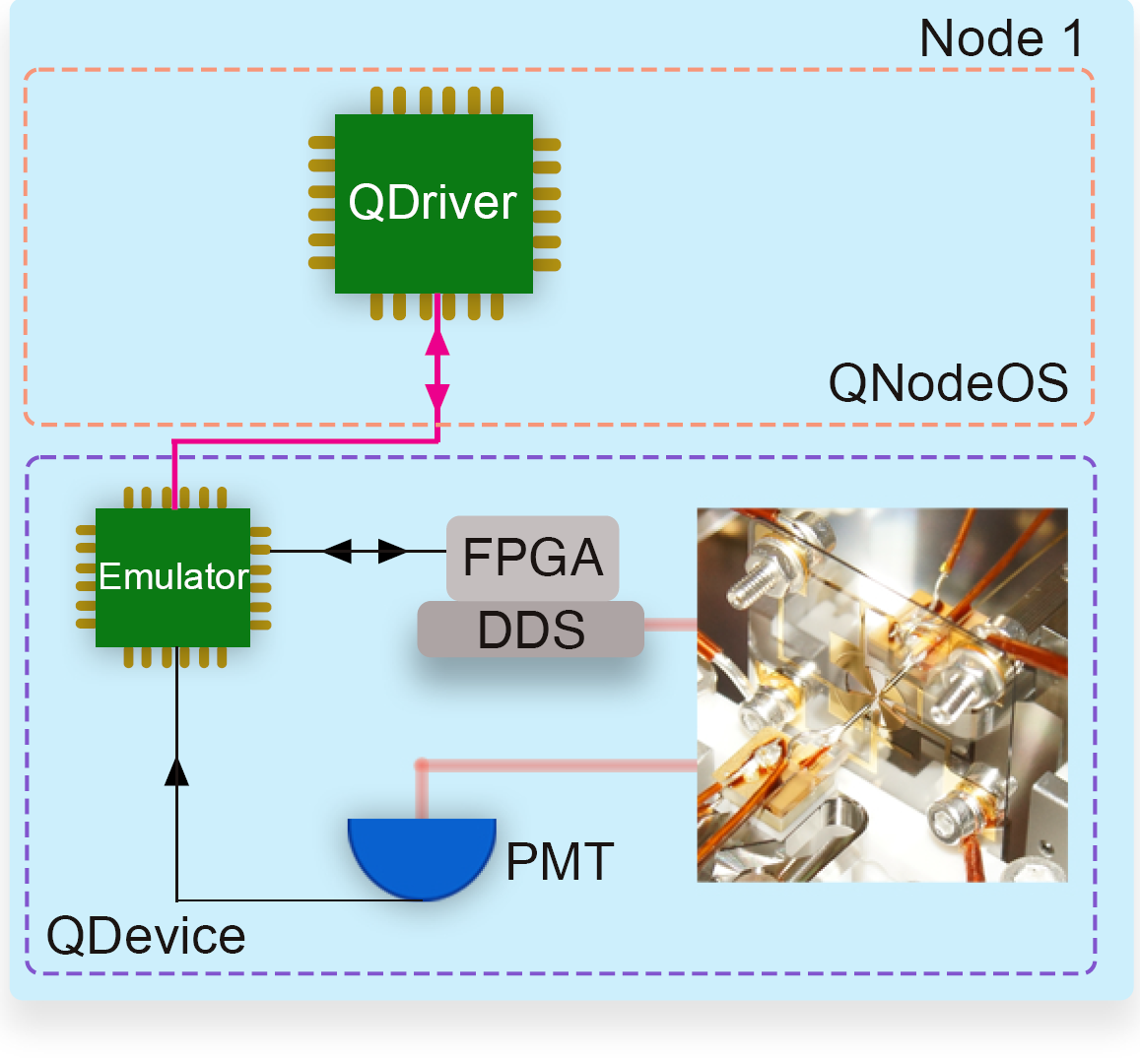}
\caption{\textbf{Trapped-ion QDevice implementation.} Schematic of our implementation of QNodeOS on a single-node setup in which the QDevice contains a single trapped-ion qubit. The QNPU QDriver is implemented on a field-programmable gate array (FPGA) that connects to its QDevice via a serial peripheral interface (SPI) (\cref{sec:methods}). The setup consists of an emulator that translates between SPI messages and TTL signals, experimental control hardware that includes an FPGA and direct digital synthesis (DDS) modules, a trapped-ion qubit~\cite{teller2023integrating} under ultra-high vacuum (\cref{fig:fig1}), and a photomultiplier tube (PMT) that registers atomic fluorescence.}
\label{fig:fig5}
\end{figure}

\paragraph{Demonstration of Multitasking}
We also validate QNodeOS's multitasking capability by the first concurrent execution of two quantum applications on a quantum network: the DQC application, and a single-node local gate tomography (LGT) application on the client (\cref{fig:fig4}a). The two programs for the client are started in the CNPU at the same time (two CNPU processes, subject to CNPU scheduler), which means that the QNPU continuously receives subroutines for both programs from the CNPU (two QNPU processes and corresponding subroutines, subject to QNPU scheduler). This leads to a multitasking challenge directly on the QNPU to schedule the different subroutines received (\cref{fig:fig4}b). Since the client has only one qubit, the multitasking of DQC and LGT never results in both programs having a quantum state alive on the client; therefore, multitasking should not affect the fidelity of LGT. We observe interleaved execution of DQC quantum blocks and LGT quantum blocks on the client node (\cref{fig:fig4}b). The LGT application produces a quantum result (fidelity, \cref{fig:fig4}c) equal to that in the scenario where we run LGT on its own (not interleaved by DQC circuit executions), as expected.

We further test multitasking by scaling up the number of programs executed concurrently, up to 5 DQC and 5 LGT programs running on the client at the same time. The interleaved execution of subroutines of different programs increases device utilization (fraction of time spent on executing physical instructions) on the client QDevice compared to the same scenario but with multitasking disabled (\cref{fig:fig4}d). As expected, we observe that LGT subroutines were scheduled to be executed in between DQC subroutines, resulting in lower client QDevice idle time. When multitasking 1 DQC and 1 LGT program, we observe 1 or 2 subroutines in between DQC iterations in most cases (LGT subroutine duration ~2.4 ms, \cref{sec:multitasking-scaling}). We observe cases where both server and client QDevice remain idle, which could be improved in part by smarter CNPU-QNPU scheduling algorithms: (1) both the client and server wait until the start of the next network schedule time-bin (time-bin length 10 ms) (2) the client QNPU finishes a subroutine for user process P, but must wait until the CNPU sends the next subroutine for P (up to 150 ms for 1 DQC and 1 LGT program, but less (up to only 8 ms) when more applications are running, since there are more CNPU processes independently submitting subroutines), (3) the client is ready to perform entanglement generation for DQC, but the next time-bin starts only at some future time $t$, preventing activation of the network process. The scheduler activates a user process which runs a LGT circuit, which completes at some time $>t$, delaying the start of the DQC network process, even though the server node was ready at $t$.

\section{Outlook}
We designed and implemented the first architecture allowing high-level programming and execution of quantum network applications. To deploy our system onto nodes separated by several kms it would be desirable to merge both the CNPU and the QNPU onto one system board, ideally with mutual access to a shared memory to avoid ms delays in their communication. Such a merge would also allow the definition of a joint classical-quantum executable and processes, opening further doors to reduce latencies by a better scheduling control.

Our work provides a framework for a new domain of computer science research into programming quantum network applications on quantum processors including: novel real-time~\cite{ramamritham_scheduling_1994} scheduling algorithms for classical-quantum processes, compile methods for quantum network applications, or novel programming language concepts including entanglement to make software development even easier, thus advancing the vision to make quantum network technology broadly available.

\section{Methods}
\label{sec:methods}

\paragraph{QDevice Model}

The QDevice includes a physical quantum device, which can initialize and store quantum bits (qubits) which are individually identified by a physical address, apply quantum gates, measure qubits, and create entanglement with QDevices on other nodes (either entangle-and-measure, or entangle-and-keep~\cite{dahlberg_2019_egp}). The QDevice exposes the following interface to QNodeOS (\cref{sec:appendix-qdevice}): number of qubits available, and the supported physical instructions that QNodeOS may send. Physical instructions include qubit initialization, single- and two-qubit gates, measurement, entanglement creation, and a `no-op' for do nothing. Each instruction has a corresponding response (including entanglement success or failure, or a measurement outcome) that the QDevice sends back to QNodeOS.

QNodeOS and the QDevice interact by passing messages back and forth on clock ticks at a fixed rate (100 kHz in our NV implementation, 50 kHz in the trapped-ion implementation). During each tick, at the same time (1) QNodeOS sends physical instruction to QDevice, (2) QDevice can send a response (for a previous instruction). Upon receiving an instruction, the QDevice performs the appropriate (sequence of) operations (e.g. a particular pulse sequence in the AWG). An instruction may take multiple ticks to complete, where the QDevice returns the response (success, fail, outcome) during the first clock tick following completion. The QDevice handles an entanglement instruction by performing (a batch of) entanglement generation attempts~\cite{pompili_2022_experimental} (synchronized by the QDevice with the neighboring node's QDevice). 

\paragraph{QNodeOS Architecture}

QNodeOS consists of two layers: CNPU and QNPU (\cref{fig:fig2}a, \cref{sec:architecture}, Supplementary). Processes on the QNPU are managed by the Process Manager, and executed by the local processor. Executing a user process means executing NetQASM~\cite{dahlberg_2022_netqasm} subroutines (quantum blocks) or that process, which involves running classical instructions (including flow control logic) on the QNPU's local processor, sending entanglement requests to the network stack, and handling local quantum operations by sending physical instructions to the QDriver (\cref{fig:fig2}a). Executing the network process means asking the network stack which request (if any) to handle and sending the appropriate (entanglement generation) instructions to the QDevice. 

A QNPU process can be in the following states (\cref{fig:process-states} in Supplementary for state diagram): idle, ready, running and waiting. A QNPU process is running when the QNPU processor is assigned to it. The network process becomes ready when a network schedule time-bin starts; it becomes waiting when it finished executing and waits for the next time-bin; it is never idle. A user process is ready when there is at least one NetQASM subroutine pending to be executed; it is idle otherwise; it goes into the waiting state when it requests entanglement from the network stack (using NetQASM entanglement instructions~\cite{dahlberg_2022_netqasm}) and is made ready again when the requested entangled qubit(s) are delivered. 

The QNPU scheduler oversees all processes (user and network) on the QNPU, and chooses which ready process is assigned to the QNPU processor. CNPU processes can run concurrently, and their execution (order) is handled by the CNPU scheduler. The QNPU scheduler operates independently and only acts on QNPU processes. CNPU processes can only communicate with their corresponding QNPU processes. Since multiple programs can run concurrently on QNodeOS, the QNPU may have multiple user processes that have subroutines waiting to be executed at the same time. This hence requires scheduling on the QNPU.

Processes allocate qubits through the Quantum Memory Management Unit (QMMU), which manages virtual qubit address spaces for each process, and translates virtual addresses to physical addresses in the QDevice. The QMMU can also transfer ownership of qubits between processes, for example from the network process (having just created an entangled qubit), to a user process that requested this entanglement. The Network Stack uses Entanglement Request (ER) sockets (opened by user programs through QNPU API once execution starts) to represent quantum connections with programs on other nodes. The Entanglement Management Unit (EMU) maintains all ER sockets and makes sure that entangled qubits are moved to the correct process.

\paragraph{NV QDevice Implementation}

The two-node network employed in this work includes the nodes “Bob” (server) and “Charlie” (client) (separated by 3 meters) described in~\cite{pompili_2021_multinode,hermans2022qubit,pompili_2022_experimental}. For the QDevice, we replicated the setup used by~\cite{pompili_2022_experimental}, which mainly consists of: an Adwin-Pro II~\cite{adwin} acting as the main orchestrator of the setup; a series of subordinate devices responsible for qubit control, including laser pulse generators, optical readout circuits and an arbitrary waveform generator (Zurich Instruments HDAWG~\cite{zurich_instruments_hdawg_2019}). The quantum physical device, based on NV centers, counts one qubit for each node. The two QDevices share a common 1 MHz clock for high-level communication and their AWGs are synchronized at sub-nanosecond level for entanglement attempts.

We address the challenge of limited memory lifetimes by employing dynamical decoupling (DD).  While waiting for further physical instructions to be issued, DD sequences are used to preserve the coherence of the electron spin qubit~\cite{de_lange_universal_2010}. DD sequences for NV-centers can prolong the coherence time ($T_{\text{coh}}$) up to hundreds of ms~\cite{hermans2022qubit} or even seconds~\cite{abobeih_2018_one_sec}. In our specific case, we measured $T_{\text{coh}}$=13(2) ms for the server node, corresponding to ~1300 DD pulses. The discrepancy to the state-of-the-art for similar setups is due to several factors. To achieve such long $T_{\text{coh}}$, a thorough investigation of the nuclear spin environment is necessary to avoid unwanted interactions during long DD sequences, resulting in an even more accurate choice of interpulse delay. Other noise sources include unwanted laser fields, the quality of microwave pulses and electrical noise along the microwave line.  

A specific challenge arises at the intersection of extending memory lifetimes using DD, and the need for interactivity: to realize individual physical instructions, many waveforms realizing are uploaded to the Arbitrary Waveform Generator (AWG), where the QDevice decodes instructions sent by QNodeOS into specific preloaded pulse sequences. This results in a waveform table, containing 170 entries. The efficiency of the waveforms is limited by the AWG's waveform granularity that corresponds to steps that are multiples of 6.66 ns, having a direct impact on the $T_{\text{coh}}$. We are able to partially overcome this limitation through the methods described in~\cite{corna_efficient_2021}. Namely, each preloaded waveform, corresponding to one single instruction, has to be uploaded 16 times in order to be executed with sample precision. To not fill up the waveform memory of the device, we apply the methods in~\cite{corna_efficient_2021} only to the DD pulses that are played while the QDevice waits for an instruction from the QNPU, whereas the instructed waveforms (gate/operation + first block of XY8 DD sequence) are padded according to the granularity, if necessary.
The physical instructions supported by our NV QDevice is given in~\cref{sec:qdevice-nv}.

\paragraph{NV QNPU Implementation}

The QNPUs for both nodes are implemented in C++ on top of FreeRTOS~\cite{freertos}, a real-time operating system for microcontrollers. The stack runs on a dedicated MicroZed~\cite{microzed}---an off-the-shelf platform based on the Zynq-7000 SoC, which hosts two ARM Cortex-A9 processing cores, of which only one is used, clocked at 667 MHz. The QNPU was implemented on top of FreeRTOS to avoid re-implementing standard OS primitives like threads and network communication. FreeRTOS provides basic OS abstractions like tasks, inter-task message passing, and the TCP/IP stack. The FreeRTOS kernel---like any other standard OS---cannot however directly manage the quantum resources (qubits, entanglement requests and entangled pairs), and hence its task scheduler cannot take decisions based on such resources. The QNPU scheduler adds these capabilities (\cref{sec:qnpu_impl_scheduler}).

The QNPU connects to peer QNPUs via TCP/IP over a Gigabit Ethernet interface (IEEE 802.3 over full-duplex Cat 5e). The communication goes via two network switches (Netgear JGS524PE, one per node). The two QNPUs are time-synchronized through their respective QDevices (granularity 10 $\mu$s), since these already are synchronized at the $\mu$s-level (common 1Mhz clock).

The QNPU interfaces with the QDevice's ADwin-Pro II through a 12.5 MHz SPI interface, used to exchange 4-byte control messages at a rate of 100 kHz.  

\paragraph{NV CNPU Implementation}

The CNPUs for both nodes are a Python runtime executing on a general-purpose desktop machine (4 Intel 3.20 GHz cores, 32 GB RAM, Ubuntu 18.04). The choice of using a high-level system was made as the communication between distant nodes would ultimately be in the ms-timescales, and this allows for ease of programming the application. The CNPU machine connects to the QNPU via TCP over a Gigabit Ethernet interface (IEEE 802.3 over full-duplex Cat 8, average ping RTT of 0.1 ms), via the same single network switch as mentioned above (one per node), and sends application registration requests and NetQASM subroutines over this interface (10 to 1000 bytes, depending on the length of the subroutine). CNPUs communicate with each other through the same two network switches.

\paragraph{Scheduler Implementation}

We use a single Linux process (Python) for executing programs on the CNPU. CNPU `processes' are realized as threads created within this single Python process. When running multiple programs concurrently, a pool of such threads is used. Scheduling of the Python process and its threads is handled by the Linux OS. Each thread establishes a TCP connection with the QNPU in order to use the QNPU API (including sending subroutines and receiving their results) and executes the classical blocks for its corresponding program. 
Both the CNPU and QNPU maintain processes for running programs. The CNPU scheduler (standard Linux scheduler, see above) schedules CNPU processes, which indirectly controls in which order subroutines from different programs arrive at the QNPU. The QNPU scheduler handles subroutines of the same process priority on a first-come-first-served (FCFS) basis, leading however to executions of QNPU processes not in the order submitted by the CNPU (\cref{sec:multitasking-scaling}).

Using only the CNPU scheduler is not sufficient since (1) we want to avoid millisecond delays needed to communicate scheduling instructions across CPNU and QNPU, (2) user processes need to be scheduled in conjunction with the network process (meeting the challenge of scheduling both local and network operations), which is only running on the QNPU, and (3) QNPU user processes need to be scheduled with respect to each other, (e.g. a user process is waiting after having requested entanglement, allowing another user process to be run; as observed in the multitasking demonstration). 

\paragraph{Sockets and the Network Schedule}
In an ER Socket, one node is a `creator' and the other a `receiver'. As long as an ER socket is open between the nodes, an entanglement request from only the creator suffices for the network stack to handle it in the next corresponding time-bin, i.e. the `receiver' can comply with entanglement generation even if no request has (yet) been made to its network stack.

\paragraph{Trapped-ion Implementation}

The experimental system used for the trapped-ion implementation is discussed in~\cite{teller2023integrating,teller2021heating} and is described in detail in~\cite{teller_measuring_2021}. The implementation itself is described in~\cite{fioretto_towards_2020}. We confine a single \CaPlus ion in a linear Paul trap; the trap is based on a 300 µm thick diamond wafer on which gold electrodes have been sputtered. The ion trap is integrated with an optical microcavity composed of two fiber-based mirrors, but the microcavity is not used here. The physical-layer control infrastructure consists of C++ software; Python scripts; a pulse sequencer that translates Python commands to a hardware description language for a field-programmable gate array (FPGA); and hardware that includes the FPGA, input triggers, direct digital synthesis (DDS) modules, and output logic.

QNodeOS provides physical instructions through a development FPGA board (Texas Instruments, LAUNCHXL2-RM57L75) that uses a serial peripheral interface (SPI). We programmed an additional board (Cypress, CY8CKIT-14376) that translates SPI messages into TTL signals compatible with the input triggers of our experimental hardware.
The implementation consisted of sequences composed of seven physical instructions: initialization, $R_x(\pi)$, $R_y(\pi)$, $R_x(\pi/2)$, $R_y(\pi/2)$, $R_y(-\pi/2)$, and measurement. First, we confirmed that message exchange occurred at the rate of 50 kHz as designed. Next, we confirmed that we could trigger the physical-layer hardware. Finally, we implemented seven different sequences. Each sequence was repeated $10^4$ times, which allowed us to acquire sufficient statistics to confirm that our QDriver results are consistent with operation in the absence of the higher layers of QNodeOS.

\paragraph{Metrics}

Both classical and quantum metrics are relevant in the performance evaluation: The quantum performance of our test programs is measured by the fidelity $F(\rho,\ket{\tau})$ of an experimentally obtained quantum state $\rho$ to a target state $\ket{\tau}$ where $F(\rho,\ket{\tau}) = \bra{\tau}\rho\ket{\tau}$, estimated by quantum tomography~\cite{paris_quantum_2004}. Classical performance metrics include device utilization $T_{\text{util}} = 1 - T_{\text{idle}} / T_{\text{total}}$ where $T_{\text{idle}}$ is the total time that the QDevice is not executing any physical instruction, and $T_{total}$ is the duration of the whole experiment excluding time spent on entanglement attempts (see below).

\paragraph{Experiment Procedure NV Demonstration}

Applications are written in Python using the NetQASM SDK~\cite{dahlberg_2022_netqasm} (code in~\cref{sec:app_source}), with a compiler targeting the NV flavour~\cite{dahlberg_2022_netqasm}, as it includes quantum instructions that can be easily mapped to the physical instructions supported by the NV QDevice. The client and server nodes independently start execution of their programs by invoking a Python script on their own CNPU, which then spawns the threads for each program. During application execution, the CNPUs have background processes running, including QDevice monitoring software.

A fixed network schedule is installed in the two QNPUs, with consecutive time-bins (all assigned to the client-server node pair) with a length of 10 ms (chosen to be equal to 1000 communication cycles between QNodeOS and QDevice as in Ref.~\cite{pompili_2022_experimental}) to assess the performance without introducing a dependence on a changing network schedule.  During execution, the CNPUs and QNPUs record events including their timestamps. After execution, corrections are applied to the results (see below) and event traces are used to compute latencies.

\paragraph{Delegated Quantum Computation}

Our demonstration of DQC (\cref{fig:fig3}) implements the effective single-qubit computation $\ket{\psi} = H \circ R_z(\alpha) \circ \ket{+}$ on the server, as a simple form of blind quantum computing (BQC) that hides the rotation angle $\alpha$ from the server, when executed with randomly chosen $\theta$, and not performing tomography. The remote entanglement protocol utilized is the single-photon protocol~\cite{cabrillo1999creation,bose1999proposal,hermans2023entangling} (\cref{sec:qdevice-nv}).

\paragraph{Filtering}

Results, with no post-selection, are presented including known errors that occur during the tomography single-shot readout (SSRO) process (\cref{fig:fig3}b, blue) (details on the correction Supplementary of~\cite{pompili_2021_multinode}). We also report the post-selected results in which data are filtered based on the outcome of the Charge-Resonance check~\cite{robledo2010control} after one application iteration (\cref{fig:fig3}b, purple). This filter enables the elimination of false events, specifically when the emitter of one of the two nodes is not in the right charge state (ionization) or the optical resonances are not correctly addressed by the laser fields after the execution of one iteration of DQC.

Additional filtering (\cref{fig:fig3}b latency filter) is done on those iterations that showed latency not compatible with the combination of $T_{\text{coh}}$ of the server and the average entangled state fidelity. For this filter, a simulation (using a depolarizing model, based on the measured value $T_{\text{coh}}$, \cref{sec:dqc-simulation}) was used to estimate the single qubit fidelity (given the entanglement fidelity measured above) as a function of the duration the server qubit stays live in memory in a single execution of the DQC circuit (\cref{fig:fig3}a). This gives a conservative upper bound of the duration as 8.95 ms, to obtain a fidelity of at least 0.667. All measurement results corresponding to circuit executions exceeding 8.95 ms duration were discarded (146 out of 7200 data points). 

Other main sources of infidelity, that are not considered in this analysis of the outcome, include, for instance, the non-zero probability of double excitation for the NV center~\cite{hermans2023entangling}. During entanglement generation, the NV center can be re-excited, leading to the emission of two photons that lower the heralded entanglement fidelity. The error can be corrected by discarding those events that registered, in the entanglement time-window, a photon at the heralding station (resonant Zero-Phonon Line photon) and another one locally at the node (off-resonant Phonon-Side Band photon). 

Finally, the dataset presented in~\cref{fig:fig3}b (not shown chronologically) was taken in “one shot” to prove the robustness of the physical layer, therefore no calibration of relevant experimental parameters was performed in between, leading to possible degradation of the overall performance of the NV-based setup.

The single qubit fidelity is calculated with the same methods as in~\cite{iuliano2024qubit}, measuring in the state $\ket{i}$ and in its orthogonal state $\ket{-i}$, provided that we expect the outcome $\ket{i}$, whereas the two-qubit state fidelity is computed taking into account only the same positive-basis correlators (XX, YY, ZZ).

\paragraph{Multitasking: Delegated Computation and Local Gate Tomography}

In the first multitasking evaluation, we concurrently execute two programs on the client: a DQC-client program (interacting with a DQC-server program on the server) and a Local Gate Tomography (LGT) program (on the client only) (\cref{fig:fig4}). The client CNPU runtime executes the threads executing the two different programs concurrently. The client QNPU has two active user processes, each continuously receiving new subroutines from the CNPU, which are scheduled with respect to each other and the network process.

Estimates of the fidelity (\cref{fig:fig4}b) include same corrections as in the Supplementary of~\cite{pompili_2021_multinode} To assess the quantum performance of the LGT application, we used a mocked entanglement generation process on the QDevices (executing entanglement actions without entanglement) to simplify the test: weak-coherent pulses on resonance with the NV transitions, that follow the regular optical path, are employed to trigger the CPLD in the entanglement heralding time-window. This results in comparable application behavior for DQC (comparable rates and latencies, \cref{sec:mocked_entanglement}) with respect to multitasking on QNodeOS.

\paragraph{Multitasking: QDevice Utilization when scaling number of programs}

We scale the number of programs being multitasked (\cref{fig:fig4}d): We observe how the client QNPU scheduler chooses the execution order of the subroutines submitted by the CNPU. DQC subroutines each have an entanglement instruction, causing the corresponding user process to go into the waiting state when executed (waiting for entanglement from the network process). The QNPU scheduler schedules another process [(56\%, 81\%, 99\%) for (N=1, N=2, N>2)] of the times that a DQC process is put into the waiting state (demonstrating that the QNPU schedules independently from the order in which the CNPU submits subroutines). The number of consecutive LGT subroutines (of any LGT process; LGT block execution time ~2.4 ms) that is executed in between DQC subroutines is 0.83 for N=1, increasing for each higher N until 1.65 for N = 5, showing that indeed idle times during DQC are partially filled by LGT blocks (\cref{sec:multitasking-scaling}).

Device utilization (see Metrics above) quantifies only the utilization factor in between entanglement generation time windows to fairly compare the multitasking and the non-multitasking scenario. In both scenarios, the same entanglement generation processes are performed, which hence have the same probabilistic durations in both cases. To avoid inaccurate results due to this probabilistic nature, we exclude the entanglement generation time windows in both cases.

\section{Acknowledgements}
This research was supported by the Quantum Internet Alliance through the European Union's Horizon 2020 program under grant agreement No. 820445 and from the Horizon Europe program grant agreement No. 101080128. S.W. acknowledges support from an NWO VICI grant. M.I., A.R.-P.M. and R.H. also acknowledge funding from the Dutch Research Council (NWO) through the Spinoza Prize 2019 (No. SPI 63-264). This project has also received funding from the European Research Council (ERC) under the European Union’s Horizon 2020 research and innovation programme (grant agreement No. 852410). H.B.O. acknowledges support from the joint research program ``Modular quantum computers'' by Fujitsu Limited and Delft University of Technology, co-funded by the Netherlands Enterprise Agency under project number PPS2007.

\section{Author contributions}
S.W. conceived the project; A.D., M.S. and S.W. developed the initial ideas; C.D.D., B.V., W.K., M.S., I.R., P.P. and S.W. designed the QNodeOS architecture; C.D.D., B.V., W.K., I.R., G.M.F., T.S., H.J., M.S. and P.P. implemented the OS; M.I. prepared the NV setup, with help from A.R.-P.M., J.F., H.B.O and N.D.; C.D.D., M.I., B.V., W.K., I.R., T.E.N., R.H. and S.W. devised the experiments; M.I. and B.V. performed the NV experiments, analyzed the data and discussed the results with all the authors.; D.F., M.T., P.F. prepared the trapped-ion setup; C.D.D., D.F., M.T. performed the trapped-ion integration tests; B.V. analyzed the delegated computation protocol, with the help of D.L., L.M., and H.O.; M.I., B.V., T.E.N and S.W. wrote the main text with input from all authors; C.D.D., M.I., B.V., W.K., P.P., T.E.N. and S.W. wrote the Supplementary Material; T.H.T., P.P., T.E.N., R.H. and S.W. supervised the research; S.W. supervised the collaboration.

\section{Data availability}
The datasets that support this manuscript and the software to analyze them are available at \url{https://doi.org/10.4121/6aa42f05-6823-4848-b235-3ea19e39f4ae}. The application software development kit used for writing program code is open-sourced on GitHub~\cite{netqasm_sdk}. The QNodeOS source code is not currently open source.

\printbibliography

\title{Supplementary Material for ``Design and demonstration of an operating system for executing applications on quantum network nodes''}

\maketitle

\begin{appendices}

\AppendixTOC

\section{Design Considerations and Challenges}
\label{sec:design-consid:challenges}

We start with providing additional information for some of the design considerations and challenges for an operating system for executing applications on a quantum network node.

\subsection{Application Paradigm}
\label{sec:design-consid:challenges:application}

Our architecture is primarily meant to \emph{enable the execution of quantum network applications in the quantum memory stage~\cite{wehner_2018_stages} and above}. That is, applications that require the use of a quantum processor that can manipulate and store quantum bits (qubits). For simpler applications in the prepare-and-measure and entanglement generation stages~\cite{wehner_2018_stages}, e.g. quantum key distribution~\cite{bb84Original,ekert_1991_e91}, where the quantum states are immediately measured by the nodes, our system can also be used, but it would be sufficient to realize a system implementing a quantum network stack and classical processing only.

\paragraph{Separated programs}

Recall that a quantum networking application consists of multiple programs, each running on one of the end nodes, where for ease of explanation we will assume we are executing an application between two nodes, i.e. a client and a server. Each node in the network runs its own independent \emph{\ac{QNodeOS}}, on which the node's program is executed. The two programs may interact with each other via message passing and entanglement generation, where both types of interactions are managed by the node's \ac{QNodeOS}. Next to interaction via the programs, the nodes may exchange additional classical messages which are not part of the program itself, for example, in order to enable the realization of a network stack~\cite{dahlberg_2019_egp} managing entanglement generation between the nodes.

Classical blocks of code consist of instructions for local classical operations and classical message passing. Quantum \emph{blocks of code} consists of
\begin{inlinelist}
\item quantum operations (initialization, quantum gates, measurement), 
\item low-level classical control logic (branching on classical variables and loops), as well as 
\item instructions to make entanglement between remote nodes. 
\end{inlinelist}
We remark that classical and quantum instructions may require many actions by the underlying \ac{QNodeOS} (and quantum system controlled by it) in order to be fulfilled: it is the goal of such instructions to abstract away aspects of the underlying system.

Classical blocks of code may depend on quantum ones via classical variables generated during the quantum execution (such as measurement results, notification of entanglement generation, and information on the state of the quantum system such as the availability of qubits). Similarly, quantum blocks may depend on variables set by the classical blocks (such as messages received from remote network nodes). Finally, quantum blocks may themselves depend on other quantum blocks via qubits in the quantum memory. 

\paragraph{Performance metrics}

Next to classical metrics, such as utility (see `Methods, Metrics`), throughput or latency~\cite{stankiewicz_commag}, the successful execution of quantum network applications is governed by quantum metrics, which are unique to quantum networks and not present in classical networks. Such quantum network-specific metrics include fidelity (see `Methods, Metrics`), or the probability of success in executing an application, where the latter depends directly on the fidelity of the quantum states prepared.

\paragraph{Mode of Execution}

There exist quantum applications and functionalities, where one pair of programs is executed only once, e.g. a simple example of quantum teleportation~\cite{bennett_1993_teleportation}. As in quantum computing, however, some quantum network applications~\cite{wehner_2018_stages} are expected to succeed only with a specific \emph{probability of success} $p_{\rm succ}$ when executed once. The application is then typically executed many times in succession in order to gather statistics (for example to amplify $p_{\rm succ}$). A common use case for executing the same application repeatedly also occurs when evaluating the performance of a system (as we do here), where the goal is to estimate quantum performance metrics, such as the probability of success or the fidelity (see `Methods, Metrics`).
When executing the same application multiple times, the programmer can choose to launch many instances of the same program at once if multitasking is possible (see below), or to write one program which repeatedly executes the node's part of the program, asking for a successive execution of the application.

\subsection{Interactive Classical-Quantum Execution}

Let us elaborate further on the relation, and differences, between the execution of quantum network applications, and the execution of quantum computing applications: One could envision building a system for executing quantum network applications on top of a simpler system for the execution of quantum computing programs, as long as the latter can be \emph{augmented with networking instructions to generate entanglement}: in essence one quantum block can be seen as one quantum computing program. Such a block may realize mid-circuit measurements by the classical control logic allowed within one quantum block, or error correction. Error correction could in this paradigm be realized both by classical control logic allowed within one quantum block, or by considering the error correction itself as part of the \ac{QDevice} (see \cref{sec:appendix-arch-node_system}) which then only exposes logical qubits and operations to \ac{QNodeOS}, instead of physical qubits and operations. In that sense, one may think of the interactivity required between classical and quantum operations as taking place not only at a higher level, but also stemming from the fact that classical messages are used to create a new interaction between \emph{separate} quantum programs, while in quantum computing we have only one single program.

\subsection{Different Hardware Platforms} 

It is desirable that an architecture for executing quantum network applications is as much as possible platform-independent, specifying a specific \ac{HAL} that allows interfacing with different (quantum) platforms. We consider as a quantum processor system the \ac{QDevice} model (see `Methods, QDevice Model`), exposing a set of physical instructions addressing specific qubits (see~\cref{sec:sec:appendix-qdevice-instructions-operands}). These physical instructions may be dependent on the type of quantum hardware, e.g. NV in diamond, or trapped ions, and
\begin{inlinelist}
\item include instructions for initializing and measuring qubits on the chip,
\item moving the state of a qubit to another location in the quantum memory
\item performing quantum gates, as well as
\item to make attempts at entanglement generation at the physical layer~\cite{pompili_2022_experimental}.
\end{inlinelist}

Quantum processors in general offer two types of qubits (see e.g.~\cite{dahlberg_2019_egp}): \emph{communication qubits} which can be used to generate entanglement with remote nodes next to other quantum operations, as well as \emph{storage qubits} which cannot be used to generate entanglement and only for implementing local quantum operations. We remark that on near-term quantum processors, the types of operations also depends on the connectivity of the qubits. That is, not all (pairs of) qubits may allow the same set of quantum operations to be performed on them.

To later enable compile time optimization, it is desirable that quantum hardware furthermore exposes the capabilities of the quantum chip:
\begin{inlinelist}
\item the number of qubits,
\item the type of each qubit, 
\item the memory lifetime of the qubits,
\item the physical instructions that can be performed on on the qubit(s) and
\item the average quality of these instructions.
\end{inlinelist}

\subsection{Timescales}

Quantum network programs are meant to be executed between distant nodes, meaning the communication times between them are in the \emph{millisecond} regime. We remark that the same is \emph{not true for networked or distributed quantum computing }: if the goal is to combine several less powerful quantum processors via a network into one more powerful quantum computing cluster, then it is advisable to place the individual processors as close to each other as possible, in order to minimize the time needed to (1) exchange messages, and (2) generate entanglement between processors. Thus, apart from the execution of applications following a different paradigm (see the main part of the paper, \cref{fig:fig1}),
the case of distributed quantum computing also has different timescales than quantum networking. Of course, it is conceivable that in the future, one may also link distant quantum computers into more powerful quantum computing clusters via quantum internet infrastructure.


\subsection{Scheduling Network Operations}

In order for two neighbouring quantum network nodes to produce heralded entanglement between them, they need to simultaneously perform an action to trigger entanglement generation (at the physical layer, \emph{synchronized to nanosecond precision}). This means neighbouring quantum network nodes need to perform a network operation (entanglement generation) in a \emph{very specific} time slot in which they make an attempt to generate entanglement. Such time slots are generally aggregated into larger time bins, corresponding to making batches of attempts in time slots synchronized at the physical layer. We refer to e.g. Ref.~\cite{pompili_2022_experimental} for background information on the physical layer of entanglement generation in quantum networks, and the readers with a background in computer science to e.g. Ref.~\cite{dahlberg_2019_egp} for a detailed explanation of scheduling of entanglement generation in quantum networks.

In short, network operations in quantum networks need to be executed by the node at very specific time bins. These time bins cannot be determined by the quantum node itself. Instead selection of time bins for a specific quantum operation require agreement with the neighbouring node~\cite{dahlberg_2019_egp} (and more generally with the quantum network when the end-to-end entanglement is made via intermediary network nodes) by means of a network schedule, e.g. determined by a (logically) centralized controller, see Ref.~\cite{skrzypczyk_2021_arch}.

\subsection{Scheduling Local Operations versus Scheduling Network Operations}

For computer scientists, we provide further information on the inability to execute at the same time both local as well as networked quantum operations on present-day quantum processors. At a high-level, present-day quantum processor can be seen as both a quantum \ac{CPU}/memory, as well as network device at the same time. Physical properties of the device and its control at the level of experimental physics, prohibits the usage of the quantum processors for both network and \ac{CPU}/memory functions at the same time. A good example is given by the system of \ac{NV} centers in diamond~\cite{kalb_2017_entanglement,humphreys_2018_delivery}: the communication qubit, i.e. the network device, of the \ac{NV} quantum processor system is given by its electron spin. Further storage qubits may be available by the surrounding nuclear spins in the diamond material. However, such nuclear spins cannot easily be addressed without involving the electron spin, prohibiting their use as a separate processor that is independent from the use as a network device. 

It is conceivable that in the future, two devices could be used~\cite{vardoyan_2022_netarch}: one \emph{quantum processor as a network device} (but not as a device for execution of general quantum gates and measurements), and a another \emph{quantum processor performing only local quantum operations} (but not as a device for long-distance networking). The network device could produce entanglement with distant quantum nodes (which may be taking many \emph{milliseconds} to conclude successfully), and only once such entanglement is ready inject it into the second quantum processor. The latter may still involve short-distance entanglement generation between the network device and the second quantum processor, which however is very fast at short distances. This way the time that the second quantum processor would be blocked by networking operations would diminish significantly.

\subsection{Multitasking}

When executing quantum network applications, multitasking is well motivated in order to increase the utility of the system. Multitasking (or time sharing) is a well-established concept in classical operating systems (see e.g.~\cite[Section 1.4]{silberschatz_book_2014}) that allows the concurrent execution of multiple programs. For the reader from physics, we summarize some of these concepts in order to give context, and then reflect on what these imply in our setting.

In order to allow for multitasking, operating systems typically employ a notion of processes (or threads~\cite[Chapter 4]{silberschatz_book_2014}, or tasks~\cite[Section 3.1]{silberschatz_book_2014}), where a process is created whenever a program starts, and the process forms an instance of the program being executed on the system. Multitasking (time sharing) thus refers to the concurrent execution of multiple processes at once, where it is possible to have multiple processes for the same program, corresponding to the execution of several instances of the program in parallel. We remark that the term concurrent thereby refers to the fact that the processes are existing in the system at the same time, while---due to the fact that they need to share limited resources (e.g. a \ac{CPU} or other devices)---not all of them may be running at the same time.

Allowing multitasking requires the system to include a number of additional features:

\subsubsection{Managing Processes}

At a high-level, multitasking requires the system to keep track of the currently running processes, which means that when program starts executing, a process must be registered in the system. Since the system needs to decide which process can be executed at what time, i.e., which process can be given access to the necessary resources to allow its execution, the system needs to keep track of the state of the process, which typically includes
\begin{inlinelist}
\item whether it is ready for execution, 
\item currently running, or 
\item whether it cannot currently be executed since it is waiting e.g. for other processes.
\end{inlinelist}

In the case of executing quantum network applications, different parts of the application require different resources in order to run: classical blocks need the classical processor (\ac{CPU}) and potentially network device present in a \ac{CNPU}, while quantum blocks require the quantum processor (\ac{QDevice}). It is desirable for our system that both resources can be used concurrently. That is, two different processes should be able to execute a classical block (on the \ac{CNPU}), and a quantum block (on the \ac{QNPU}) at the same time.

\subsubsection{Memory Management Unit}

A program typically relies on the ability to store classical variables (in a classical memory), as well as quantum variables (the state of qubits in a quantum memory). Such variables are stored in a classical and quantum memory device (here, the quantum processor), respectively. In order to allow multiple concurrent processes at the same time, the system needs to keep track of which part of the classical and quantum memory is assigned to which process. This concept is known broadly as memory management~\cite[Section 1.7]{silberschatz_book_2014} in classical operating systems.

In order to allow multitasking of quantum network applications, we thus require a \ac{QMMU} (next to standard ways of performing classical memory management). The \ac{QMMU} is responsible for the following tasks:

\paragraph{Qubit information handling} 

A \ac{QMMU} has knowledge of the physical qubits available on the underlying quantum hardware, and may keep any other information about said qubits, such as the qubit type (communication or storage qubit) and qubit lifetime. Physical qubits thereby refer to both qubits realized at the device level, e.g. in the electron spin states of the \ac{NV} center in diamond, or at a logical level where quantum error correction~\cite{lidar2013quantum} is used to protect the quantum memory, i.e. one logical qubit is created by performing error-correcting using many device level qubits. A \ac{QMMU} should allow such physical qubits to be assigned to different owners, i.e. different processes, or the operating system itself. 

\paragraph{Transfer of qubit ownership}

The \ac{QMMU} may also allow a transfer of ownership of the qubits from one owner to the other, such as for example from a network process which makes entanglement to a user process. 

\paragraph{Quantum memory virtualization}

A \ac{QMMU} may also provide abstractions familiar to classical computing such as a virtual address space, where the applications refer to virtual qubit addresses that are then translated to physical qubit addresses. This virtual address space avoids the situation in which physical qubit addresses must be bound at compile time, particularly limiting when allowing multiple applications to concurrently run on the same node. This would allow the transparent moving of qubits in a quantum memory in the future (for example moving them from a processor to a memory-only device while the process is waiting, e.g., for a message from a remote node). We remark however that the noise in present-day quantum devices means that any such move introduces a significant amount of additional noise to the quantum state that may prevent the successful execution of the application.

\paragraph{Qubit memory lifetime management}

Advanced forms of a \ac{QMMU} may also cater to the limitations of near term quantum devices, by matching memory lifetime requirements specified by the application code to the capabilities of the underlying qubits, as well their topology, i.e., taking into account which two qubits allow two-qubit gates to be performed on them directly. While one cannot measure the decoherence of a qubit during a general program execution on the quantum level, the \ac{QMMU} could also take into account additional information from the classical control system to signal to the application that a qubit has become invalid.

\subsubsection{Scheduler}

When multitasking, we need to decide which process should be executed at what time. This concept is referred to as  scheduling in classical operating systems~\cite[Section 2.4]{tanenbaum_operating_2005},~\cite[Section 3.2]{silberschatz_book_2014}. We first discuss design considerations for scheduling when executing quantum network applications, and then reflect on how scheduling may be realized at different levels of the operating system for the quantum network nodes. 

\paragraph{General considerations}

We first provide three general considerations for completeness, which are not specific to the execution of quantum network applications but apply to all system in which several resources (such as the \ac{QDevice} and a classical \ac{CPU}) can be used (largely) independently of each other:
\begin{enumerate}
\item \emph{Local quantum computation}: in addition to quantum networking, a node's resources must also be reserved for local quantum gates, which are integral parts of quantum network applications.
\item \emph{Multitasking}: for a node to be shared by multiple users, the scheduler should not allocate all the available resources to a single application indefinitely, and instead it should be aware of the presence of multiple applications.
\item \emph{Inter-block dependencies}: quantum and classical processing blocks of an application may depend on results originating from other blocks, and thus cannot be scheduled independently.
\end{enumerate}

\paragraph{Quantum network considerations}

Two specific considerations stand out in the domain of quantum networking:
\begin{enumerate}
\item \emph{Synchronized network schedule}: due to the bilateral nature of entanglement, each node needs to have its quantum networking activity synchronized with its immediate neighbors. This means that while the scheduler at each \ac{QNodeOS} node runs independently of each other, nodes must take into account the network schedule which defines when the node needs to perform networking actions with its neighbouring node. 
\item \emph{Limited memory lifetimes}: the performance of quantum networking applications depends on both classical as well as quantum metrics. Once qubits are initialized, or entanglement has been created, the limited lifetime of present-day quantum memories implies that execution must be completed by a specific time in order to achieve a desired level of quantum performance. 
\end{enumerate}

\paragraph{Quantum/classical performance metrics trade-off}

The best quantum performance is reached when the entire quantum network system (all nodes) are reserved for the execution of one single quantum network application. That is, programs are executed in a serial fashion and no multitasking is performed that could introduce delays which negatively impact the quantum performance. However, this approach does not in general achieve the best utilization of the system. 

While our implementation makes use of a simple priority based scheduler, we remark that our work opens the door to apply more advanced forms of schedulers in the future. In particular, the fact that execution quality degrades over time suggests using forms of real-time schedulers for quantum network applications (taking inspiration from the extensive work on this topic in classical systems, see e.g. Ref.~\cite{liu_1973_scheduling}).  We remark that a programmer (or compiler) could provide advise on such (soft) deadlines, for example in the form of a lookup table that includes suggestions for deadlines for a desired level of quantum performance, based on the capabilities provides by the underlying hardware systems (e.g. memory lifetimes, expected execution time of quantum blocks), and the network (e.g. rate and quality (fidelity) of the entanglement that can be delivered). This advise could then be used by the scheduler to inform its scheduling decisions.

We remark that determining precise deadlines (e.g. when too much time has elapsed for the qubits to yield a specific probability of success) is in general a computationally expensive procedure, sometimes estimated in practice by a repeated simulation of the execution. It is an interesting open question to find (heuristic) efficient methods to approximate a performance prediction. We remark that there is no way in quantum mechanics to measure the current quality of a qubit or operation during the ongoing execution, and such qualities are determined by performing estimates independently of the program execution itself. Of course, \ac{QNodeOS} could itself engage in such estimates when idling in order to update its knowledge of the capabilities of the quantum hardware.

To allow for potentially time-consuming classical pre- and post-processing, it is natural to apply such deadlines not for the entirety of the application, but for the period between initializing the qubits and terminating the quantum part of the execution. While outside the scope of this work, we remark that this type of scheduling offers to inspire new work in a form of ``quantum soft-real time'' scheduling, where deadlines may occasionally be missed at the expense of reduced application performance (success probability), to maximize the overall (averaged) performance of the system in which
applications are typically executed repeatedly. 

\paragraph{Scheduling at different system levels}

Above we discussed scheduling at the level of processes, corresponding to executions of program instances. A system may realize scheduling at different levels, including
\begin{enumerate}
\item \emph{Classical versus quantum processes}: The system may sub-divide processes into classical processes (executing classical code blocks), and quantum user processes (executing quantum code blocks). In this case, these can be scheduled independently (provided inter-dependencies are taken into account). 
\item \emph{Scheduling of quantum blocks}: The system may further sub-divide quantum processes into smaller units to allow different quantum code blocks of the same process to be scheduled independently.
\item \emph{Scheduling of individual operations}: The level of operating systems is not typically concerned with the scheduling of individual operations, which is instead taken care of by the underlying \ac{CPU}. We remark that while we do not envision this type of scheduling to be part of such a system in the future, but rather be relegated to control hardware in a microarchitecture for quantum nodes as e.g. in 
\textcite{fu_2017_microarch}, our current realization of \ac{QNodeOS} achieves a simple form of instruction schedule by populating an instruction queue in software due to the absence of a suitable low-level microarchitecture.
\end{enumerate}
\clearpage
\section{QNodeOS Design and Implementation}
\label{sec:design}

We proceed with a more detailed description of the \ac{QNodeOS} architecture and implementation, where (for the reader's convenience) we include some information already found in the main part of the paper. Recall, that a quantum network application is realized by running separate programs, one on each end node of the quantum network that takes part in the quantum application. The individual programs interact with each other only via classical message passing and entanglement generation. Each program itself consists of classical and quantum blocks of code (see~\cref{sec:design-consid:challenges:application}) which require execution in the quantum memory for the application to succeed.

\subsection{QNodeOS Architecture}
\label{sec:appendix-arch}

\subsubsection{Quantum Network Node System}
\label{sec:appendix-arch-node_system}

Within a quantum network, one can distinguish between two main types of node: \emph{end nodes}, which are used by the users to execute quantum network applications, and \emph{intermediate nodes}, that perform routines necessary to connect two or more end nodes. We refer the reader with a background in computer science to~\cite{vanMeter_book} for a gentle introduction to quantum networks. 

We remark that \ac{QNodeOS}---a real-time system for quantum network nodes---is designed to be deployed on end and intermediary nodes, where \ac{QNodeOS} use on intermediary nodes can be restricted to facilitate entanglement generation over the network via a (series) of intermediate nodes. As the focus of this paper is the execution of quantum network application, we focus here on running \ac{QNodeOS} on end nodes.

In our model, as depicted in~\cref{fig:fig2}a, we divide the functions of a node into three high-level components:

\begin{itemize}
\item a \emph{\ac{CNPU}}, on which classical blocks of code are executed. The \emph{CNPU} is required at end nodes, and requires classical computing hardware (including a classical \ac{CPU}), as well as a classical network device to allow the exchange of messages with the \ac{CNPU} of remote nodes. While quantum networking programs can in principle be developed and compiled outside of the \ac{CNPU}), the \emph{CNPU} may also realize a user environment where quantum networking programs (refer to \cref{sec:design:network-application}) are developed and compiled, and where program results are stored; 
\item a \emph{\ac{QNPU}}, which receives quantum blocks from the \ac{CNPU} and entanglement generation requests from peer nodes, and manages execution on the quantum physical device; 
\item a \emph{\ac{QDevice}}---the quantum physical device---consisting of a quantum processor, a quantum network device, and a quantum memory, where actual quantum computations and communications take place. In present-day quantum hardware implementations, the same device acts as a quantum processor, a network device and a memory.
\end{itemize}

In summary, in our design a quantum network program starts on the general-purpose \ac{OS}, i.e. a \ac{CNPU}, which runs classical code blocks internally, and offloads quantum code blocks to the \ac{QNPU}. The \ac{QNPU} runs the quantum code blocks, relying on the underlying quantum device, i.e., \ac{QDevice}, to execute the actual quantum operations. 

\ac{CNPU} and \ac{QNPU}---while both being capable of performing non-quantum operations---are conceptually separate components, with the main difference being that the \ac{QNPU} is expected to meet real-time requirements (to enable entanglement generation) and perform its arbitration tasks within set deadlines, whereas the \ac{CNPU} does not need to provide such guarantees. This is because the QNPU should adhere to a network schedule which imposes real-time requirements. \ac{CNPU}, \ac{QNPU} and \ac{QDevice} have a classical connection to their counterpart at the remote node, where the \ac{QDevice} also has an additional optical fiber connection to the quantum network to perform quantum operations.

An implementation of the quantum network node could have these three top-level components (\ac{CNPU}, \ac{QNPU} and \ac{QDevice}) deployed on three physically distinct environments, or group some of them on the same chip or board. Furthermore, classical and quantum code blocks can be run on a single system, provided that this system has a connection to the quantum device to execute the actual quantum instructions. However, in the interest of a simpler implementation, where each system has a scoped responsibility, we opted to map classical and quantum blocks onto two distinct environments. Classical blocks are run on a system that features a fully-fledged \ac{OS} (here, Linux), with access to high level programming languages (like C++ and Python) and libraries. Quantum blocks are delegated to the \emph{\ac{QNPU}}, which is a system capable of interpreting quantum code blocks and managing the resources of a quantum device. 

We note that the \ac{QNPU} itself is an entirely classical system that interacts with the quantum hardware (the \ac{QDevice}). At the moment, our implementation of the \ac{QNPU} is fully software, including the instruction processor. In general, the system may be implemented entirely in software running on a classical \ac{CPU}, or parts of its functionality may be implemented in classical hardware, e.g.~\ac{FPGA} (see the description of the trapped-ion platform implementation in~\cref{sec:trapped-ion-platform}) or \ac{ASIC}.

\subsubsection{Quantum Network Programs}
\label{sec:design:network-application}

A quantum networking user program is what a programmer writes on the \ac{CNPU}, in a high-level language, through the use of some \ac{SDK}. Classical code blocks can in principle be programmed in any language yielding an executable suitable to run on the \ac{CNPU}. Fully-classical code blocks---which include local processing and communication with other end nodes---often produce input data for the next quantum code blocks. That is, a classical code block typically precedes a quantum code block whose instructions depend on external data coming from a remote end node. In the future, quantum blocks could include real-time execution constraints, for example, a deadline by which execution should be completed in order to reach a specific application performance while the quantum memory has a limited memory lifetime.

\paragraph{NetQASM} To express quantum code blocks, we make use of \emph{\ac{NetQASM}}~\cite{dahlberg_2022_netqasm} as an instruction set for quantum network programs, which is described in detail in~\cite{dahlberg_2022_netqasm}. 
Before this work, \ac{NetQASM} has only ever been used to execute quantum network programs on simulated quantum network nodes, and has never been realized on hardware to execute quantum network applications.

The instruction set used in \ac{NetQASM} for the quantum code blocks is similar to other \ac{QASM} languages (see e.g. Refs.~\cite{cross_2017_qasm,khammassi_2018_cqasm,fu_2019_eqasm}), but it is extended to include instructions for quantum networking. We emphasize that \ac{NetQASM} is not a strict requirement of \ac{QNodeOS}, and other ways to express quantum code blocks could be used in other implementations. The instruction set of this language should support both computational and networking quantum instructions, as well as simple classical arithmetic and branching instructions to be used for real-time processing on the \ac{QNPU}. It is the compiler's task to transform high-level blocks for the \ac{QNPU} into \ac{NetQASM} blocks. 

\ac{NetQASM} defines a notion of \ac{NetQASM} subroutines, where each subroutine corresponds to a quantum block of code, specified by the compiler or programmer. We therefore use the term \emph{quantum block} to refer to a \emph{NetQASM subroutine} in the remainder of this text. A full list of operands that can appear in a \ac{NetQASM} subroutine is given in~\cite[Appendix B]{dahlberg_2022_netqasm}. \ac{NetQASM} assumed subroutines would be executed on a form of \ac{QNPU} (without specifying an architecture for the \ac{QNPU}), potentially using a form of shared memory with \ac{CNPU}. In the absence of a shared memory, \ac{NetQASM} allowed classical variables inside subroutines to be kept on the \ac{QNPU}, and accessed read-only by the \ac{CNPU} via the \ac{NetQASM} interface (see below). The \ac{CNPU} can also specify classical constants for the use inside subroutines, as part of submitting a subroutine to the \ac{QNPU}.

We use here the \ac{NetQASM} \ac{SDK}~\cite{netqasm_sdk} to write programs, where the \ac{SDK} compiles a quantum network program, written in Python, into a series of classical and quantum code blocks. This \ac{SDK} was previously used to express programs on a simulated quantum network~\cite{squidASM}.

\paragraph{NetQASM Interface}

Our interface between the \ac{CNPU} and the \ac{QNPU} (Section~\ref{sec:QNPU-api}) includes the \ac{NetQASM} interface 
defined in~\cite[Appendix A]{dahlberg_2022_netqasm}. This interface in particular allows the \ac{CNPU} to register a program on the \ac{QNPU}, submit \ac{NetQASM} subroutines, and access the results of said subroutines.

\subsubsection{Program Processing Pipeline}

\paragraph{CNPU Processing}

When a program start execution on the \ac{CNPU}, a new \ac{CNPU} process is created. As we separate the \ac{CNPU} from the \ac{QNPU} in our implementation, it is natural to rely on the properties of an existing classical operating system to take care of this function. In our implementation, we start a single program on the \ac{CNPU} which then creates a thread (using standard Linux thread library~\cite{linux_threads} for each \ac{CNPU} process. The classical blocks belonging to the \ac{CNPU} program are executed locally on the \ac{CNPU}. These may involve some form of coordination with the remote \ac{CNPU} of the user program, as well as pre- or post-processing of the results coming from \ac{NetQASM} subroutines. While this can also be done later, when the program starts it will typically also establish a TCP/IP connection with the program running on the remote \ac{CNPU} leading to the establishment of a TCP/IP socket that will be used for classical application level communication between the \acp{CNPU}.

The \ac{CNPU} then registers the program on the \ac{QNPU}. Later, \ac{NetQASM} subroutines of these programs are sent from the \ac{CNPU} to the \ac{QNPU} through the \emph{\ac{NetQASM} interface}. 

\paragraph{QNPU Processing}

When a program is registered with the \ac{QNPU} by the \ac{CNPU}, the \ac{QNPU} creates a user process to store program data and execution state. The \ac{QNPU} also keeps track of \ac{NetQASM} subroutines belonging to the user process, which may be submitted only later, as well as other run-time data analogous to what a typical process control block contains, useful for the execution of the program. As depicted in~\cref{fig:fig2}a, a subroutine can, in general, be composed of three classes of instructions:

\begin{itemize}
\item \emph{Quantum operations}: quantum physical operations, to be performed on the underlying quantum device;
\item \emph{Classical logic}: arithmetic and branch instructions, to be executed in-between quantum operations, useful to store results of quantum operations and to perform responsive decision-making;
\item \emph{Entanglement requests}: requests to generate an entangled qubit pair with a remote node in the network.
\end{itemize}
Classical logic is processed locally on the \ac{QNPU}, and potentially results in the update of a process's data. This data includes \ac{NetQASM} variables capturing measurement results, for example, that may latter be conveyed to the \ac{CNPU}. 

When the user process starts on the \ac{QNPU} an \ac{ER} socket (see~\cref{sec:design_er_socket}) is established with the remote \ac{QNPU} that is used to associate later entanglement requests with the specific user process. Entanglement requests contained in the \ac{NetQASM} subroutines are forwarded to the quantum network stack,  which stores them together with other requests coming from network peers. Entanglement generation requests coming from other nodes in the network are received on the quantum network stack through the \emph{\ac{QNetStack} interface}.

Quantum instructions are sent to the \ac{QDevice} through the \ac{QDriver}, which provides an abstraction of the \ac{QDevice} interface. The \ac{QDriver} translates \ac{NetQASM} instructions into physical instructions suitable to the underlying physical platform.

\paragraph{QDevice Processing}

Physical instructions are executed on the \ac{QDevice}, the quantum processing and networking unit. The \ac{QDevice} processing stack heavily depends on the underlying physical platform---for instance, \ac{NV} centers in diamond, or Trapped Ions.

As we remarked in~\cref{sec:appendix-arch-node_system}, a \ac{QDevice} has two communication channels with its direct neighbors: a \emph{classical channel}, used for low-level synchronization of the entanglement generation procedure and other configuration routines, and a \emph{quantum channel}, typically an optical fiber, through which qubits can travel.

\subsection{QNPU Stack}
\label{sec:design:qnpu_stack}

\begin{figure}
\begin{center}
\includegraphics[width=\linewidth]{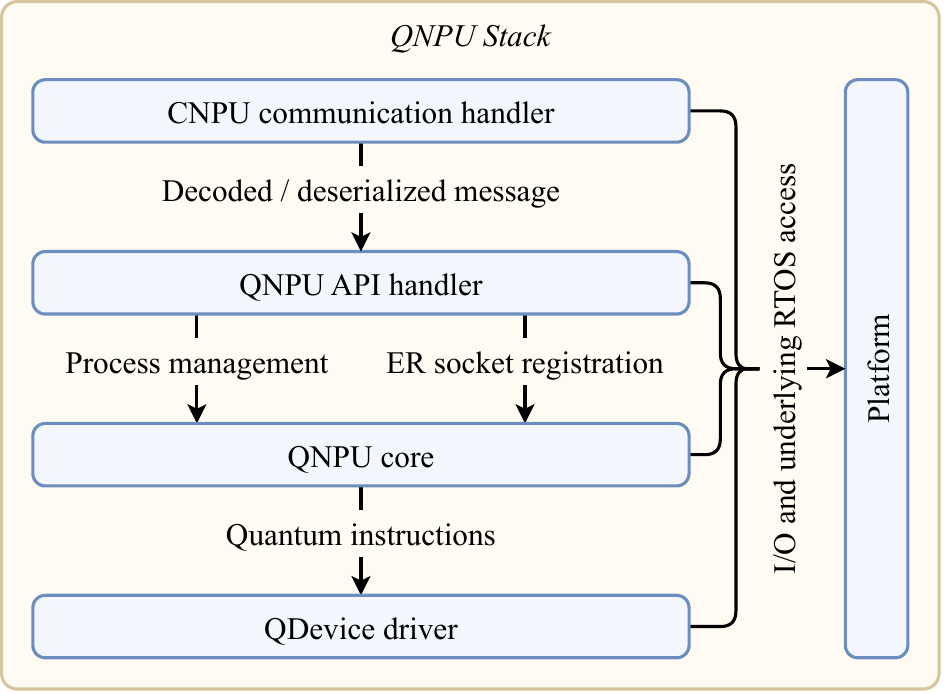}
\end{center}
\caption[]{\ac{QNPU} stack. The \emph{\ac{QNPU} \acs{API} handler} and \emph{the \ac{QNPU} core} form the central processing layers, and are independent of the underlying quantum physical platform and of the device where \ac{QNodeOS} runs. The \emph{\ac{CNPU} communication handler} translates protocol-specific messages from the \ac{CNPU} into \acs{API} calls. The \emph{\ac{QDevice} driver} (or \ac{QDriver}) abstracts the \ac{QDevice} hardware. The \emph{Platform} layer abstracts the hardware where \ac{QNodeOS} runs, and is accessible to all other layers. Note that three other \ac{API} types are implemented, i.e. \emph{control}, \emph{management}, and \emph{operations}. Control \ac{API} is used for the network schedule, while management and operations \ac{API} are for operational purposes. }
\label{fig:qnodeos-stack}
\end{figure}

\ac{QNodeOS} is a system consisting of multiple abstraction layers, as depicted in \cref{fig:qnodeos-stack}. It is designed to be platform-independent, i.e., independent of the underlying quantum physical platform (quantum hardware) controlled by \ac{QNodeOS}, where connections to different realizations of \ac{QDevice} are captured by a platform-dependent \ac{QDriver}. The implementation of \ac{QNodeOS} itself is of course dependent on the classical physical platform(s) on which \ac{QNodeOS} is implemented, including the physical interface between the \ac{CNPU} and \ac{QNPU}.

\subsubsection{QNPU API}
\label{sec:QNPU-api}

At the center of the stack lie the \emph{the \ac{QNPU} \ac{API} handler} layer and the \emph{the \ac{QNPU} core} layer. The \ac{API} handler is responsible for listening to system calls made to the the \ac{QNPU} \ac{API}, and to relay these calls to the appropriate component inside of the core layer. Such system calls may originate from the \ac{CNPU} via the \emph{CNPU communication handler}, see again~\cref{fig:qnodeos-stack}. 

The \ac{QNPU} \ac{API} is the central engine for managing the execution of local quantum operations and entanglement requests, and manages the hardware resources of the \ac{QDevice}. The \ac{QNPU} \ac{API} exposes services to:
\begin{itemize}
\item \emph{Register and deregister a program on the \ac{QNPU}}; This is part of the \ac{NetQASM} interface (see~\cref{sec:design:network-application}).
\item \emph{Add a quantum block (subroutine) for a user process}; This is again part of the \ac{NetQASM} interface.
\item \emph{Open an \ac{ER} socket with a remote node}.
\item \emph{Control to configure the quantum network stack}, i.e., to configure the network schedule; This is used for the interaction with a network controller that sets network-wide entanglement schedules, as presented in Ref.~\cite{skrzypczyk_2021_arch}.
\item \emph{Perform management and operations functions.}
\end{itemize}
The topmost horizontal layer is the \emph{\ac{CNPU} communication handler}, which implements a \emph{protocol wrapper} around \ac{NetQASM}.
We implement this wrapper protocol using EmbeddedRPC~\cite{erpc} for the on-the-wire definition of the messages (including (de-)serialization)). The communication handler translates protocol-specific messages into \ac{API} calls for the \ac{QNPU}. EmbeddedRPC allows to decouple the interface definition and (de-)serialization from the underlying transport layer. We note that only the transport layer is implementation-specific, which depends on the devices where \ac{CNPU} and \ac{QNPU} are implemented and on what the physical interface between them looks like.\footnote{TCP/IP for now, shared memory in the future.}

The \emph{\ac{QDevice} driver} (\ac{QDriver}) layer, at the bottom of the stack, provides an abstraction of the \ac{QDevice} hardware, and its implementation depends on the nature of the \ac{QDevice} itself, and on the physical communication interface between \ac{QNPU} and \ac{QDevice}. Two \ac{QDevice} implementations may differ in a variety of factors, including what quantum physical platform they feature and what digital controller interfaces with the \ac{QNPU}.

Lastly, the vertical \emph{Platform} layer provides \ac{SoC}-specific abstractions for the \ac{QNPU} to access the physical resources of the platform it is implemented on, including I/O peripherals, interrupts controllers and timers. Additionally, if the \ac{QNPU} is implemented on top of a lower-level operating system, this layer gives access to system calls to the underlying \ac{OS}. The Platform layer is vertical to indicate that it can be accessed by all other \ac{QNPU} layers.

Porting the \ac{QNPU} to a different \ac{SoC} (or similar hardware) boils down to implementing a new platform layer. Deploying the \ac{QNPU} on a different \ac{QDevice}, instead, requires both a new \ac{QDriver} and a compiler---on the \ac{CNPU}---that emits quantum instructions supported by the specific \ac{QDevice}.

\subsection{Processes}
\label{sec:design:processes}

A quantum network program starts on the \ac{CNPU}---there, the \ac{CNPU} environment compiles it into classical and quantum code blocks, and creates a new process associated with the program. In the future, an optimized compilation ahead of execution could produce an executable that includes further information (such as execution deadlines depending on the device's memory lifetimes, as mentioned at the beginning of~\cref{sec:design:network-application}). The \ac{CNPU} then registers the program with the \ac{QNPU} (through the \ac{QNPU}'s end node \ac{API}, see \cref{sec:design:qnpu_stack}), which, in turn, creates its own process associated with the registered program. The process on the \ac{CNPU} is a standard \ac{OS} process, which executes the classical code blocks and interacts, (that is: communicating NetQASM subroutines and their results between \ac{CNPU} and \ac{QNPU}), with the counterpart process on the \ac{QNPU}. This interaction can be done by means of a shared memory (and when no shared memory is physically realized: by an exchange of messages~\cite{dahlberg_2022_netqasm}). On the \ac{QNPU}, a process encapsulates the execution of quantum code blocks of a program with associated context information, such as process owner, process number (ID), process state, and process priority.

In the near-term test applications we execute, the execution time of a program is typically dominated by that of quantum blocks, as entanglement generation is a time-consuming operation. Without advanced quantum repeaters~\cite{RepeaterSurvey}, its duration grows exponentially with the distance between the nodes. For this reason, we focus on the scheduling of quantum blocks only, and thus we only discuss \ac{QNPU} processes (also referred to as \emph{user processes}) from this point onward. Again, this does not exclude that in a future iteration of the design \ac{CNPU} and \ac{QNPU} could be merged into one system, and therefore classical and quantum blocks would be scheduled jointly.

\subsubsection{Process Types and Their Interaction}

\paragraph{QNPU user processes}

The \ac{QNPU} allocates a new \emph{user process} to each quantum network program registered by the \ac{CNPU}. A user process is the program's execution context, and consists of \ac{NetQASM} blocks and other context information---the process control block---including process number (ID), process owner, process state, process scheduling priority, program counter, and pointers to process data structures. Process state and priority determine how processes are scheduled on the \ac{QNPU}. A user process becomes active (ready to be scheduled) as soon as the \ac{QNPU} receives a quantum code block from the \ac{CNPU}. Multiple user processes---relative to different \ac{CNPU} programs---can be concurrently active on the \ac{QNPU}, but only one can be running at any time. A running user process executes its quantum code block directly, except for entanglement requests, which are instead submitted to the quantum network stack and executed asynchronously.

\paragraph{QNodeOS network process}

The \ac{QNPU} also defines \emph{kernel processes}, which are similar to user processes, but are created when the system starts (on boot) and have different priority values. Currently, the only existing kernel process is the \emph{network process}. The network process, owned by the \ac{QNetStack}, handles entanglement requests submitted by user processes, coordinates entanglement generation with the rest of the network, and eventually returns entangled qubits to user processes. The activation of the network process is dictated by a network-wide entanglement generation schedule. Such a schedule defines when a particular entanglement generation request can be processed, and therefore it has intersecting entries on adjacent nodes (given that entanglement is a two-party process). The schedule can be computed by a centralized network controller~\cite{skrzypczyk_2021_arch} or by a distributed protocol~\cite{dahlberg_2019_egp}. In our design, the network process follows a \emph{time division multiple access schedule}, computed by a centralized network controller (as originally proposed by \textcite{skrzypczyk_2021_arch}) and installed on each \ac{QNodeOS} node (see~\cref{sec:arch-networking}).

\begin{figure}
\begin{center}
\includegraphics[width=\linewidth]{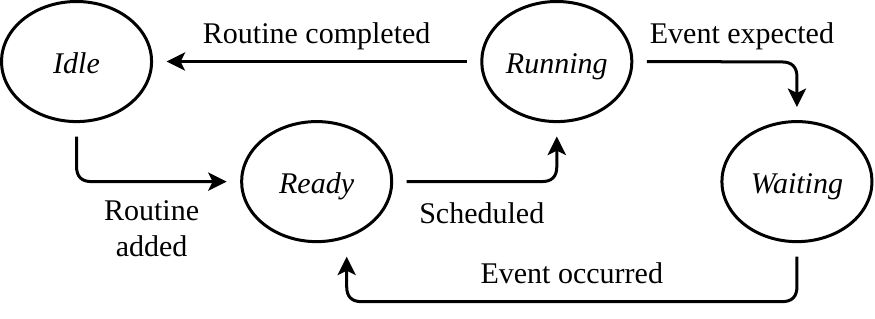}
\end{center}
\caption[]{Process state diagram. An \emph{idle} process becomes \emph{ready} when a block for that process is loaded onto the \ac{QNPU} (from the \ac{CNPU}). A ready process becomes \emph{running} when it is scheduled. A running process goes back to idle if all blocks are completed, or transitions to \emph{waiting} if it expects an event to occur before it can proceed. A waiting process becomes ready again when the expected event occurs.}
\label{fig:process-states}
\end{figure}

\paragraph{QNPU process states}

A \ac{QNPU} process can be in any of the following states:
\begin{inlinelist}
\item \emph{Idle}: when the \ac{CNPU} has registered a program and the \ac{QNPU} has spawned a process, but it has not received a block yet;
\item \emph{Ready}: when it has (at least) one block, sent from the \ac{CNPU}, and can be scheduled and run;
\item \emph{Running}: when it is running on the \ac{QNPU} and has the quantum processor and the quantum network device assigned to it;
\item \emph{Waiting}: when it is waiting for some event to occur.
\end{inlinelist}
\Cref{fig:process-states} shows the possible process states and the valid state transitions. A process transitions from idle to ready when one block gets added. A ready process transitions to running when the the \ac{QNPU} scheduler assigns it to the processor. A running process transitions to waiting when it has to wait for an event to occur, and transitions from waiting to ready when the event occurs---for instance, a process could be waiting for an \ac{EPR} pair to be generated, and become ready again when the pair is established. Finally, a process goes back to the idle state when all its blocks have been completed.

\paragraph{Inter-process communication}

At the moment, the \ac{QNPU} does not allow for any explicit inter-process communication. The only indirect primitive available to processes to interact with one another is \emph{qubit ownership transfer}, used when a process produces a qubit state which is to be consumed by another process. Most notably, the quantum network stack kernel process transfers ownership of the entangled qubits that it produces to the process which requested the \ac{EPR} pairs.

\paragraph{Process concurrency}

The strict separation between local quantum processing and quantum networking is a key design decision in \ac{QNodeOS}, as it helps us address the scheduling challenge, see~\cref{sec:design-consid:challenges}. A user process can continue executing local instructions even after it has requested entanglement. Conversely, networking instructions can execute asynchronously of local quantum instructions. This is important in a quantum network, since entanglement generation must be synchronized with the neighboring node (and possibly the rest of the network~\cite{skrzypczyk_2021_arch}). Additionally, separating user programs into user processes also allows \ac{QNodeOS} to schedule several programs \emph{concurrently}.

\begin{figure}
\centering
\includegraphics[width=\linewidth]{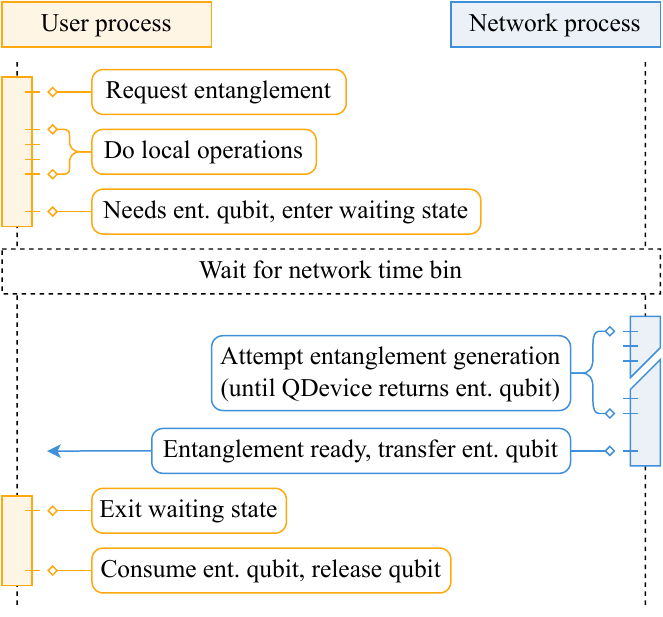}
\caption{Flow of execution between a user process requesting entanglement and the network process responsible for generating entanglement. The user process starts by asynchronously issuing an entanglement request. Once issued, it is free to continue with other local operations or classical processing. Once it reaches a point in its execution where entanglement is required the process enters the waiting state. The network process is scheduled once the appropriate time bin (as determined by the network schedule) starts. Once running, it attempts entanglement generation until entanglement success (or until a set timeout). The entangled qubit is then transferred to the user process. This unblocks the process which consumes the entanglement and releases the qubit. In our experiments, the process always immediately waits after requesting entanglement (no local operations are done in between).}
\label{fig:process-flow}
\end{figure}

\paragraph{Process flow}

\cref{fig:process-flow} illustrates the typical control flow between a user process and the network process. User processes are free to execute any non-networked instructions independently of the network process and other user processes. Once the program reaches a point in its execution where an entangled qubit is required, the process enters the waiting state and is flagged as waiting for entanglement. When the network process is scheduled, it issues network instructions and generates entanglement as requested by the user process. Once an entangled pair is generated by the network process, the qubit is handed over to the waiting user process. When all the entangled pairs that the user process was waiting for are delivered, the user process becomes ready and can start running again.

\subsubsection{Process Scheduling}
\label{sec:design:scheduling}

At present, the the \ac{QNPU} scheduler does not give any guarantees on when a process is scheduled---for that, one would need to define concrete real-time constraints to feed to the scheduler. Instead, the current version of the \ac{QNPU} implements a best-effort scheduler, which selects processes on the basis of their \emph{priority}, and does not allow preemption. In particular, the network process is assigned the highest priority, and is activated whenever the network schedule specifies entanglement should be made in the next time-bin~\cite{skrzypczyk_2021_arch}. 

As already mentioned, \ac{QNodeOS} defines the concept of \emph{user} processes and \emph{kernel} processes, with the \ac{QNetStack} process being the only kernel process at the moment. User processes are released (i.e., they become ready) asynchronously---when a process block is loaded, or when they leave the waiting state---while the \ac{QNetStack} process is released periodically---at the beginning of each time bin of the network schedule (although the period of time bins can vary). Given that generating an \ac{EPR} pair on a link requires that both nodes attempt entanglement simultaneously, the \ac{QNPU} assigns the \ac{QNetStack} process a priority higher than any user process. This ensures that, at the beginning of each time bin of the network schedule, the \emph{priority-based} process scheduler can assign the \ac{QNetStack} process as soon as the processor is available, and thus a node can start attempting entanglement with its neighbor as soon as possible and minimize wasted attempts on the neighbor node.

\begin{figure}
\begin{center}
\includegraphics[width=\linewidth]{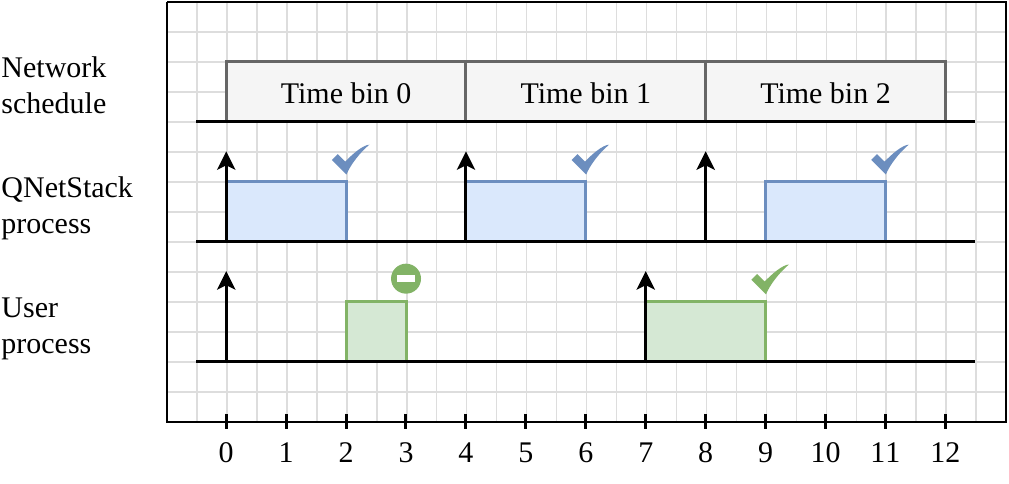}
\end{center}
\caption[]{Snapshot of a hypothetical execution of a user process and the \ac{QNetStack} process. The higher-priority \ac{QNetStack} process is activated at the start of each time bin of the network schedule, and it is assigned to the processor as soon as it is available. The lower-priority user process gives precedence to the \ac{QNetStack} process when they become ready at the same time, but, when it is running on the processor, it is not preempted if the \ac{QNetStack} process becomes ready while the user process is running. Black arrows represents a moment where the process goes into the ready state and the green stop sign (at time $3$) represents a process going into the waiting state.}
\label{fig:scheduling-impl}
\end{figure}

\Cref{fig:scheduling-impl} exemplifies a snapshot of a hypothetical execution of a user process and the \ac{QNetStack} process. The latter is activated at the beginning of a time-bin assigned to networking, and is scheduled as soon as the processor is available---for instance, at times $0$ and $4$ it is scheduled immediately, while at time $8$ it is scheduled after one time unit, as soon as the running process yields. The user process becomes ready at time $0$---at which point the \ac{QNetStack} process is ready as well and has highest priority, meaning the network process is scheduled; then it is scheduled at time $2$, as soon as the \ac{QNetStack} process completes; then it goes into waiting state at time $3$ because the user process requested entanglement and it waits for the entanglement to be established; finally it becomes ready again at time $7$---and it is scheduled immediately given that no other processes are running.

To avoid context switching overhead, potentially leading to degraded fidelity, the \ac{QNPU} scheduler is \emph{cooperative}. That is, once a process is scheduled, it gets to run until it either completes all of its instructions or it blocks waiting for entanglement. Allowing process preemption would need a definition of critical section and could potentially impact the quality of the affected qubit states. Moreover, the lack of a preemption mechanism could potentially result in low-priority user processes hogging the processor at the expense of high-priority entanglement generation attempts. On the other hand, if entanglement instructions always consume the entirety of the time bin, the \ac{QNetStack} process would be immediately assigned the processor each time it relinquishes it, causing low-priority user processes to starve. To at least mitigate the second issue, we made sure that the number of consecutive entanglement attempts performed by the \ac{QDevice} within one single entanglement instruction is always less than how many would fit in a time bin, so as to leave some slack for low-priority user processes to run.

\subsubsection{Networking}
\label{sec:arch-networking}

The network stack \ac{QNetStack} is based on the existing stack~\cite{dahlberg_2019_egp}, including the link layer \ac{QEGP}~\cite{dahlberg_2019_egp}.
However the main difference between the \ac{QNetStack} implemented on the \ac{QNPU} and the original design of the protocols lies in how the \ac{QEGP} processes the outstanding entanglement requests. \ac{QEGP}~\cite{dahlberg_2019_egp} employed the concept of a distributed queue to sort and schedule entanglement requests on one node by coordinating with the counterpart node on the other end of the link, to ensure that both nodes would be servicing the same entanglement request at any given time. This synchronization is necessary because different entanglement requests may require different \ac{EPR} pair fidelities, in which case \ac{QEGP} would issue different \ac{QDevice} entanglement instructions. However, link-local request scheduling becomes more complicated if nodes have more than just one link. In that case, entanglement requests would be better scheduled at a level where network-wide request schedules are known. 

\paragraph{Network Schedule}
The \ac{QEGP} protocol implemented on the \ac{QNPU} transitioned~\cite{pompili_2022_experimental} from scheduling entanglement requests via a pairwise agreed upon distributed queue, to deferring this task to a \emph{logically centralized control plane}, whereby a node's schedule can be computed on the basis of the whole network's needs by a (logically) centralized controller (see e.g.~\cite{skrzypczyk_2021_arch}). This means that the network stack of the nodes convey their demands for end-to-end entanglement generation to the central controller, who then makes a \emph{network schedule}, which is communicated back to the nodes. 

All nodes divide time into time-bins, where the central controller employs a scheduling algorithm to assign either network actions (or no actions) to time-bins. That is, the term network schedule refers to a schedule, i.e. allocation of resources over time, of time-bins at the nodes, where a time-bin may be marked for networking activities (entanglement generation) or be left empty (to be used arbtirarily to execute local operations). 
Given that entanglement generation requires a non-deterministic amount of attempts and time, time bins are computed to be large enough to accommodate the average run time of an entanglement generation instruction. 
We remark that the node functions internally as a higher timing granularity than a time-bin allocated by the network scheduler, that is, it can execute other operations (such as for example local quantum operations) also within a time-bin allocated by the network schedule, provided entanglement is made early.

\begin{figure}
\begin{center}
\includegraphics[width=\linewidth]{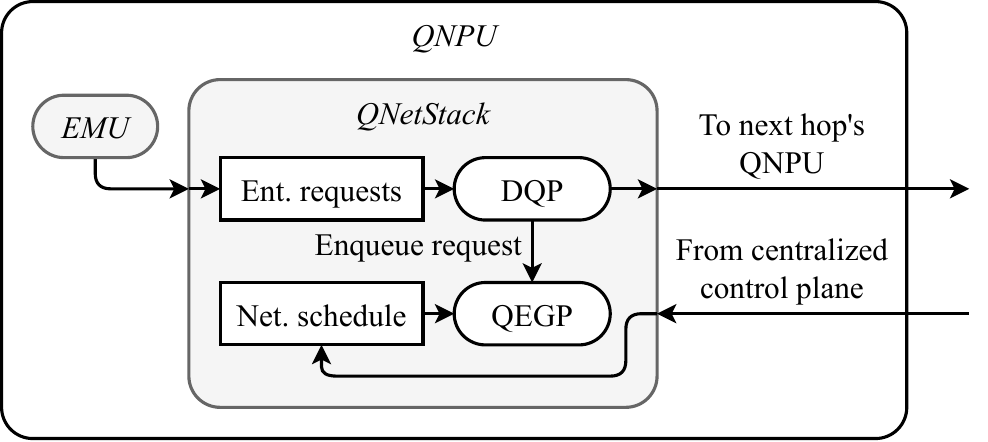}
\end{center}
\caption[]{Internal components and data structures of the \acf{QNetStack}. Entanglement requests are received through the \acf{EMU}, while the network schedule is installed by a centralized control plane. 
\acf{QEGP} maps such requests onto the network schedule to produce the correct entanglement instructions. While not needed on our 2 node implementation, a \ac{DQP} (which is a simplified version of the \ac{DQP} presented in~\cite[Section 5.2.1]{dahlberg_2019_egp}) could forward entanglement requests to the next hop's \acf{QNPU} to realize a network layer protocol such as~\cite{kozlowski_2020_qnp}.}.
\label{fig:qnetstack-impl}
\end{figure}

Once the node received the network schedule from the controller, the network schedule is used to satisfy all outstanding end-to-end entanglement requests, and is used by \ac{QEGP} to produce the correct \ac{QDevice} instructions at any point in time. Whenever a time bin is assigned to networking to two neighboring nodes, the nodes attempt entanglement generation over their shared link in order to realize the \ac{QEGP} link layer protocol. 
\Cref{fig:qnetstack-impl} shows internal components and data structures of the \ac{QNetStack} as it is implemented on the \ac{QNPU}. Entanglement requests received by the \ac{EMU} are forwarded by \ac{QNP} to the next hop's \ac{QNPU} system. Entanglement requests and network schedule---the latter installed by a logically centralized control plane---are used by \ac{QEGP} to produce the correct entanglement instructions to populate the \ac{QNetStack} process's block at each activation of the process.

\paragraph{ER Socket}
\label{sec:design_er_socket}
The concept of an \ac{ER} socket is inspired by that of a classical network socket, in that it defines the endpoint of an entanglement generation request, and is used by the \ac{QNPU}'s quantum network stack to set up network tables and to establish connections with its peers. 
We remark that the current realization of the \ac{ER} Socket (see below) is a proof of concept implementation opening future computer science research, and does not aim to prevent misuse if different users had access to the same node. 
A program can request from QNodeOS the opening of an \ac{ER} socket with a program on a remote node. An \ac{ER} socket is identified by the tuple \texttt{(node\_id, er\_socket\_id, remote\_node\_id, remote\_er\_socket\_id)}. The other program (on the other node) must open its own corresponding \ac{ER} socket (i.e with values \texttt{(remote\_node\_id, remote\_er\_socket\_id, node\_id, er\_socket\_id)}) on its own QNodeOS. A request for opening an \ac{ER} socket is executed by the CNPU, by asking the QNPU (through the QNPU API) to open the socket. The QNPU then registers the ER socket with the quantum network stack (provided it did not yet exist), and the CNPU also keeps a reference using the tuple as an identifier. The program can then use this socket for requests. The network stack only handles requests for entanglement between two nodes if the corresponding \ac{ER} sockets are opened on both nodes.

Programs are themselves responsible for coordinating the ER socket IDs. Using these IDs allow the same node pair to open multiple pairs of ER sockets, which may be used by different applications or inside the same application. Socket IDs must be unique within the node. \ac{ER} sockets are typically opened at the start of a program, and live (and may be used multiple times) until the program finishes.

Programs use the \ac{ER} socket to submit entanglement requests to the network stack. This is done through NetQASM instructions (\texttt{create\_epr} and \texttt{recv\_epr}) that refer to the \ac{ER} socket in their operands. One program must execute a \texttt{create\_epr} instruction and the other a \texttt{recv\_epr} instruction (to be coordinated by the programs themselves). The program executing the \texttt{create\_epr} instruction is treated by the network stack as the \emph{initiator} and the program executing \texttt{recv\_epr} the \emph{receiver}. 
Upon receiving an entanglement request, the network stacks of the two nodes communicate between each other in order to coordinate entanglement generation. The initator node always initiates this communication. The receiver node always accepts the entanglement initiative as long as the corresponding \ac{ER} socket is open. This means that the receiver node agrees with entanglement generation as soon as the initiator node has submitted an entanglement request (through its \texttt{create\_epr}), even if the receiver node itself has not yet submitted its corresponding request (through its \texttt{recv\_epr}). On the receiver node, the generated entangled qubit will remain in memory until it gets asked for by a user process executing this \texttt{recv\_epr}.

\subsubsection{Multitasking}

Multi-tasking forms an essential element of our architecture already at the level of scheduling the network process in relation to any user process, to address the challenges inherent in the way entanglement is produced at the physical layer, requiring agreement on a network schedule (see Section~\ref{sec:arch-networking}, and main paper). For this important reason, the \ac{QNPU} is designed to arbitrate between these two processes (see \cref{fig:process-flow}), and to manage the resources being used by each of them. \emph{Multitasking}, hence, is a fundamental requirement for a system managing the hardware of a quantum network node, especially while such hardware has only limited resources available. 

To further increase the utility of the system, we also allow the multi-tasking of user processes (main paper):
Like in most operating systems, these tasks, which on the \ac{QNPU} are encapsulated into processes, can sometimes necessitate a resource which is not immediately available---for instance, a free qubit, or a qubit entangled with a remote one. Maximizing the utilization of the quantum device is also one of the goals of \ac{QNodeOS}, whose design allows multiple processes, user and kernel, to be active concurrently, so that whenever one is in a waiting state, another one can potentially be scheduled to use the quantum device.  This design aspect is relevant for quantum networking nodes, as the execution of the local program is often waiting, both for classical messages from remote nodes, as well as the generation of entanglement. 

Lastly, multitasking is an important feature for systems that are to be shared by multiple users, and that offer each user the possibility to run multiple programs concurrently. The multitasking capabilities of \ac{QNodeOS} are also aimed at improving the average throughput and latency of user programs.

\subsection{QNodeOS Components and Interfaces}
\label{sec:appendix-qnodeos-core}

\begin{figure*}
\begin{center}
\includegraphics[width=\linewidth]{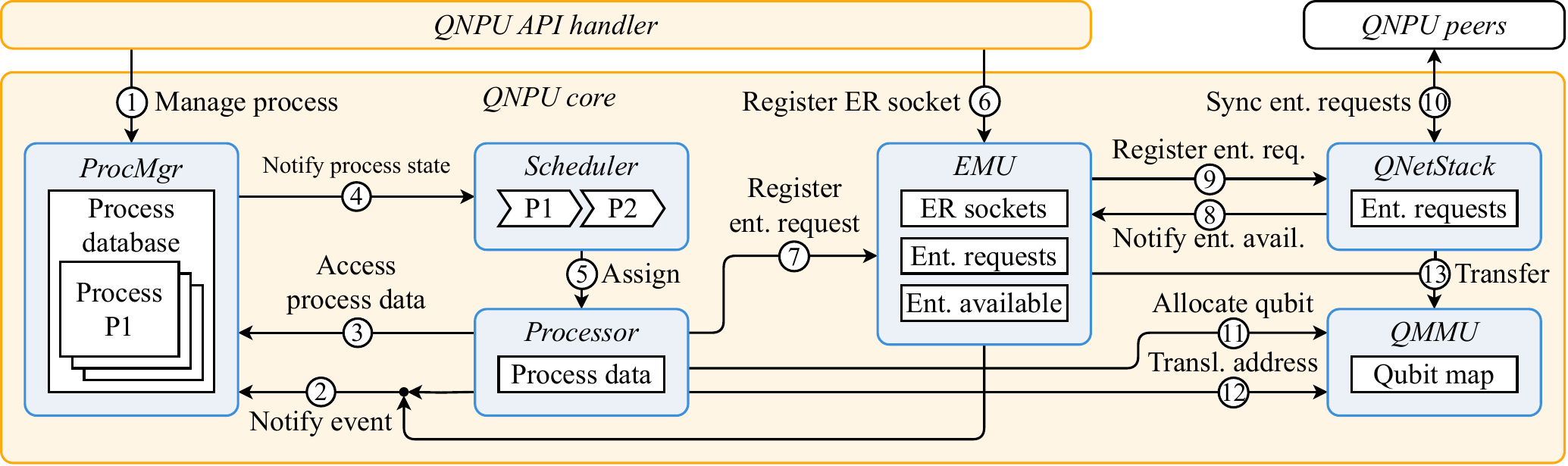}
\end{center}
\caption[]{\acf{QNPU} core components and internal interfaces. The core layer includes:
\begin{inlinelist}
\item a \emph{process manager} (ProcMgr), which owns and manages access to \ac{QNPU} processes;
\item a \emph{scheduler}, responsible for selecting the next process to be run;
\item a \emph{processor}, which processes blocks' instructions;
\item an \emph{\ac{EMU}}, which keeps a list of entanglement requests and available entangled qubits;
\item a \emph{\ac{QNetStack}}, whose responsibility is to coordinate with peer nodes to schedule quantum networking instructions;
\item a \emph{\ac{QMMU}}, which keeps a record of allocated qubits.
\end{inlinelist}
}
\label{fig:qnodeos-core}
\end{figure*}

We provide here additional details on the components of the \ac{QNPU} architecture and their interfaces. \Cref{fig:qnodeos-core} gives an overview of all the components of the \ac{QNPU}. The \emph{process manager} marshals accesses to all user and kernel processes. The \emph{scheduler} assigns ready processes to the \emph{processor}, which runs quantum instructions through the underlying \ac{QDevice}, processes classical \ac{NetQASM} instructions locally, and registers entanglement requests with the \emph{\ac{EMU}}. The \ac{EMU} maintains a list of \ac{ER} sockets and entanglement requests, forwards the latter to the \emph{quantum network stack}, which, in turn, registers available entangled qubits with the \ac{EMU}. Finally, the \emph{\ac{QMMU}} keeps track of used qubits, and transfers qubit ownership across processes when requested.

\subsubsection{Process Manager}

The process manager owns \ac{QNPU} processes and marshals accesses to those. Creating a process, adding a block to it and accessing the process's data must be done through the process manager. Additionally, the process manager is used by other components to notify \emph{events} that occur inside the \ac{QNPU}, upon which the state of one of more processes is updated. Process state updates result in a notification to the scheduler.

\paragraph{Interfaces}

The process manager exposes interfaces for three services:
\begin{itemize}
\item Process management (interface 1 in \cref{fig:qnodeos-core}): to create and remove processes, and to add quantum blocks to them. When the user registers a program, the the \ac{QNPU} \ac{API} Handler uses the process manager to create a \ac{QNPU} user process. The returned process ID can be later used to add a block to that process, or to remove the process once all its blocks are fully processed.
\item Event notification (interface 2 in \cref{fig:qnodeos-core}): to notify an event occurred inside the \ac{QNPU}, including the addition of a block, the completion of a block, the scheduling of the process, the hitting of a \emph{Waiting} condition (see \cref{fig:process-states}), and the generation of an entangled qubit destined to the process. Some events trigger follow-up actions---for instance, when a process that was waiting for an event becomes ready, it gets added to the queue of ready processes maintained by the scheduler.
\item Process data access (interface 3 in \cref{fig:qnodeos-core}): to access a process's blocks and its classical memory space, mostly used while running the process (through the processor).
\end{itemize}

\subsubsection{Scheduler}

The \ac{QNPU} scheduler registers processes that are ready to be scheduled, and assigns them to the \ac{QNPU} processor when the latter is available. Ready processes are stored in a \emph{prioritized ready queue}, and processes of the same priority are scheduled with a first-come-first-served policy.

\paragraph{Interfaces}

The scheduler only exposes one interface for process state notifications (interface 4 in \cref{fig:qnodeos-core}), used by the process manager to signal when a process transitions to a new state. When a \ac{QNPU} process transitions to the ready state, it is directly added to the scheduler's prioritized ready queue. When a process becomes idle, or is waiting for an event to happen, the scheduler simply registers that the processor has become available.

\subsubsection{Processor}

The \ac{QNPU} processor handles the execution of \ac{QNPU} user and kernel processes, by running classical instructions locally and issuing quantum instructions to the \ac{QDriver}. It is also responsible for multitasking by means of process manager. While executing a process, the processor reads its blocks and accesses (reads and writes) its classical memory. The processor implements a specific instruction set architecture dictated by the \ac{NetQASM} language of choice.

\paragraph{Interfaces}

The processor exposes one interface for processor assignment (interface 5 in \cref{fig:qnodeos-core}), used by the \ac{QNPU} scheduler to activate the processor, when it is idling, and assign it to a \ac{QNPU} process.

\subsubsection{Entanglement Management Unit}

The \acf{EMU} contains a list of open \emph{\ac{ER} sockets} and a list of \emph{entanglement requests}, and keeps track of the \emph{available entangled qubits} produced by the quantum network stack. Received entanglement requests are considered valid only if an \ac{ER} socket associated to such requests exists. Valid requests are forwarded to the quantum network stack. Entangled qubit generations are notified as events to the process manager.

\paragraph{Interfaces}

The \ac{EMU} exposes interfaces for three services:
\begin{itemize}
\item \ac{ER} socket registration (interface 6 in \cref{fig:qnodeos-core}): to register and open \ac{ER} sockets belonging to a program, and to set up internal classical network tables and to establish classical network connection.
\item \ac{ER} registration (interface 7 in \cref{fig:qnodeos-core}): to add entanglement requests to the list of existing ones, to be used when matching produced entangled qubits with a process that requested them.
\item Entanglement notification (interface 8 in \cref{fig:qnodeos-core}): to register the availability of an entangled qubit, produced by the quantum network stack, and to link it to an existing entanglement request.
\end{itemize}

\subsubsection{Quantum Network Stack}

The quantum network stack on the \ac{QNPU} closely follows the model presented by
\textcite{dahlberg_2019_egp} which is based on the classical \ac{OSI} network stack model for the purpose of the separation of responsibilities. In particular, \emph{data link layer} is part of the quantum network stack on the \ac{QNPU}. The \emph{physical layer} is implemented on the \ac{QDevice}, the \emph{application layer} is part of the \ac{CNPU}, and all remaining layers are not currently part of the stack.

The quantum network stack component has an associated \emph{\ac{QNPU} kernel process}, created statically on the \ac{QNPU}. However, this process's block is dynamic: the instructions to be executed on the processor depend on the outstanding entanglement generation requests received from \ac{EMU} and network peers.

\paragraph{Interfaces}

The quantum network stack exposes interfaces for two services:
\begin{itemize}
\item Entanglement request registration (interface 9 in \cref{fig:qnodeos-core}): to add entanglement requests coming from the \ac{EMU} to the list of existing ones, which are used to fill in the quantum network stack process's block with the correct quantum instructions to execute.
\item Entanglement request synchronization (interface 10 in \cref{fig:qnodeos-core}): similar to the entanglement request registration interface, but to be used to synchronize (send and receive) requests with \ac{QNodeOS} network peers.
\end{itemize}

\subsubsection{Quantum Memory Management Unit}

\acf{QMMU} receives requests for \emph{qubit allocations} from \ac{QNPU} processes, and manages the subsequent usage of those. It also translates \ac{NetQASM} \emph{virtual qubit addresses} into physical addresses for the \ac{QDevice}, and keeps track of which process is using which qubit at a given time. In general, a \ac{QMMU} should take into account that the topology of a quantum memory determines what operations can be performed on which qubits, and thus allow processes to allocate qubits of a specific type upon request. An advanced \ac{QMMU} could also feature algorithms to move qubits in the background---that is, without an explicit instruction from a process's block---to accommodate a program's topology requirements while not trashing the qubits being used by other \ac{QNPU} processes. Such a feature could prove crucial to increase the number of processes that can be using the quantum memory at the same time, and to enhance multitasking performances.

\paragraph{Interfaces}

The \ac{QMMU} exposes interfaces for three services:
\begin{itemize}
\item Qubit allocation and de-allocation (interface 11 in \cref{fig:qnodeos-core}): a running process can ask for one or more qubits, which, if available, are allocated by the \ac{QMMU}, and the physical addresses of those are mapped to the virtual addresses provided by the requesting process.
\item Virtual address translation (interface 12 in \cref{fig:qnodeos-core}): before sending quantum instructions to the \ac{QDriver}, the processor uses virtual qubit addresses specified in \ac{NetQASM} to retrieve physical addresses from the \ac{QMMU}, and then replaces virtual addresses with physical addresses in the instructions for the \ac{QDriver}.
\item Qubit ownership transfer (interface 13 in \cref{fig:qnodeos-core}): qubits are only visible to the process that allocates them. However, in some cases, a process may wish to transfer some if its qubits to another one. A notable example is the quantum network process transferring an entangled qubit to the process that will use it.
\end{itemize}

\subsection{QNPU implementation: scheduler}
\label{sec:qnpu_impl_scheduler}
The \ac{QNPU} scheduler is an important component of our \ac{QNodeOS} architecture, and deals with scheduling of QNPU processes. The QNPU is implemented on FreeRTOS~\cite{freertos}, which itself includes a scheduler. On FreeRTOS, code is organized into tasks, which can be seen as separate threads or processes. These tasks are scheduled concurrently by FreeRTOS based on priority. In our implementation, we realize QNPU components and interfaces (hence including the QNPU scheduler) as FreeRTOS tasks. We configured task priorities such that the components with tight interaction with the QDevice (\ac{QDriver}, quantum network stack, QNPU processor) have highest priority.
We stress the difference between the FreeRTOS scheduler and our QNPU process scheduler. 
The QNPU scheduler schedules QNPU processes based on their status and priorities, which are independent of the priorities assigned by the FreeRTOS scheduler.
The FreeRTOS hence runs on a different layer: it makes sure the QNPU components (including QNPU scheduler, processor, \ac{QDriver}) run concurrently. The QPNU scheduler runs on the level of QNPU processes. Whenever the FreeRTOS scheduler activates the FreeRTOS task realizing the QNPU scheduler, the QNPU scheduler then schedules the process with the highest priority on a first come first serve basis, by adding it to the processing queue of the relevant resource (e.g. QNPU processor) and generating an interrupt leading to the execution of the QNPU processor task on FreeRTOS (and consequently the execution of the process).

\subsection{QDevice Interface}
\label{sec:appendix-qdevice}

The implementation of a \ac{QDevice} depends on a number of factors. Most importantly, the physical signals that are fed to the quantum processing and networking device (and those that are output from the device) are specific to the nature of the device itself. Different qubit realizations require different digital and analog control. For instance, manipulating the state of a spin-based qubit (e.g., in a \ac{NV} center processor) and that of an atom qubit (e.g., in a trapped ion processor) are two physical processes that vastly differ in a number of complex ways.

For \ac{QNodeOS} to be portable to a diverse set of quantum physical platforms, there needs to be a common \emph{\ac{QDevice} interface} that \ac{QNodeOS} can rely on, and that each \ac{QDevice} instance can implement as it is most convenient for the underlying quantum device. This interface \begin{inlinelist} \item needs to be \emph{general}, \item to be able to \emph{express all quantum operations} that different quantum devices might be capable of performing, and \item \emph{abstract}, so that two different implementations of a well-defined qubit manipulation operation can be expressed with the same instruction on \ac{QNodeOS}.\end{inlinelist} Nevertheless, an interface that is too general could result in a high implementation complexity on the \ac{QDevice}, as it might have to transform high-level instructions in a series of native operations on the fly. Other than complexity of implementation, a very high-level set of \ac{QDevice} instructions might compromise the compiler's ability to optimize a program for a certain physical platform, as reported by~\textcite{murali_2019_fullstack}.

\subsubsection{Design Choices}

Defining a set of instructions to express abstract quantum operations as close as possible to what different quantum physical platforms can natively perform is---to some extent---an open problem. Nonetheless, we have made an effort to specify an interface which is a good compromise between generality and expressiveness. The \ac{QDevice} interface is essentially a set of instructions that \ac{QNodeOS} expects a \ac{QDevice} to implement. To be precise, a \ac{QDevice} might implement a subset of the interface, according to what native physical operations it can perform. The CNPU compiler must then have knowledge about the set of instructions implemented by the underlying \ac{QDevice}, so that it can decompose instructions that are not natively supported.

Even though this interface does not impose any formal timing constraints, it is important to note that a \ac{QDevice} implementation that tries to guarantee more or less deterministic instruction processing latencies can prove more beneficial to the real-time requirements of the \ac{QNPU}. Particularly, it would be advisable to time-bound the processing time of operations whose duration is by nature probabilistic---most notably, those involving entanglement generation. Creating an \ac{EPR} pair may involve a varying number of attempts. Sometimes, if the remote node becomes unresponsive for some time, the number of necessary attempts can increase by a large amount. Capping the number of attempts could, for instance, provide a more deterministic maximum processing latency for entanglement instructions, which in turn might help \ac{QNodeOS} react more timely to temporary failures or downtime periods of remote nodes. Not to mention that unbounded entanglement attempts affect the state of other qubits in memory, because of both passive decoherence and cross-qubit noise.

\subsubsection{QDevice synchronization}
\label{sec:qdevice-sync}
The QDevice receives physical instructions from QNodeOS, acts on them, and returns a response. For entanglement instructions, the QDevice must first synchronize with the QDevice on the other node (using classical communication). If the other QDevice is busy, (e.g. it is still trying to pass the CR check, see~\cref{sec:qdevice-nv} and \cite{pompili_2022_experimental}), synchronization fails, and an \texttt{ENT\_SYNC\_FAIL} response is returned (see~\cref{tab:qdevice-return-values}).

\subsubsection{Instructions and Operands}
\label{sec:sec:appendix-qdevice-instructions-operands}

\begin{table}[t]
\begin{tabularx}{\linewidth}{>{\ttfamily}l X}
\toprule
\normalfont{Instruction} & Description \\
\midrule
INI & Initialize a qubit to default state \\
SQG & Perform a single-qubit gate \\
TQG & Perform a two-qubit gate \\
AQG & Perform a gate on all qubits \\
MSR & Measure a qubit in a specified basis \\
ENT & Attempt entanglement generation \\
ENM & Attempt entanglement and measure qubit \\
MOV & Move qubit state to another qubit \\
SWP & Swap the state of two qubits \\
ESW & Swap qubits belonging to two \ac{EPR} pairs \\
PMG & Set pre-measurement gates \\
\bottomrule
\end{tabularx}
\caption[]{Summary of \ac{QDevice} instructions defined in the \ac{QDevice} interface. A specific \ac{QDevice} might implement a subset of these, depending on the underlying quantum physical device and on other design constraints.}
\label{tab:qdevice-instructions}
\end{table}

\Cref{tab:qdevice-instructions} lists the complete set of instructions defined in the \ac{QDevice} interface. Instructions can have operands, whose range of valid values depends on the underlying \ac{QDevice}. For instance, an operand that specifies which qubit to apply an operation to can only have as many valid values as there are physical qubits in memory. Details for each instruction and its operands are given below.

\paragraph{Qubit Initialization (\texttt{INI})}

The \texttt{INI} instruction brings a qubit to the $\ket{0}$ state. On some physical platforms, single-qubit initialization is not possible, thus this instruction initializes all qubits to the $\ket{0}$ state.

\medskip \noindent
\begin{tabularx}{\linewidth}{>{\ttfamily}l X}
\toprule
\normalfont{Operand} & Description \\
\midrule
qubit & Physical address of the qubit to initialize, ignored on platforms where single-qubit initialization is not possible \\
\bottomrule
\end{tabularx}

\paragraph{Single-Qubit Gate (\texttt{SQG})}

The \texttt{SQG} instruction manipulates the state of one qubit. The gate is expressed as a rotation in the Bloch sphere.

\medskip \noindent
\begin{tabularx}{\linewidth}{>{\ttfamily}l X}
\toprule
\normalfont{Operand} & Description \\
\midrule
qubit & Physical address of the qubit to manipulate \\
axis  & Rotation axis, can be X, Y, Z or H (support is \ac{QDevice}-dependent) \\
angle & Rotation angle (granularity and range are \ac{QDevice}-dependent) \\
\bottomrule
\end{tabularx}

\paragraph{Two-Qubit Gate (\texttt{TQG})}

The \texttt{TQG} instruction manipulates the state of two qubits. The gate is expressed as a controlled rotation, with one qubit being the control and the other one being the target.

\medskip \noindent
\begin{tabularx}{\linewidth}{>{\ttfamily}l X}
\toprule
\normalfont{Operand} & Description \\
\midrule
qub\_c & Physical address of the control qubit \\
qub\_t & Physical address of the target qubit \\
axis   & Rotation axis, can be X, Y, Z or H (support is \ac{QDevice}-dependent) \\
angle  & Rotation angle (granularity and range are \ac{QDevice}-dependent) \\
\bottomrule
\end{tabularx}

\paragraph{All-Qubit Gate (\texttt{AQG})}

The \texttt{AQG} instruction manipulates the state of all available qubits. The gate is expressed as a rotation in the Bloch sphere.

\medskip \noindent
\begin{tabularx}{\linewidth}{>{\ttfamily}l X}
\toprule
\normalfont{Operand} & Description \\
\midrule
axis  & Rotation axis, can be X, Y, Z or H (support is \ac{QDevice}-dependent) \\
angle & Rotation angle (granularity and range are \ac{QDevice}-dependent) \\
\bottomrule
\end{tabularx}

\paragraph{Qubit Measurement (\texttt{MSR})}

The \texttt{MSR} instruction measures the state of one qubit in a specified basis. A qubit measurement is destructive---that is---the qubit has to be reinitialized before it can be used again.

\medskip \noindent
\begin{tabularx}{\linewidth}{>{\ttfamily}l X}
\toprule
\normalfont{Operand} & Description \\
\midrule
qubit & Physical address of the qubit to measure \\
basis & Measurement basis, can be X, Y, Z, H (support is \ac{QDevice}-dependent) \\
\bottomrule
\end{tabularx}

\paragraph{Entanglement Generation (\texttt{ENT})}

The \texttt{ENT} instruction performs a series of entanglement generation attempts, until one succeeds, or until a maximum number of attempts is reached (the behavior is \ac{QDevice}-dependent).

\medskip \noindent
\begin{tabularx}{\linewidth}{>{\ttfamily}l X}
\toprule
\normalfont{Operand} & Description \\
\midrule
nghbr & Neighboring node to attempt entanglement with, if the local \ac{QDevice} has multiple quantum links \\
fid & Target entanglement fidelity (granularity and range are \ac{QDevice}-dependent) \\
\bottomrule
\end{tabularx}

\paragraph{Entanglement Generation With Qubit Measurement (\texttt{ENM})}

The \texttt{ENM} instruction performs a series of entanglement generation attempts followed by an immediate measurement of the local qubit, until one succeeds, or until a maximum number of attempts is reached (the behavior is \ac{QDevice}-dependent).

\medskip \noindent
\begin{tabularx}{\linewidth}{>{\ttfamily}l X}
\toprule
\normalfont{Operand} & Description \\
\midrule
nghbr & Neighboring node to attempt entanglement with, if the local \ac{QDevice} has multiple quantum links \\
fid & Target entanglement fidelity (granularity and range are \ac{QDevice}-dependent) \\ basis & Measurement basis, can be X, Y, Z, H (support is \ac{QDevice}-dependent) \\
\bottomrule
\end{tabularx}

\paragraph{Qubit Move (\texttt{MOV})}

The \texttt{MOV} instruction moves the state of one qubit to another qubit. A qubit move renders the state of the source qubit undefined, and the qubit has to be reinitialized before it can be used again.

\medskip \noindent
\begin{tabularx}{\linewidth}{>{\ttfamily}l X}
\toprule
\normalfont{Operand} & Description \\
\midrule
qub\_s & Physical address of the source qubit \\
qub\_d & Physical address of the destination qubit \\
\bottomrule
\end{tabularx}

\paragraph{Qubit Swap (\texttt{SWP})}

The \texttt{SWP} instruction swaps the state of two qubits.

\medskip \noindent
\begin{tabularx}{\linewidth}{>{\ttfamily}l X}
\toprule
\normalfont{Operand} & Description \\
\midrule
qub\_1 & Physical address of the first qubit \\
qub\_2 & Physical address of the second qubit \\
\bottomrule
\end{tabularx}

\paragraph{Entanglement Swap (\texttt{ESW})}

The \texttt{ESW} instruction results in two qubits belonging to two \ac{EPR} pairs to have their roles swapped.

\medskip \noindent
\begin{tabularx}{\linewidth}{>{\ttfamily}l X}
\toprule
\normalfont{Operand} & Description \\
\midrule
qub\_1 & Physical address of the first qubit \\
qub\_2 & Physical address of the second qubit \\
\bottomrule
\end{tabularx}

\paragraph{Pre-Measurement Gates Setting (\texttt{PMG})}

The \texttt{PMG} instruction allows for a set of (up to) 3 rotations to be performed before a qubit measurement (\texttt{MSR} or \texttt{ENM}). If the axis of the second rotation is orthogonal to the axis of the first and the third rotation, these gates can be used to perform a qubit measurement in an arbitrary basis, given that most likely a \ac{QDevice} can natively measure in a limited set of bases.

\medskip \noindent
\begin{tabularx}{\linewidth}{>{\ttfamily}l X}
\toprule
\normalfont{Operand} & Description \\
\midrule
axes & Combination of orthogonal axes to use for the three successive rotations, can be X--Y--X, Y--Z--Y and Z--X--Z (support is \ac{QDevice}-dependent) \\
ang\_1 & Rotation angle of the first gate, relative to the first axis in \texttt{axes} (granularity and range are \ac{QDevice}-dependent) \\
ang\_2 & Rotation angle of the second gate, relative to the second axis in \texttt{axes} (granularity and range are \ac{QDevice}-dependent) \\
ang\_3 & Rotation angle of the third gate, relative to the third axis in \texttt{axes} (granularity and range are \ac{QDevice}-dependent) \\
\bottomrule
\end{tabularx}

\paragraph{No operation (\texttt{NOP})}

The \texttt{NOP} instruction does not result in any operation on the \ac{QDevice}.

\subsubsection{Return values}
\Cref{tab:qdevice-return-values} lists the possible return values that the QDevice sends back to QNodeOS as a response to a physical instruction.

\begin{table*}[htpb]
    \centering
    \begin{tabularx}{\textwidth}{|l|l|X|}
    \hline
    Physical Instruction & Return values & Description \\ 
    \hline
    \texttt{INI}, \texttt{SQG}, \texttt{TQG}, \texttt{AQG}, \texttt{PMG} & \texttt{SUCCESS} & Always successful \\
    \texttt{MSR} & \texttt{SUCCESS\_0} or \texttt{SUCCESS\_1} & Measurement outcome is 0 or 1* \\ 
    \texttt{ENT} & \texttt{SUCCESS\_<state>} & Entanglement generation was successful; the state is <state>$^\dagger$ \\
    \texttt{ENM} & \texttt{SUCCESS\_<state>\_<outcome>} & Entanglement generation was successful; state was <state>$^\dagger$ and outcome is <outcome> (0 or 1) \\
    \texttt{ENT, ENM} & \texttt{ENT\_FAILURE} & Entanglement generation was attempted and failed \\
    \texttt{ENT, ENM} & \texttt{ENT\_SYNC\_FAILURE} & Entanglement generation was not attempted since synchronization failed (other node is busy) \\
    \hline
    \end{tabularx}
    \caption{*Measurements are always in the Z basis, where outcome 0 corresponds to $\ket{0}$ and outcome 1 to $\ket{1}$. $^\dagger$ Possible states depend on the implementation. For NV these are \texttt{PSI\_PLUS} and \texttt{PSI\_MINUS}, see~\cref{sec:qdevice-nv}.}
    \label{tab:qdevice-return-values}
\end{table*}

\clearpage
\section{QDevice Implementations}
\subsection{NV Center Platform}
\label{sec:qdevice-nv}

\begin{figure}
    \centering
    \includegraphics[width=0.8\linewidth]{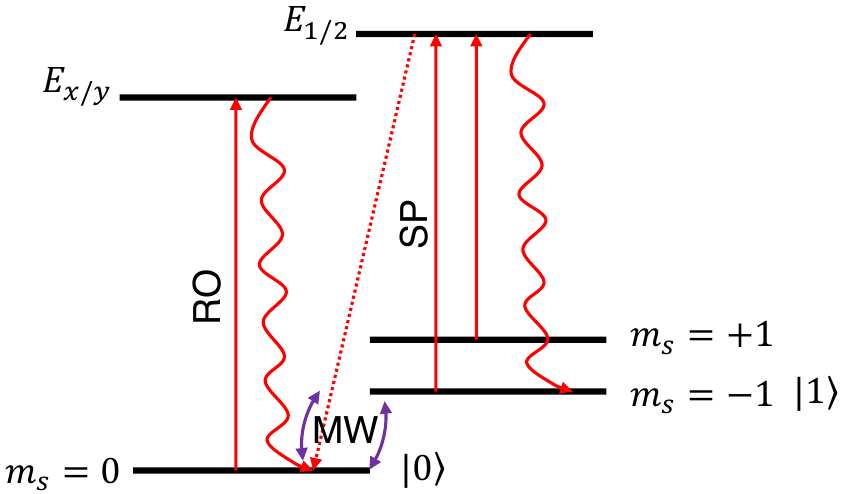}
    \caption{Energy structure of \ac{NV}$^-$ at 4\,K. The ground state of the \ac{NV} splits into three distinct levels (Zeeman splitting). The optical transitions are spin-selective. The excited states are represented as one, but they are non-degenerate when lateral strain is applied. We denote as \acf{RO} the transition $\ket{0}\rightarrow E_{x/y}$ and as \acf{SP} the transition $\ket{1}\rightarrow E_{1/2}$. The wiggly lines represent the photoluminescence when such transitions are excited, whereas the dashed lines represent the decay via metastable states that is used for initialization of the qubit state into $\ket{0}$. \acf{MW} pulses enables the transfer of population between the two states of the qubit, allowing for quantum information processing.}
    \label{fig:nv_levels}
\end{figure}

The \ac{QDevice} employed for the benchmark experiments is constituted by an \ac{NV} center processor. We use the \ac{NV} center in its negatively charged state (called \ac{NV}$^-$) for quantum information processing. \ac{NV}$^-$ is a spin-1 system, whose ground states are non-degenerate in the presence of an external magnetic field, see \cref{fig:nv_levels}~\cite{doherty_2013}. We employ the $m_s=0$ as our $\ket{0}$ state for the qubit, while for the $\ket{1}$ we can choose one of $m_s=\pm 1$. Details on how the choice is made will follow in the next section. The \ac{NV} can be optically excited resonantly (637\,nm) and off-resonantly (typically 532\,nm), and it emits in 3\% of the cases single photons (\ac{ZPL} photons), while the remaining part is constituted by the emission of a photon and a phonon (\ac{PSB}). The optical transitions are spin-selective, as shown in~\cref{fig:nv_levels}. In the presence of lateral strain and external DC field (Stark effect), the excited states of the \ac{NV} split apart, maintaining their spin-selective properties. In this work, we use a natural lateral strain between 2\,GHz and 5\,GHz. The cycling transition denoted as \ac{RO} in \cref{fig:nv_levels} is used to emit single photons (\ac{ZPL}) for entanglement generation and to read out the state of the qubit (fluorescence in the \ac{PSB}). From the excited states, the \ac{NV} can also decay through metastable states (not shown in \cref{fig:nv_levels}). The preferable decay from such metastable states is the $m_s=0$ state. In this way, it is possible to optically initialize the qubit state to $\ket{0}$ (dashed line in~\cref{fig:nv_levels}), with fidelity above 99\%, when on-resonantly exciting the \ac{SP} transition and averaging for long enough to ensure a spin-flip. In our experiments, we apply a laser field on resonance with the \ac{SP} transition at 500\,$n$W for 1.5\,$\mu$s for fast initialization during entanglement attempts, whereas a slow initialization (10\,$n$W for 100\,$\mu$s) is used for single-qubit gates experiments (like local tomography). On the other hand, while exciting the \ac{RO} transition, decays to $m_s=\pm 1$ are also possible, but they present longer cyclicity. This feature is relevant for the optical read-out of the qubit state, which can be done in a single shot and is discussed in the following sections.

In our demonstration, the server has an external magnetic field of $B_z=189$\,mT aligned along the symmetry axis $z$ of the \ac{NV}, while the client experiences $B_z=23$\,mT. The magnetic field is applied via permanent magnets placed both inside and outside the high-vacuum chamber of our closed-cycle cryostats. Fluctuations of the magnetic field are observed on the order of nT on a timescale of hours, therefore they do not constitute a limitation for our demonstration. We also measured a perpendicular component of the permanent magnetic field for both setups of $\sim$1\,mT. Such misalignment becomes crucial for the coherence time of the electron spin qubit, as in the interaction with the surrounding nitrogen nuclear spins, the off-axis hyperfine interaction terms become non-negligible and the decoupling of the electron spin is harder~\cite{doherty_2013}. Notably, the server node is in the regime of ``high magnetic field''. In the level structure depicted in \cref{fig:nv_levels}, this means that the $m_s=-1$ ground state crossed the $m_s=0$ state (at $\sim$\,100mT), and the optical transitions for the \ac{SP} are well separated, such that a double laser field with proper detuning is necessary to correctly address both of them.

\subsubsection{Single Node Operations}

In this section, details on how to operate a single node for quantum information processing are given. The physical setup is the one employed for the demonstration in Ref.~\cite{pompili_2022_experimental}.

\paragraph{Charge-resonance check}

To use the \ac{NV} as a processing node, it is necessary to guarantee that it is in the correct charge state and the laser fields are on resonance with the transitions. Before executing any instructions coming from the \ac{QNPU}, both nodes go through the so-called \ac{CR} check. We apply resonant fields for 100\,$\mu$s on both the \ac{RO} (1\,$n$W) and \ac{SP} (10\,$n$W) transitions and we monitor the fluorescence. If the number of photons exceeds the threshold (25 for the client and 60 for the server for our experiments), the node is considered ready to accept instructions from the \ac{QNPU} and can proceed with synchronization with the other node (for multinode instructions). The threshold is set considering the brightness of each \ac{NV}. The success is considered valid for 100\,$m$s. After this time, if no instructions arrive, the \ac{CR} check is repeated. In case the number of photons is below the threshold, we distinguish two cases: (1) the counts are between the success threshold and a second threshold called Repump: we repeat the \ac{CR} check and tune the frequency of the red lasers, as they might not address the transitions correctly; (2) the number of counts is below the Repump threshold (set at 15 for the client and 25 for the server): this means that the \ac{NV} might be in the dark charge state (\ac{NV}$^{0}$) due to ionization. To restore the charge state, in the next round of \ac{CR} check we first illuminate with off-resonant green laser (20\,$\mu$W for 50\,$\mu$s), or, for the client node only, with yellow light (575\,nm, 35\,$n$W for 300\,$\mu$s) on resonance with the \ac{ZPL} transition of \ac{NV}$^{0}$~\cite{baier_2020}. This is necessary because we additionally apply an external DC field to the \ac{NV} on the client node. We, indeed, exploit the Stark effect to tune the \ac{RO} transition to be the same as the server's one \cite{basssett_2011}. In this way, we can ensure photon indistinguishability in frequency that is crucial for entanglement generation. The typical DC field used for this work is $\sim$2V, modulated via an error signal that is computed on the \ac{CR} check counts, acting as a \ac{PID} loop.

The \ac{CR} check is repeated after an experiment iteration. This round is utilized to validate the experiment and post-select the results based on success or failure of this procedure, as discussed in \cref{sec:methods}.

\paragraph{Single qubit gates}

To manipulate the state of the electron spin qubit, microwave pulses are on resonance with the transition $\ket{0}\rightarrow\ket{1}$ are employed. For the server node, the $m_s=-1$ state is used as $\ket{1}$ and the resonance frequency is 2.4\,GHz. The client node utilizes the $m_s=+1$, with a resonance frequency of 3.5\,GHz. The choice of the $\ket{1}$ is made based on the gate fidelity.

We use skewed-Hermite \acf{MW} pulses~\cite{warren_1984,vandersypen_2005} with high Rabi frequency ($\sim$10\,MHz), which generates an alternating magnetic field capable of manipulating the state of the qubit. The characterizing values for the two nodes are reported in Table \ref{tab:mw_pulse}. The measured infidelity on a single \ac{MW} pulse is below 1\%. Instructed by the \ac{QNPU}, we performed local quantum tomography on both the server and the client, showing high fidelity. One example is reported in \cref{fig:tomography-no-multitasking}.

\begin{table}[]
    \centering
    \begin{tabular}{|r|c|c|}
    \hline
 &Client & Server\\
\hline
        Duration $\pi$ rotation & 200\,$n$s & 190\,$n$s\\
        Amplitude $\pi$ rotation & 0.78 & 0.89\\
        Skewness $\pi$ rotation & -1.5e$^{-9}$ & -3.5e$^{-9}$\\
        Duration $\pi/2$ rotation & 150\,$n$s&  100\,$n$s\\
        Amplitude $\pi/2$ rotation & 0.38 & 0.56\\
        Skewness $\pi/2$ rotation & -1.2e$^{-8}$& -7.1e$^{-9}$ \\
        Power & 42\,W & 42\,W\\
        \hline
    \end{tabular}
    \caption{Characterizing values for the \ac{MW} pulses. Other rotation angles have the same duration and skewness as the $\pi$ pulse, and the amplitudes scale accordingly. The rotation axes are obtained by changing the phase of the pulse. With the current setup configuration, only rotations along $\hat{x}$ and $\hat{y}$ axes are feasible, so $\hat{z}$ rotations are compiled as combinations of gates along $\hat{x}$ and $\hat{y}$.}
    \label{tab:mw_pulse}
\end{table}

\paragraph{Dynamical decoupling}

Once \ac{MW} pulses are set up with high fidelity, it is possible to implement \ac{DD} sequences that increase the coherence time of the electron spin qubit. \ac{DD} sequences are especially crucial in our experiments when the latency of the \ac{QNPU} is long (milliseconds timescale), like in the \ac{DQC} demonstration. The characterizing parameter for a \ac{DD} sequence is the time delay between the $X$ and $Y$ pulses. To optimize it, we swept the interpulse delay, at the sample precision of our Arbitrary Waveform Generator (0.42\,$n$s, Zurich Instruments HDAWG), while playing the effective single-qubit computation of the \ac{DQC} protocol instructed by the \ac{QNPU}, as explained in \cref{sec:methods}, on both the client and the server. In doing so, we added an extra waiting time of 5\,$m$s between the initialization of the qubit into the superposition state and the subsequent gates to mimic the real-case scenario of the \ac{DQC}. In this way, we are able to set the optimal interpulse delay, obtaining a single-qubit fidelity of 0.96(2) for the server and 0.88(2) for the client.

\paragraph{Single-shot readout}

When a measurement instruction arrives from the \ac{QNPU}, this is translated by the physical layer as a Single-Shot Readout measurement. To assign a state to the qubit, we can use the \ac{RO} optical transition. The \ac{RO} laser field is on for $\sim$10\,$\mu$s at 1\,$n$W. This will produce fluorescence only if the \ac{NV} is in the $\ket{0}$ state. If no photons are detected while the laser is on, the outcome is assigned to the $\ket{1}$ state. The fidelity of the measurement process is defined as $F=1/2(F_{0|0}+F_{1|1})$, where $F_{0|0}$ ($F_{1|1}$) represents the fidelity of measuring $\ket{0}$ ($\ket{1}$) when the qubit is prepared in $\ket{0}$ ($\ket{1}$). For our experiments, we obtain 0.841(4) and 0.997(1) respectively for the client, and 0.912(3) and 0.995(1) for the server, achieving above 0.90 of process fidelity.

\subsubsection{Entanglement generation}
\label{sec:NVentanglement}

The entanglement request from the \ac{QNPU} is translated into executing a single-photon protocol. The communication qubit on each node is initialized in the state $\sqrt{\eta} \ket{0} + \sqrt{1-\eta} \ket{1}$, where $\eta$ represents the bright state population. For maximum state fidelity, the condition $\eta_C p_C \approx \eta_S p_S$ applies, where $\eta_{C(S)}$ is the bright state population of the client (server) and $p_{C(S)}$ is the photon detection probability of the client (server). In this work, $\eta_{C}=0.07$ and $\eta_S = \eta_C \frac{p_C}{p_S} = 0.04$. The choice of $\eta_C$ is a trade-off between entangled state fidelity and entanglement generation rate. The produced entangled state is (non-deterministically) one of two Bell states $\ket{\Psi^\pm} = \frac{1}{\sqrt{2}}(\ket{01} \pm e^{i\Delta\theta}\ket{10})$, based on which detector clicked at the heralding station. The phase $\Delta\theta$ is actively stabilized~\cite{pompili_2021_multinode} before the execution of the entanglement request, via a combination of homodyne interference, for the global phase of the network, and a heterodyne interference, to stabilize the local phase at each node. Pauli-correction gates, based on the state prepared, are issued from the server \ac{QNPU} to its \ac{QDevice} to obtain $\ket{\Phi^+}$: an $X_{\pi}$ gate if the generated Bell-state is $\ket{\Psi^+}$ and an $X_{\pi}$ gate followed by a $Z_{\pi}$ gate (decomposed into X and Y gates) for $\ket{\Psi^-}$. As preparation for the experiment, we verified the entanglement generation, instructed by the \acp{QNPU} and using the same method as in Ref.~\cite{pompili_2022_experimental}, achieving a fidelity of $0.72(\pm 0.02)$ for the $\ket{\Phi^+}$ state. The Bell corrections done through the server \ac{QNPU} take up to 0.16 ms for $\ket{\Psi^+}$ and up to 0.49\,ms for $\ket{\Psi^-}$. On the other hand, generating entanglement without the \ac{QNPU} and with no Pauli correction, we achieve an average fidelity of 0.74($\pm$0.03), with $\eta_{C}=0.1$ and $\eta_S=0.06$. The choice of different $\eta$ values is due, in the first place, to speed up the rate of such a measurement. It shows, however, better performance with respect to the instructed version, which is due to the fact that the Pauli-correction instruction comes with a latency.
\subsection{Trapped-Ion Platform}
\label{sec:trapped-ion-platform}

\subsubsection{Setup}

The trapped-ion \ac{QDevice} implementation faces different challenges than the \ac{NV} \ac{QDevice} implementation. Trapped-ion state preparation, gate operations, and readout occur on longer timescales (between microseconds and milliseconds) than the corresponding operations for \ac{NV} centers. As a result, latencies introduced by \ac{QNodeOS} are insignificant, and we do not expect the use of \ac{QNodeOS} to reduce fidelities of local gate operations or entangling operations on the trapped-ion \ac{QDevice}. On the other hand, trapped ions are typically manipulated using control sequences that are compiled for a given set of parameters and uploaded to hardware. (Here, sequences consist of pulses of \ac{TTL} signals, analog voltages, and radio frequency or microwave signals, some of which are phase-referenced to one another. These pulses typically control the laser and microwave fields with which ions are manipulated.) The challenge here is that decision making within \ac{QNodeOS} must take place further up the network stack and is not compatible with pre-compiled sequences.

We address this challenge by exploiting a triggering capability within our pre-compiled sequences (which are written as Python scripts and then translated to a hardware description language for a \ac{FPGA}). A sequence can contain labels that act as memory pointers; at any point in a given sequence, a function can jump to one of these labels, at which point execution continues starting at that label. Thus, we can structure a sequence as a list of possible subsequences---each of which corresponds to a physical instruction or some part thereof. This list is preceded by a control subsequence. Input triggers from \ac{QNodeOS} cause the control subsequence to jump to a certain subsequence representing a physical instruction. After the subsequence---that is, the physical instruction---is implemented, the sequence returns to the control subsequence, where it waits for another input trigger.

A second challenge is the compatibility between \ac{QNodeOS} and physical-layer hardware. The \ac{QDriver} for \ac{QNodeOS} is implemented with a development \ac{FPGA} board (Texas Instruments LAUNCHXL2-RM57L~\cite{ti_launchxl2_rm57l_2024}) that sends messages via \ac{SPI}. Our physical layer hardware, however, is not compatible with serial communication protocols. We bridge this gap with an emulator board (Cypress, CY8CKIT-14371~\cite{infineon_cy8ckit_143a_2024}). The emulator board requests and reads \ac{SPI} messages from \ac{QNodeOS} and, based on the message, generate \ac{TTL} signals that are sent as input triggers to the physical layer hardware. The emulator also receives outputs from the physical layer hardware: it monitors whether the hardware is available for new commands or busy, and it collects measurement results and passes them back to \ac{QNodeOS}.
In this case, the measurement result consists of \ac{TTL} signals from the \ac{PMT} detecting ion fluorescence. When a counter value on the emulator board exceeds a certain preset threshold, the ion state is registered as the qubit state $\ket{0}$ and otherwise as $\ket{1}$.

\subsubsection{Testing the QDriver}

Tests were carried out using a trapped-ion setup designed for integration with a fiber-based cavity~\cite{teller2023integrating, teller2021heating}. The qubit states consisted of the $4^2 S_{1/2}$ and $3^2 D_{5/2}$ manifolds of $^{40}$Ca$^+$, hereafter referred to as $\ket{0}$ and $\ket{1}$. The cavity was not used in these tests, which focused on single-qubit operations. The cavity was designed to enable ion-photon entanglement, which we plan to implement in future work through physical instructions from \ac{QNodeOS}. Our primary goal in these tests was to verify that the \ac{QNodeOS} hardware, the emulator, and the physical-layer hardware could work together.

Initial tests confirmed that messages were being exchanged at the programmed clock rate of 50\,kHz and that hardware pulses in the physical layer were triggered correctly via the emulator. Next, the following seven tests were implemented:

\begin{enumerate}
\item Initialization of the ion in a specific Zeeman state via Doppler cooling and optical pumping;
\item a bit flip around the X axis via a $\pi$ pulse with phase 0 on the 729\,nm quadrupole transition of $^{40}$Ca$^+$;
\item a bit flip around the Y axis via a $\pi$ pulse with phase $\pi/2$ on the 729 nm transition;
\item preparation of a superposition state via a $\pi/2$ pulse with phase 0 on the 729\,nm transition;
\item readout of a qubit eigenstate in the Y basis via a $\pi$ rotation around X followed by a $\pi/2$ pulse with phase $\pi/2$ around X on the 729\,nm transition;
\item readout of a superposition state in the X basis via a $\pi/2$ rotation around Y followed by a $\pi/2$ pulse around X on the 729\,nm transition;
\item measurement of the ion’s electronic state via fluorescence at 397\,nm in the presence of an 866\,nm repump, following preparation of a superposition state.
\end{enumerate}

Operations are considered to be correctly realized from the point of \ac{QNodeOS}, but do contain errors at the quantum level. Results for the tests (numbers above) were as follows:
\begin{enumerate}
\item The ion was detected in the target initial state $\ket{0}$ in 98.4\% of trials;
\item Following the X-axis bit flip operation, the ion was detected in $\ket{1}$ in 96\% of trials;
\item Following the Y-axis bit flip operation, the ion was detected in $\ket{1}$ in 95\% of trials;
\item A projective measurement determined that the $^{40}$Ca$^{+}$ ion was in $\ket{0}$ 52\% of the time and in $\ket{1}$ 48\% of the time;
\item A projective measurement determined that the $^{40}$Ca$^{+}$ ion was in $\ket{0}$ 54\% of the time and in $\ket{1}$ 46\% of the time;
\item The ion was detected in $\ket{0}$ 93\% of the time;
\item The ion population was found to be in $\ket{0}$ 37\% of the time and in $\ket{1}$ 63\% of the time.
\end{enumerate}

These results were consistent with the performance of the physical-layer hardware in the absence of \ac{QNodeOS}. (Note that Doppler cooling had not been optimized and that magnetic-field drifts at the time were not properly compensated for. Gate operations with much higher fidelities are typically achieved in trapped-ion experiments, but here our focus was on verifying the electronic signaling.) No problems or inconsistencies with the electronic signaling were identified. 

A next step will be to implement a more sophisticated processing of \ac{PMT} \ac{TTL} signals by the emulator board in order to identify when the ion has been delocalized due to a background-gas collision; in that case, additional laser cooling will be implemented that returns the ion to Doppler-limited temperatures. Such a step is a typical part of compiled physical-layer sequences but should now be implemented within \ac{QNodeOS} as part of the physical instruction for qubit initialization.
\clearpage
\section{Delegated Quantum Computation (DQC) experiment on NV}
\label{sec:delcomp}

\subsection{Procedure}

We execute the application in a tomography way to establish \ac{QNodeOS} the quantum performance metric (\cref{fig:fig3}b, where we use $P_c$ to refer to the client program, and $P_s$ to the server program): The client \ac{CNPU} initiates $P_c$ with fixed $(\alpha, \theta)$. This results in a single \ac{CNPU} process, a single \ac{QNPU} process, and opening of an \ac{ER} socket (see \cref{sec:design_er_socket}) with the server node. At the same time, the server \ac{CNPU} initiates $P_s$ resulting in single \ac{CNPU} process, a single \ac{QNPU} process, and opening of an \ac{ER} socket with the client node. The client and the server programs execute the subroutines in \cref{fig:fig3}c, looping 1200 times: both immediately start the second iteration once the first is completed. After the 1200th iteration, both client and server stop their respective \ac{CNPU} and \ac{QNPU} processes. Source code including compiled NetQASM subroutines is available in \cref{sec:app_source}. We repeat 6 times for $(\alpha, \theta) \in \{\pi/2, \pi\} \times \{\pi/4, \pi/2, \pi\}$ for a total of 7200 executions of the circuit depicted in \cref{fig:fig3}a. We expect $\ket{\psi}$ to be either $\ket{-Y}$ (for $\alpha = \pi/2$) or $\ket{-Z}$ (for $\alpha = \pi$). To estimate the resulting $\ket{\psi}$ per $(\alpha, \theta)$, the contents of S2 (containing the server qubit measurement) in the server loop is was varied such that we obtained 600 measurement outcomes in basis $\ket{+Y}$ ($\ket{+Z}$) and 600 measurement outcomes in the corresponding orthogonal basis $\ket{-Y}$ ($\ket{-Z}$) for $\alpha = \pi/2$ ($\pi$).

Since our experiments are conducted on two \ac{NV} nodes that are directly connected, we install a constant network schedule with time-bins of 10\,ms in which all time-bins are assigned to networking. This allows us to assess the performance of executing quantum network applications without introducing a dependence on changing network schedules. This means the network process is made ready at the start of each such time-bin, although may not instruct the \ac{QDevice} to make entanglement if no requests for entanglement have been made.

\subsection{Definitions}

The result of a single \ac{DQC} circuit execution (\cref{fig:fig3}a) is a single-qubit state $\rho_{\text{DQC}}$ on the server. The success of running \ac{DQC} can be expressed as the fidelity of $\rho_{\text{DQC}}$ compared to the expected state (in case of no noise) $\ket{\psi}$ (\cref{fig:fig3}a). In the following we will call this fidelity the \ac{DQC} fidelity, or $F_{\text{DQC}}$.

The value of $F_{\text{DQC}}$ is affected the most by (1) the fidelity $F_{\text{EPR}}$ of the entangled pair created between the client and server, and (2) the \textit{qubit memory time} $t_{\text{mem}}$, which is the time that the server qubit must remain in memory (from entanglement success until measurement). The latter depends on the time at which the client sends a message to the server (\cref{fig:fig3}). We refer to the two-qubit maximally entangled Bell states as $\ket{\Phi^+} = \left(\ket{00} + \ket{11}\right)$, and $\ket{\Psi^{\pm}} = \left(\ket{01} \pm \ket{10}\right)$, where $\Phi^+ = \proj{\Phi^+}$ and $\Psi^{\pm} = \proj{\Psi^{\pm}}$.

\subsection{Post-Selection Based on Latency}
\label{sec:post-selection-latency}

In our experiments, the server qubit memory time $t_{\text{mem}}$ has a significant variance across executions of the \ac{DQC} circuit. In some iterations, there were huge spikes in latencies, which skew the results significantly. An upper bound $t_{\max}$ (see \cref{sec:dqc-simulation}) was used to filter out results from iterations in which $t_{\text{mem}}$ was larger than $t_{\max}$. This resulted in filtering 146 out of 7200 data points. We note that for computing $F_{\text{DQC}}$, we applied the latency filter on top of the \ac{SSRO} and \ac{CR} filters (see Methods). For the processing time analysis (below), however, we applied only the latency filter directly to all 7200 original data points.

\subsection{Simulation}
\label{sec:dqc-simulation}

A simulation (using NetSquid~\cite{coopmans_2021_netsquid}) of the \ac{DQC} application was performed in order to estimate the expected $F_{\text{DQC}}$ on our \ac{NV} setup, and to establish a suitable value for $t_{\max}$ (used in latency post-selection).

We emphasize that this simulation is a heuristic to find $t_{\max}$, and does not aim to predict the performance to full accuracy. All runs for which latencies were less than $t_{\max}$ were ultimately used to assess the performance from data, not using this simulation.

The simulation contains the following steps, where we 
used the model explained in Ref.~\cite{pompili_2021_multinode}:
\begin{enumerate}
    \item Start with a density matrix $\rho_{\text{EPR}}$ describing the approximate state of the \ac{EPR} pair just after entanglement success.
    \item Apply operations representing the local gates on both the client and server, including the measurement on the client qubit. These operations are assumed to be perfect (no noise).
    \item Apply depolarizing noise to the server qubit for a duration of $t_{\text{mem}}$, using the decoherence formula $e^{-\left(t_{\text{mem}}/T_{\text{coh}}\right)^n}$ where $T_{\text{coh}}$ was set to 13\,ms and $n=1.67$. These values are obtained via fitting experimental data from prior tests.
    \item Calculate the fidelity between the final server qubit state and the expected state $\ket{\psi}$.
\end{enumerate}

Based on the parameters of the setup when the \ac{DQC} experiment was performed, $\rho_{\text{EPR}}$ is set to
\[
\begin{bmatrix}
0.049 & 0 & 0 & 0 \\
0 & 0.437 & 0.284 & 0 \\
0 & 0.284 & 0.454 & 0 \\
0 & 0 & 0 & 0.061 \\
\end{bmatrix}
\]
which has fidelity 0.729 to the perfect $\Psi^+$ state. The setup can also produce $\Psi^-$ states but for simplicity we use only the $\Psi^+$ case here. 

\begin{figure*}[htbp]
    \centering
    \begin{subfigure}{0.48\textwidth}
        \centering
        \includegraphics[width=\linewidth]{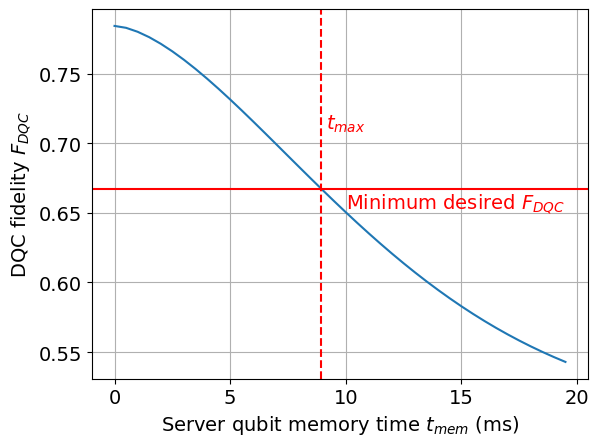}
        \caption{Expected values (based on simulation, \cref{sec:dqc-simulation}) of \ac{DQC} fidelity $F_{\text{DQC}}$ for different duration values that the server qubit must remain in memory ($t_{\text{mem}}$). The maximum allowed qubit memory time $t_{\max}$ is chosen such that application iterations that are expected to result in too low $F_{\text{DQC}}$ ($<0.667$) are filtered out.}
        \label{fig:delcomp-fidelity-idle-time}
    \end{subfigure}\hfill
    \begin{subfigure}{0.48\textwidth}
        \centering
        \includegraphics[width=\linewidth]{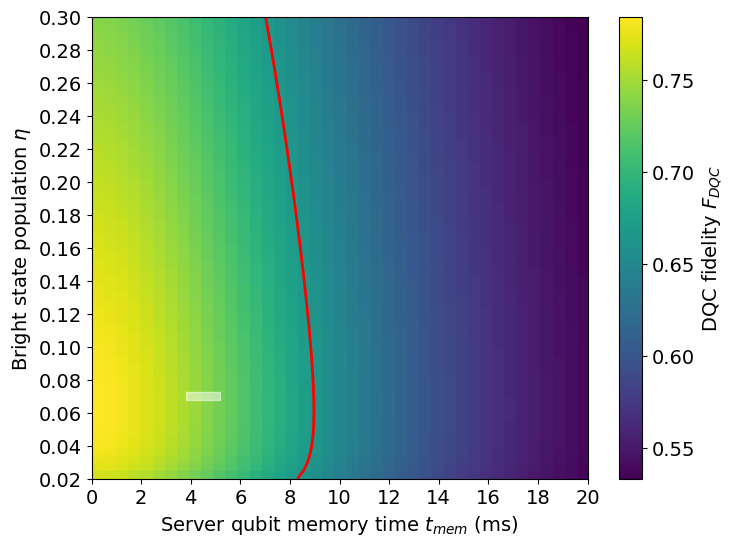}
        \caption{Expected values (based on simulation) of \ac{DQC} fidelity $F_{\text{DQC}}$ for different values of the bright state population ($\eta$) in the single click protocol, and for different duration values that the server qubit must remain in memory ($t_{\text{mem}}$). The red line indicates the threshold of 0.667 for the target fidelity. The white box represents the experimentally obtained results (we fixed $\eta = 0.07$ and observed $t_{\text{mem}} ~4.8(8)$\,ms, see \cref{fig:fig3}d).}
        \label{fig:delcomp-fidelity-alpha-colormap}
    \end{subfigure}
    \caption{Estimated fidelities based on simulation for executing \acf{DQC} on our \ac{NV} setup.}
    \label{fig:delcomp-simulation}
\end{figure*}

The simulation computes an estimate of $F_{\text{DQC}}$ for a given server qubit memory time $t_{\text{mem}}$. Since the desired minimum value for $F_{\text{DQC}}$ was 0.667, the latency threshold $t_{\max}$ was set to 8.95\,ms (\cref{fig:delcomp-fidelity-idle-time}).

\subsection{Sweep of Qubit Memory Time and Bright State Population}

As explained in \cref{sec:NVentanglement}, entanglement is created using the single-photon protocol using bright state population parameter $\eta$.\footnote{In most literature, the variable $\alpha$ is used for this parameter; here we use $\eta$ to avoid confusion with the $\alpha$ parameter of the \ac{DQC} application.} Using the simulation, we can estimate how $F_{\text{DQC}}$ would change for different values of $\eta$ and $t_{\text{mem}}$. \cref{fig:delcomp-fidelity-alpha-colormap} shows the estimated $F_{\text{DQC}}$ for different values of $\eta$ and $t_{\text{mem}}$. It indicates that for the particular setup used, increasing $\eta$ has little effect, while reducing qubit memory time does. For the \ac{DQC} experiment $\eta = 0.07$ was used.

\subsection{Processing Time and Latencies}
\label{sec:processing_time_latencies}

Here we provide a detailed breakdown of the duration of execution phases of the \ac{DQC} application, in order to gain insights into the processing times and latencies of the system for the different components.

\subsubsection{Server qubit memory time}
\label{sec:server-qubit-memory-time}

\begin{figure*}[htbp]
\centering
\includegraphics[width=0.8\linewidth]{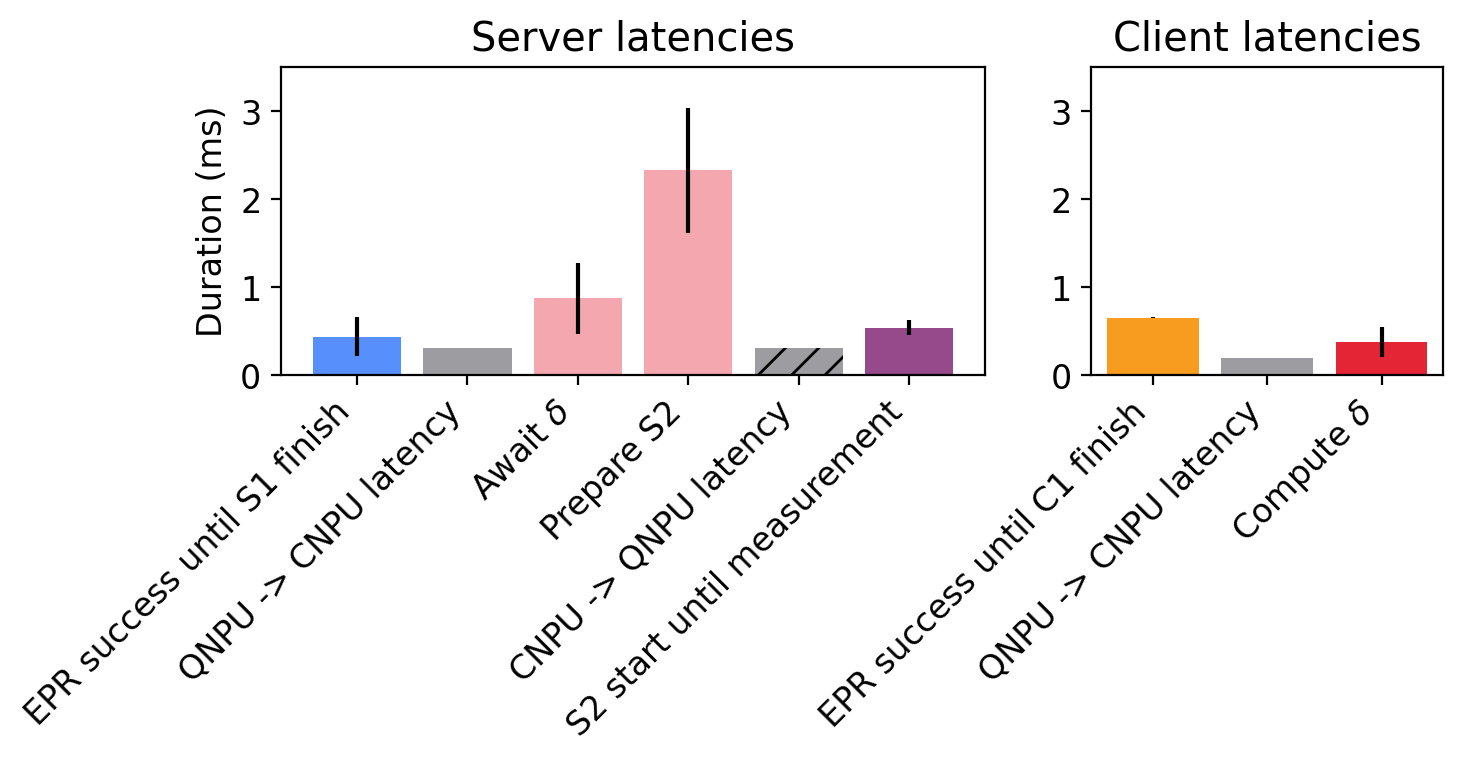}
\caption{Average latency (duration) of each of the processes happening while the server qubit remains in memory in the \ac{DQC} application. The \ac{QNPU} to \ac{CNPU} latency and \ac{CNPU} to \ac{QNPU} latency are estimated as explained in \cref{sec:cnpu-qnpu-latency-estimation}, and fixed to 0.305\,ms (server) and 0.197\,ms (client). The other latencies are the mean and variance of the corresponding processes averaged over all \ac{DQC} circuit iterations that passed the latency filter.}
\label{fig:delcomp-latencies-variance}
\end{figure*}

\Cref{fig:fig3}c shows the duration that the server qubit must remain in memory $t_{\text{mem}}$ while waiting, averaged over all \ac{DQC} circuit iterations that passed the latency filter. \Cref{fig:fig3}d shows the breakdown of $t_{\text{mem}}$ into individual segments of processing on both client and server. In \cref{fig:delcomp-latencies-variance} we show the average duration and the variance of each of these segments. The largest time is spent on preparing S2, which involves running Python code on the \ac{CNPU} and converting this (using Python) into a \ac{NetQASM} subroutine. Caching of the preparation of the \ac{NetQASM} subroutine could significantly speed up this process. In the future, further improvements could include an optimized ahead-of-time compilation step. The large variance is due to the fact that on the \ac{CNPU}, other (background) processes run simultaneously with the \ac{DQC} application process, and there is no precise control over the scheduling of these processes.

\subsubsection{Tracing}

The \ac{CNPU}, \ac{QNPU}, and \ac{QDevice} all keep track of events happening in their system, by storing a tuple $(t, e)$ where $t$ is a timestamp and $e$ the name of the event. The events that are traced on the \ac{CNPU} and \ac{QNPU} are listed in \cref{sec:traces}. A trace plot showing events in \ac{CNPU}, \ac{QNPU}, and \ac{QDevice} during a single execution of the \ac{DQC} circuit is also shown in \cref{sec:traces}.

The \ac{QNPU} timestamp granularity is 10\,$\mu$s, since that is the duration of a single \ac{QNPU} clock cycle. This clock cycle is synchronized with the clock of the \ac{QDevice}, which in turn is synchronized with the \ac{QDevice} of the other node (see \cref{sec:methods} and all paragraphs therein related to \ac{NV} implementation). This results in the two \acp{QNPU} (of the two nodes in the experiment) having synchronized clocks with 10\,$\mu$s precision. This means that the event indicating to the \acp{QNPU} that \ac{EPR} generation has succeeded happens at the same clock cycle on both \acp{QNPU}.

The \ac{CNPU} is not a real-time system (instead, it runs on a general purpose Linux \ac{OS}) and records timestamps by consulting the system clock at $\mu$s precision. These timestamps are not synchronized to the \ac{QNPU} timestamps. Furthermore, the \ac{CNPU} timestamps obtained in this way are not as consistent as the real-time clock ticks on the \ac{QNPU}. Therefore, the relative \ac{CNPU} time compared to the \ac{QNPU} time (on the same node) may fluctuate.




\subsubsection{CNPU-QNPU communication latency}
\label{sec:cnpu-qnpu-latency-estimation}

\begin{figure*}[htbp]
    \centering
    \begin{subfigure}{0.61\textwidth}
        \centering
        \includegraphics[width=\linewidth]{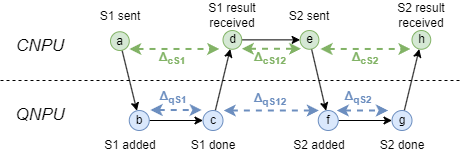}
        \caption{}
        \label{fig:subfig3}
    \end{subfigure}\hfill
    \begin{subfigure}{0.35\textwidth}
        \centering
        \includegraphics[width=\linewidth]{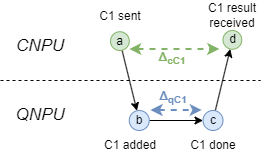}
        \caption{}
        \label{fig:subfig4}
    \end{subfigure}
    \caption{Schematic of events happening on the \ac{CNPU} and \ac{QNPU} during a single execution of \ac{DQC} on the server (a) and the client (b). Time flows to the right. The $\Delta$ variables are the time differences between events, and are used to estimate \ac{CNPU}-\ac{QNPU} communication latencies ($a\rightarrow b$, $c\rightarrow d$, $e\rightarrow f$, $g\rightarrow h$ on the server and $a\rightarrow b$, $c\rightarrow d$ on the client).
    }
    \label{fig:cnpu_qnpu_latencies}
\end{figure*}

The latency of communication between the \ac{CNPU} and \ac{QNPU} can be calculated by looking at the time between \ac{CNPU} events and \ac{QNPU} events. However, since the \ac{CNPU} timestamps are fluctuating compared to the \ac{QNPU} timestamps, we cannot use a direct comparison between \ac{CNPU} and QNPU timestamps. Instead, we look at time differences on the \ac{CNPU} and compare them to time differences on the QNPU, given that we know the order in which events occur during the DQC application execution. \cref{fig:cnpu_qnpu_latencies} shows a schematic overview of events happening on the \ac{CNPU} and the \ac{QNPU} during a single execution of the \ac{DQC} circuit. By comparing, e.g., (1) the time difference on the \ac{CNPU} between sending subroutine S1 and receiving its result with (2) the time difference on the \ac{QNPU} between receiving subroutine S1 and finishing it, we can estimate the total latency of sending S1 from \ac{CNPU} to \ac{QNPU} and receiving its result. Using this technique, we can estimate the latencies for each communication between \ac{CNPU} and \ac{QNPU}, as listed in \cref{tab:delta_diffs}. Again, since the \ac{CNPU} timestamps fluctuate compared to the \ac{QNPU} timestamps, the derived latencies fluctuate and can even be negative. However, for all derived latencies, we found that a constant function best fit the data. This verifies that the actual latency is constant as expected, and that the variance is due to the inaccuracy of \ac{CNPU} timestamps.

\begin{table*}
    \centering
    \begin{tabular}{|c|c|c|}
    \hline
    \textbf{Derived latency (fit)} & \textbf{Description} & \textbf{Value (ms)} \\ 
    \hline
    $\Delta_{cS1} - \Delta_{qS1}$   & Send S1 + receive S1 result  & $0.384$ \\
    $\Delta_{qS12} - \Delta_{cS12}$ & Receive S1 result + Send S2  & $0.609$ \\
    $\Delta_{cS2} - \Delta_{qS2}$   & Send S2 + receive S2 result  & $0.467$ \\
    $\Delta_{cC1} - \Delta_{qC1}$   & Send C1 + receive C1 result  & $0.394$ \\
    \hline
    \end{tabular}
    \caption{Derived values for \ac{CNPU}-\ac{QNPU} communication latencies. The $\Delta$ variables are observed timestamp differences on the \ac{CNPU} or \ac{QNPU}, per execution of the \ac{DQC} circuit, as shown in \cref{fig:cnpu_qnpu_latencies}. Subtracting pairs of variables from each other produces sums of two \ac{CNPU}-\ac{QNPU} communication latencies. These sums of latencies highly fluctuate per execution of the \ac{DQC} circuit, due to the inaccuracy of the \ac{CNPU} timestamps. However, the data fits a constant value, which is shown in the table and used in further analysis.}
    \label{tab:delta_diffs}
\end{table*}

Using the result from \cref{tab:delta_diffs}, we can compute bounds on the four individual latency variables of the server (we have a system of three linear equations, and we know that all latencies must be strictly non-negative):
\begin{itemize}
    \item Sending S1 from \ac{CNPU} to \ac{QNPU}: $<$ 0.242\,ms.
    \item Receiving S1 result on \ac{CNPU} from \ac{QNPU}: between 0.142 and 0.384\,ms.
    \item Sending S2 from \ac{CNPU} to \ac{QNPU}: between 0.225 and 0.467\,ms.
    \item Receiving S2 result on \ac{CNPU} from QNPU: $<$ 0.242\,ms.
\end{itemize}

In the latency breakdown of the server qubit memory time (see \cref{sec:server-qubit-memory-time}) we are only interested in the latencies that happen during the time that the server qubit is in memory. For the server these are the latencies for receiving the S1 result and sending S2. The sum of these two latencies is $\Delta_{qS12} - \Delta_{cS12} = 0.609$\,ms (see \cref{tab:delta_diffs}). For simplicity, we say that both latencies constitute half of this time, as mentioned in the caption of \cref{fig:delcomp-latencies-variance}. Similarly, for the client we are only interested in the latency of receiving the C1 result. For simplicity we take this latency to be the same as that of sending C1, i.e. we use half of $\Delta_{cC1} - \Delta_{qC1}$.

\subsubsection{Entanglement generation}

An overview of all values discussed in this section is given in \cref{tab:entanglement_stats}.

\ac{EPR} generation happens by attempting entanglement repeatedly until success. The \ac{QNPU} sends an \texttt{ENT} physical instruction (\cref{tab:qdevice-instructions}) to the \ac{QDevice}, which starts a batch of physical attempts. Each attempt takes 3.95\,$\mu$s and a batch contains 500 attempts. If a batch fails (no success after 500 attempts), the \ac{QNPU} sends another \texttt{ENT} instruction. \cref{tab:entanglement_stats} lists the average success probability per attempt and per batch that we found in the \ac{DQC} experiments. As explained in \cref{sec:NVentanglement}, the \ac{NV} \ac{QDevice} creates either a $\Psi^+$ or a $\Psi^-$ state. \Cref{tab:entanglement_stats} shows statistics on how often each of these states was created during our experiments.

\begin{figure}
\centering
\includegraphics[width=\linewidth]{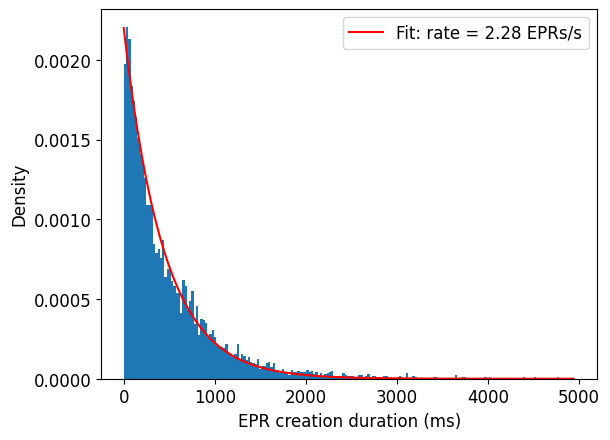}
\caption{Histogram of \ac{EPR} generation durations (time from first attempt until success) based on all \ac{EPR} generations in the \ac{DQC} experiment (using only latency-filtered data points, see \cref{sec:post-selection-latency}). The histogram shows which fraction of all durations were in a particular duration window (window width: 25\,ms). Expected \ac{EPR} generation duration follows an exponential decay, with a rate parameter of 2.28(3) successes (\ac{EPR} pairs) per second.}
\label{fig:delcomp-epr-rate}
\end{figure}

\cref{fig:delcomp-epr-rate} shows the distribution of time it takes to generate an \ac{EPR} pair in the \ac{DQC} experiment, where the average duration of such is 439\,ms. This is the duration between starting the network process and finishing it, which includes entanglement attempts until success on the \ac{QDevice} and subsequent Bell state corrections to $\Phi^+$ (see \cref{sec:qdevice-nv}). This duration corresponds to a fitted rate of 2.28(3) created \ac{EPR} pairs per second. If only the \ac{QDevice} entanglement generation is considered (i.e. without Bell state corrections and without \ac{QNPU} processing overhead), this rate is 2.37(2) \ac{EPR} pairs per second.

\begin{table*}[htpb]
    \centering
    \begin{tabular}{|r|l|}
    \hline
    \textbf{Parameter} & \textbf{Value} \\ 
    \hline
    Duration of a single entanglement attempt* & 3.95\,$\mu$s \\
    Number of attempts per batch* & 500 \\
    Average number of failed batches until success & 144 \\
    Average success probability per batch & $6.95 \times 10^{-3}$ \\
    Average success probability per attempt & $1.39 \times 10^{-5}$ \\
    Number of Psi+ states generation & 3187 (44.3\%) \\
    Number of Psi- states generation & 4013 (55.7\%) \\
    EPR generation rate (fit) (QDevice) & 2.37(2) EPRs/s \\
    EPR generation rate (fit) (QNodeOS) & 2.28(3) EPRs/s \\
    Average fraction of EPR generation time spent on sync failure & 0.18 \\
    
    \hline
    \end{tabular}
    \caption{Overview of values derived from the \ac{DQC} experiment analysis, based on all 7200 \ac{DQC} circuit executions. Entries with an asterisk (*) are values that we fixed in our experiments. The other values are observed experimental results. Average success probabilities are derived from the number of failed batches until success.
    \ac{EPR} generation rate is distinguished between \ac{QDevice} and \ac{QNodeOS}. For the \ac{QDevice}, it indicates the fitted (to an exponential decay function) time between the first \texttt{ENT} physical instruction and the first entanglement success (see \cref{sec:appendix-qdevice}). For \ac{QNodeOS}, it indicates the fitted time between the start of the network process and the end of the network process (i.e. when entanglement has been created and Bell state corrections have been applied, see \cref{sec:qdevice-nv}). Entanglement sync failures happen when one \ac{QDevice} (server or client) wants to attempt entanglement but the other \ac{QDevice} is not ready (\cref{sec:qdevice-sync}). Such sync failures were observed intermittently during a batch of entanglement attempts.}
    \label{tab:entanglement_stats}
\end{table*}

\subsubsection{Local gate durations}

As part of the \ac{DQC} execution, the \ac{QNPU} sends physical instructions to the \ac{NV} \ac{QDevice} for executing local quantum gates. In \cref{tab:gate_durations} we report on the observed durations of these gates from the perspective of the \ac{QNPU}: these durations are from the time the physical instruction is sent to the \ac{QDevice} until the corresponding result is received from the \ac{QDevice}. We note that these durations are longer than these gates would take if they were executed directly on the \ac{QDevice} (without \ac{QNodeOS}, see \cref{tab:mw_pulse}) because of two reasons: (1) the limited granularity with which the \ac{QNPU} and \ac{QDevice} communicate (rounds of 10\,$\mu$s) and (2) the fact the \ac{QDevice} interleaves \ac{DD} sequences in between sequences for the physical instruction itself, as explained in \cref{sec:methods}.

\begin{table*}
    \centering
    \begin{tabular}{|c|c|c|}
    \hline
    \textbf{Physical instruction} & \textbf{Duration (client)} & \textbf{Duration (server)} \\ 
    \hline
    Measure & 130--160\,$\mu$s & 80 - 100\,$\mu$s \\
    X90 & 80--100\,$\mu$s & 50 - 130\,$\mu$s \\
    X180 & 80--100\,$\mu$s & 10 - 130\,$\mu$s \\
    -X90 & --- & 50 - 130\,$\mu$s \\
    Y90 & 70--200\,$\mu$s & 50 - 130\,$\mu$s \\
    Y90 & --- & 50 - 130\,$\mu$s \\
    \hline
    \end{tabular}
    \caption{Duration of executing local quantum gates on the \ac{NV} \ac{QDevice} in the \ac{DQC} experiment. Durations are from sending the physical instruction from \ac{QNPU} to \ac{QDevice} until receiving the \ac{QDevice} response. The -X90 and Y90 gates were never executed in the client \ac{DQC} program.}
    \label{tab:gate_durations}
\end{table*}

\subsubsection{General experiment statistics}

\cref{tab:exp_durations} lists statistics about the overall \ac{DQC} experiment (all 7200 \ac{DQC} circuit executions combined). We confirm our hypothesis that the overwhelming fraction of time is spent on the network process, namely generating \ac{EPR} pairs. We also see that as expected, the server spends more time on user processes than the client does, since it does more local gates than the client (namely, the gates in subroutine S2).

\begin{table*}[htpb]
    \centering
    \begin{tabular}{|r|c|c|}
    \hline
    \textbf{Value} & \textbf{Client} & \textbf{Server} \\ 
    \hline
    Total experiment duration & 4243\,s & 4065\,s \\
    Time spent executing network process & 3840\,s & 3825\,s \\
    Time spent executing user processes & 5.041\,s & 7.618\,s \\
    \hline
    \end{tabular}
    \caption{Overall durations of the \ac{DQC} experiment.}
    \label{tab:exp_durations}
\end{table*}

\subsection{QNPU Network process analysis}

In this section we focus on the execution of the network process in the \ac{QNPU} as observed in the execution of \ac{DQC}. The \ac{ER} sockets~\ref{sec:design_er_socket} are designed to facilitate the generation of entanglement belonging to a pair of user processes between two different \acp{QNPU}. In particular, the \ac{ER} socket allows the \ac{QNPU} to proceed with entanglement generation, while only one node may not have issued a request for entanglement yet. 

During execution of the \ac{DQC} application, the client \ac{QNPU} has a single user process $P_c$ for its \ac{DQC} program and the server \ac{QNPU} has a single user process $P_s$ for its \ac{DQC} program. Both user processes realize the repeated execution of subroutines that jointly realize the \ac{DQC} circuit (\cref{fig:fig3}a).

In each single repetition of the \ac{DQC} circuit, $P_s$ executes first S1 and then S2, and $P_c$ executes C1. $P_s$ (in S1) and $P_c$ (in C1) execute a \ac{NetQASM} instruction for creating an entangled pair, which results in an entanglement request that is submitted to the network stack. Then, $P_c$ and $P_s$ go into the waiting state (see \cref{sec:design:processes}) until the entangled pair is delivered by the network process.

$P_c$ executes a \texttt{create\_epr} instruction and $P_s$ executes a \texttt{recv\_epr} instruction (determined by program source code, see \cref{sec:app_source}. Therefore, the client is seen as the \emph{initiator} (see \cref{sec:design_er_socket}). $P_s$ and $P_c$ open a pair of \ac{ER} sockets with each other when they start and keep it open for the whole experiment. $P_c$ and $P_s$, being on different nodes, operate independently, and may hit their entanglement request instruction at different times. Since the client is the initiator and the server the receiver, the server is always willing to handle an entanglement request with the client. So, the network stack on both client and server will handle a request for entanglement as soon as the client submitted it to its network stack, regardless of whether the server already executed the corresponding \texttt{recv\_epr} in S1.

We observe that in 3245 out of all 7200 \ac{DQC} circuit executions, the client submitted the corresponding entanglement request to its network stack (in C1) \textit{before} the server submitted its entanglement request to its own network stack (in S1), but where the server still complied by starting the network process and handling the request.

\subsubsection{Client waits for server}

From our architecture, we expect that it can happen that the client must wait for the server. This can be the case in the following scenario: The client executes C1 for \ac{DQC} circuit iteration $i$ and submits the entanglement request. Then, the next network time bin starts and the client \ac{QNPU} starts the network process. However, the server is at this time (the beginning of the time bin) still busy with executing S2 for iteration $i-1$ (in user process $P_s$). Therefore the server \ac{QNPU} cannot yet activate its own network process. Since the \ac{ER} socket with the server is open and the client is the `initiator', the client will send entanglement physical instructions to the \ac{QDevice} anyway, but the \ac{QDevice} will not be able to do actual attempts because the server \ac{QDevice} is not ready (\cref{sec:qdevice-sync}). Only when the server \ac{QNPU} completes S2, it can activate the network process, which then sends entanglement physical instructions to the \ac{QDevice}. Only at this point the \acp{QDevice} can start actual entanglement generation. We observe that it did indeed happen that the client had to wait for the server, although we observed this behaviour in only in 60 out of 7200 \ac{DQC} circuit executions.

\subsubsection{Server waits for client}

We expect that it can also happen that the server must wait for the client. This can be the case in the following scenario: The server executes S1 for \ac{DQC} circuit iteration $i$ and submits the entanglement request. Then, the next network time bin starts. However, the client did not yet hit the entanglement request in C1 for \ac{DQC} iteration $i$, so there is nothing to do for the server network process. The server hence needs to wait for the next time-bin, and check again if by now the client has submitted its entanglement request. We observe that in 1323 out of 7200 \ac{DQC} circuit executions, the server had to wait for the client.

\subsubsection{Start of network process}

We examine the start of the network process in relation to the start of a time bin. In particular, the start of the network process may be delayed if there is still a user process running. 

The network process is only activated at the beginning of a time bin. In our experiment, a time bin starts every 10\,ms and lasts 10\,ms. In most cases when the network process is activated, this activation happens very quickly after the time bin start (within 100\,$\mu$s, as some \ac{QNPU} software processing is needed). For the client \ac{QNPU}, the network process never starts more than 100\,$\mu$s after a time bin start. For the server, in 13 out of 7200 \ac{DQC} circuit executions, the network process starts more than 100\,$\mu$s after a time bin starts, since in these cases there was still a user process running. In \cref{tab:network_process_stats}, an overview of all network process statistics is given.

\begin{table*}
    \centering
    \begin{tabular}{|l|l|}
    \hline
    \textbf{Parameter} & \textbf{Value} \\ 
    \hline
    Number of times server puts \ac{EPR} request to network stack before client & 1774/7200 \\
    Number of times server starts entanglement before putting in \ac{EPR} request & 3245/7200 \\
    Number of times submitted \ac{EPR} request is handled in immediate next time bin & 5523/7200 \\
    Average number of bins that pass before request is handled & 2.33 \\
    Number of times server needs to wait for client & 1323/7200 \\
    Number of times client needs to wait for server & 60/7200 \\
    Number of times client network process starts > 100\,$\mu$s after time bin starts & 0 \\
    Number of times server network process starts > 100\,$\mu$s after time bin starts & 13 \\
    \hline
    \end{tabular}
    \caption{Statistics on the \ac{QNPU} network process behavior during the whole \ac{DQC} experiment, i.e. totalled over all 7200 \ac{DQC} circuit iterations.}
    \label{tab:network_process_stats}
\end{table*}
\clearpage
\section{Multitasking experiments on NV}

The multitasking evaluation was done in two parts:
\begin{itemize}
    \item \textbf{Quantum tomography while multitasking}: Executing a single \ac{DQC} application (on client and server) and a single \ac{LGT} application (on client only) where it was verified that the \ac{LGT} application produces expected quantum results (see~\cref{sec:multitasking-tomography}).
    \item \textbf{Scaling the number of applications}: Executing $N$ \ac{DQC} applications and $N$ \ac{LGT} applications, where the classical device utilization metric was compared with a version of \ac{QNodeOS} without multitasking, and where we investigated the behavior of the \ac{QNPU} scheduler on the client in the context of multiple programs (see~\cref{sec:multitasking-scaling}).
\end{itemize}
The network schedule was set as in the previous \ac{DQC} experiment for direct comparison.

\subsection{Mocked entanglement}
\label{sec:mocked_entanglement}

For the multitasking evaluation, we focused on the behavior of \ac{QNodeOS}, and opted not to use the standard entanglement generation procedure in our \ac{NV} \acp{QDevice} as done in the \ac{DQC} experiments (\cref{sec:delcomp}) to allow for a simpler experiment. Instead, we used a mocked entanglement generation process on the \acp{QDevice} (executing entanglement actions without entanglement): Weak-coherent pulses on resonance with the \ac{NV} transitions, that follow the regular optical path, are employed to trigger the \ac{CPLD} in the entanglement heralding time-window.

We stress that in our multitasking experiments, the exact same physical instructions are sent to the \ac{QDevice} as would be done when using real entanglement, and the exact same responses are sent back. Therefore, \ac{QNodeOS} needed to perform the same operations (including scheduling decisions) as it would have needed to do with real entanglement. Furthermore, we aimed to keep the rate of entanglement `success' in the \acp{QDevice} the same order of magnitude as that of the \ac{DQC} experiments (10.14\,\acp{EPR}/s compared to 2.37\,\acp{EPR}/s in the \ac{DQC} experiment) by keeping the mean-photon number of the weak-coherent pulse comparable to $p_C$ and $p_S$ (in the order of $\sim10^{-4}$). 

\subsection{Tomography results}
\label{sec:multitasking-tomography}

\begin{figure}
\centering
\includegraphics[width=\linewidth]{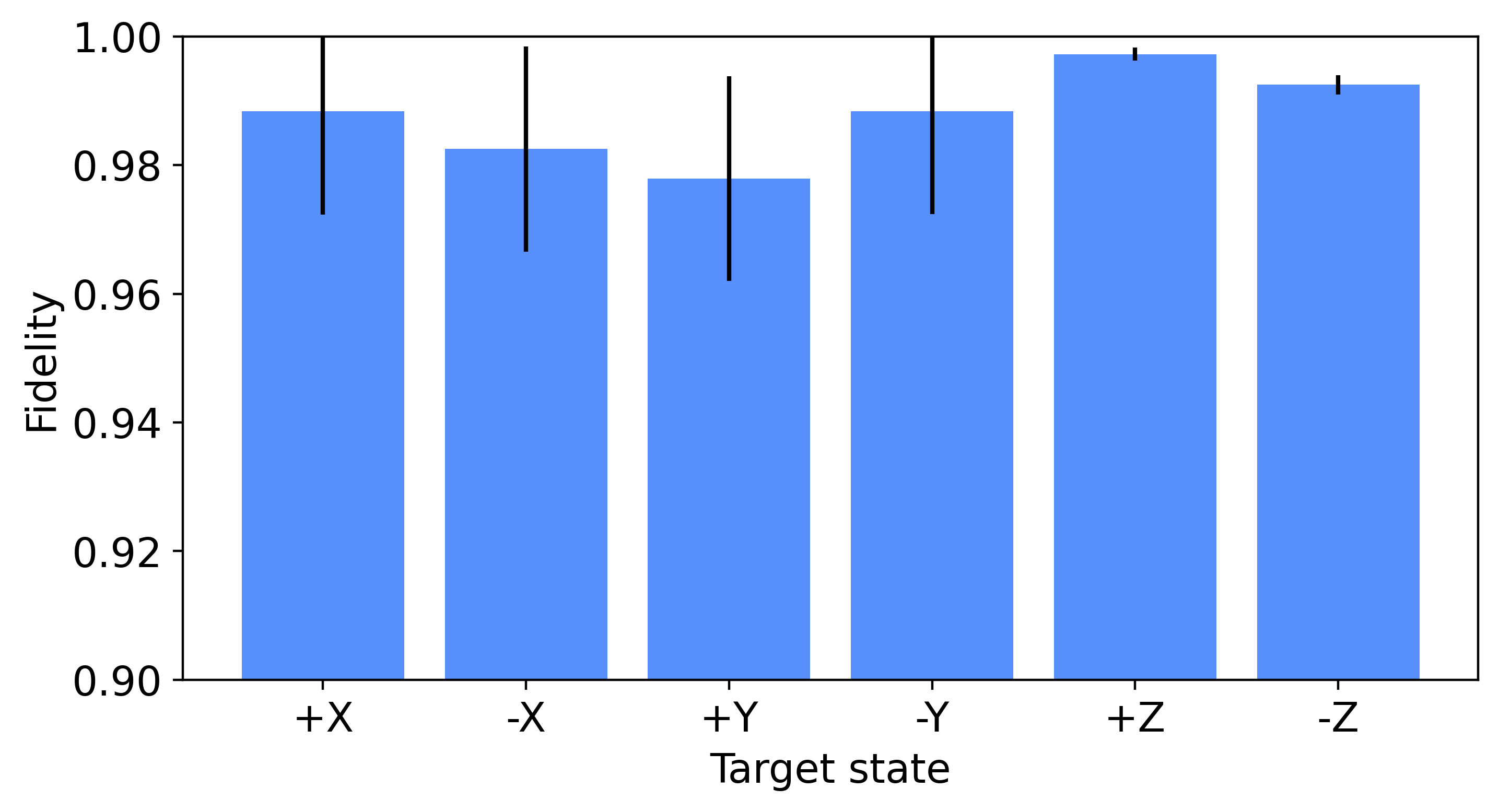}
\caption{Local Gate Tomography results on the client node in a non-multitasking scenario.}
\label{fig:tomography-no-multitasking}
\end{figure}

We perform tomography when not multi-tasking, in order to verify our expectation that multi-tasking should not affect the quantum performance of \ac{LGT}: The tomography results of the \ac{LGT} application in the multitasking scenario are given in \cref{fig:fig4}c. We also ran the same \ac{LGT} application on the client in a non-multitasking scenario. In this case, the client ran the \ac{LGT} application and there was no \ac{DQC} application run at all (the server did nothing). The tomography results of \ac{LGT} for the non-mulitasking scenario are given in \cref{fig:tomography-no-multitasking}. The results are slightly different since the multitasking experiment was done on a different day than the non-multitasking experiment. However, within error bars we verify that multitasking does not affect the quantum performance of the \ac{LGT} application.

\subsection{Scaling to more than two applications}
\label{sec:multitasking-scaling}

\subsubsection{QNPU processes and steps}

\begin{figure*}[!htb]
\centering
\includegraphics[width=\linewidth]{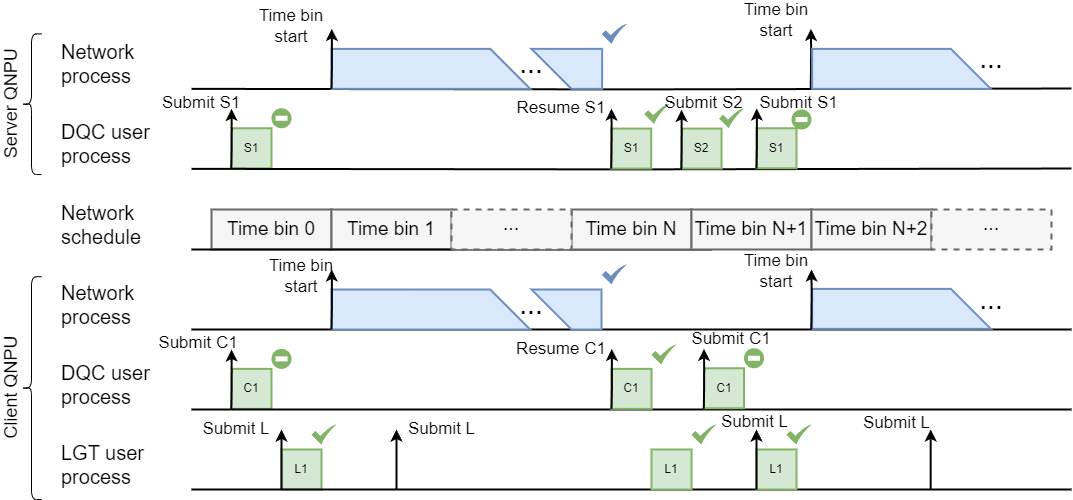}
\caption{Nominal scheduling pattern on the client and server \acp{QNPU} when multitasking 1 \ac{DQC} application (on client and server) and 1 \ac{LGT} application (on client only). Pictured is a slice of time (moving to the right) in which a whole \ac{DQC} circuit execution is realized, and 3 \ac{LGT} circuit executions. Up-arrows indicate that the process becomes \textit{ready} (either since a subroutine was submitted from the \ac{CNPU}, or because a requested entangled qubit becomes available). Green blocks are \ac{NetQASM} subroutines. Blue blocks are entanglement generation. Ticks indicate completion of a subroutine (user process) or entanglement request (network process). Stop sign means the user process goes into the \textit{waiting} state. Time not to scale. Time bin length is 10\,ms. Duration of L1 is $\approx$2.4\,ms. Duration of entanglement generation is non-deterministic. On the server \ac{QNPU} the following happens. S1 arrives from \ac{CNPU}; \ac{DQC} user process becomes \textit{ready}. \ac{DQC} user process is activated and executes S1. The entanglement instruction inside S1 is reached; entanglement request is sent to network stack; \ac{DQC} user process becomes \textit{waiting}. When time bin 1 starts, network process becomes \textit{ready}. There is a pending entanglement request, so network process is activated; \ac{QDevice} attempts entanglement until success (after non-deterministic number of time bins, blue tick). Requested entangled qubit is available: \ac{DQC} user process becomes \textit{ready} again; is activated; executes S1 until completion; becomes \textit{idle}. \ac{QNPU} receives subroutine S2 from \ac{CNPU}; activates \ac{DQC} user process; executes S2 until completion. At this point, the \ac{QNPU} completed execution of the current repetition of the \ac{DQC} circuit. \ac{QNPU} then receives again a subroutine S1 (for the next \ac{DQC} circuit iteration), and the same pattern repeats. On the client \ac{QNPU} the following happens. C1 arrives from \ac{CNPU}; \ac{DQC} user process becomes \textit{ready}. \ac{DQC} user process is activated and executes C1. The entanglement instruction inside C1 is reached; entanglement request is sent to network stack; \ac{DQC} user process becomes \textit{waiting}. L1 arrives from \ac{CNPU}; \ac{LGT} user process becomes \textit{ready}. \ac{LGT} user process is activated; fully executes L1. When time bin 1 starts, network process becomes \textit{ready}. There is a pending entanglement request, so network process is activated; \ac{QDevice} attempts entanglement until success (blue tick). While network process is active, another L1 block arrives from \ac{CNPU} (for next \ac{LGT} circuit iteration) so \ac{LGT} user process becomes \textit{ready}. \ac{LGT} user process is not activated since network process is still running. Upon entanglement success, requested qubt is available; \ac{DQC} user process is activated to complete C1. \ac{QNPU} has now completed execution of the current repetition of the \ac{DQC} circuit. \ac{LGT} user process is activated to execute L1 which was still pending. The same pattern repeats.
}
\label{fig:multitasking-default}
\end{figure*}

For the scaling evaluation, we did an experiment for each $N \in \{1, 2, 3, 4, 5\}$. For each experiment, the client \ac{CNPU} started $N$ \ac{DQC}-client programs and $N$ \ac{LGT} programs concurrently (pseudocode in \cref{src:cnpu_runner}), and the server \ac{CNPU} started $N$ \ac{DQC}-server programs. In this section we discuss the observed behavior of the client and server \acp{QNPU} during these experiments. The client \ac{QNPU} has $2N$ user processes ($N$ \ac{DQC} user processes and $N$ LGT user processes), each of which continuously receives quantum blocks in the form of \ac{NetQASM} subroutines (C1 for \ac{DQC} processes and L1 for \ac{LGT} processes). These $2N$ user processes and the single client network process are scheduled by the client \ac{QNPU} scheduler. The server has $N$ user processes (all for \ac{DQC}) which are scheduled together with the server network process by the server \ac{QNPU} scheduler. \cref{fig:multitasking-default} shows a schematic diagram of the nominal (most often occurring) pattern of scheduling.

In both S1 and C1, there is a single \texttt{create\_epr} \ac{NetQASM} instruction~\cite{dahlberg_2022_netqasm} for creating entanglement with the other node, followed by a \texttt{wait\_all} \ac{NetQASM} instruction that waits until the request entangled qubit is delivered. The \texttt{create\_epr} instruction is handled by the \ac{QNPU} processor by sending the entanglement request to the network stack. Upon executing the \texttt{wait\_all} instruction, the user process executing this subroutine (S1 or C1) goes into the waiting state (green stop sign in \cref{fig:multitasking-default}). When the network process completes (having created the entangled qubit), the user process can be resumed, finishing the subroutine (C1 or S1). 

On the server \ac{QNPU}, for each \ac{DQC} user process $U$ the following sequence is repeated:
\begin{itemize}
    \item $U$ is in the \textit{idle} state;
    \item \ac{NetQASM} subroutine S1 is submitted by the \ac{CNPU} to the \ac{QNPU}, moving $U$ to \textit{ready};
    \item $U$ is activated; S1 is executed until it hits the \texttt{wait\_all} instruction; $U$ goes into the \textit{waiting} state;
    \item The network process handles the entanglement request for S1 until \ac{EPR} creation succeeds; $U$ goes into \textit{ready} again;
    \item $U$ is activated; S1 is executed until completion; $U$ goes to \textit{idle};
    \item \ac{NetQASM} subroutine S2 is submitted by the \ac{CNPU}; $U$ goes to \textit{ready};
    \item $U$ is activated; S2 is executed until completion; $U$ goes to \textit{idle}.
\end{itemize}
The above sequence is for one execution of the \ac{DQC} circuit (\cref{fig:fig3}a), and is hence repeated many times.

On the client \ac{QNPU}, for each \ac{DQC} user process $U$ the following sequence is repeated:
\begin{itemize}
    \item $U$ is in the \textit{idle} state;
    \item \ac{NetQASM} subroutine C1 is submitted by the \ac{CNPU}, moving $U$ to \textit{ready};
    \item $U$ is activated; C1 is executed until it hits the \texttt{wait\_all} instruction; $U$ goes into the \textit{waiting} state;
    \item the network process handles the entanglement request for C1 until \ac{EPR} creation succeeds; $U$ goes into \textit{ready} again;
    \item $U$ is activated; C1 is executed until completion; $U$ goes to \textit{idle}.
\end{itemize}
The above sequence is for one execution of the \ac{DQC} circuit (\cref{fig:fig3}a), and is hence repeated many times.

On the client \ac{QNPU}, for each \ac{LGT} user process $U$ the following sequence is repeated:
\begin{itemize}
    \item $U$ is in the \textit{idle} state;
    \item \ac{NetQASM} subroutine L1 is submitted by the \ac{CNPU}, moving $U$ to \textit{ready};
    \item $U$ is activated; L1 is executed until completion; $U$ goes to \textit{idle}.
\end{itemize}
The above sequence is for one execution of the \ac{LGT} circuit (\cref{fig:fig4}a), and is hence repeated many times.

\begin{figure*}
\centering
\includegraphics[width=\linewidth]{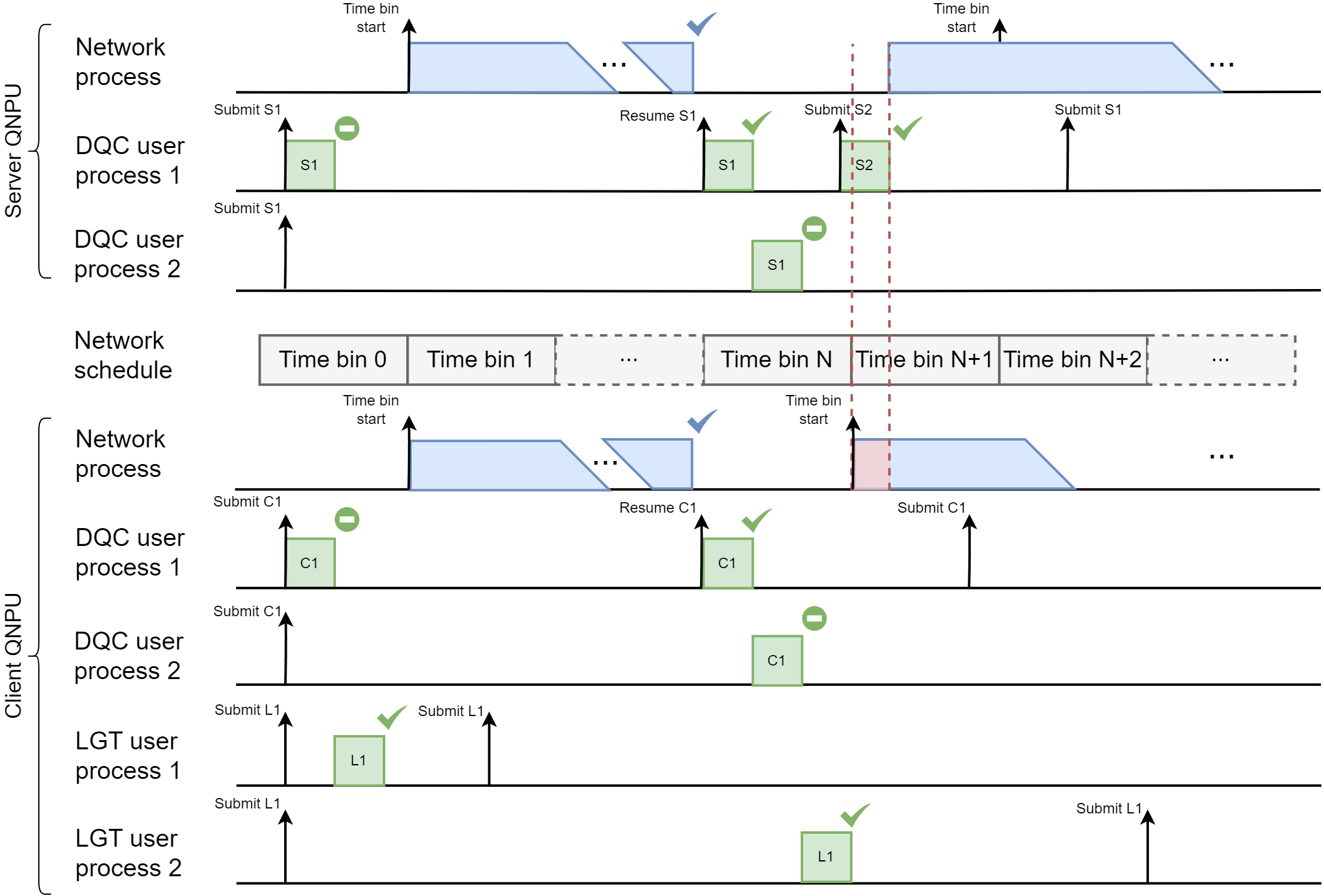}
\caption{Example scheduling pattern of scenario with 2 \ac{DQC} applications and 2 \ac{LGT} applications (the symbol and color coding is the same as in~\cref{fig:multitasking-default}). In this case, the client needs to wait (red shared area) for the server to finish S2 of \ac{DQC} user process 1, before they can do entanglement generation for \ac{DQC} user process 2. Scenario: 2 \ac{DQC} applications (A1 and A2) are concurrently executed (A1: \ac{DQC}-server program executed by server \ac{DQC} user process 1 and \ac{DQC}-client program executed by client \ac{DQC} user process 1; A2: \ac{DQC}-server program executed by server \ac{DQC} user process 2 and \ac{DQC}-client program executed by client \ac{DQC} user process 2). Client and server successfully create entanglement for some \ac{DQC} circuit execution $i$ for A1 (just after time bin $N$ starts). Client finishes C1 for user process 1, and meanwhile the server finishes S1 for user process 1. The client has completed its part of \ac{DQC} circuit execution $i+1$ for A1, but the server still needs to wait for S2 from the \ac{CNPU}. Then, the client executes C1 for user process 2, which is the start of circuit execute $j$ for A2; user process 2 becomes waiting. Meanwhile the server executes S1 for user process 2 which becomes waiting. The client needs to wait until the start of the next time bin ($N+1$) until it can activate the network process to handle the request. In the meantime, it can execute an L1 block. Time bin $N+1$ starts and the client handles the request. However, the server has received S2 for execution $i$ of A1, and starts executing it just before the time bin starts. Only after finishing it, the server can start the network process, which picks up the request for A2. While S2 is executing, the client \ac{QDevice} tries to do entanglement attempts, but gets entanglement sync failures~\cref{sec:qdevice-sync} since the server \ac{QDevice} is busy with S2.}
\label{fig:multitasking-2-apps}
\end{figure*}

\begin{figure*}
\centering
\includegraphics[width=0.8\linewidth]{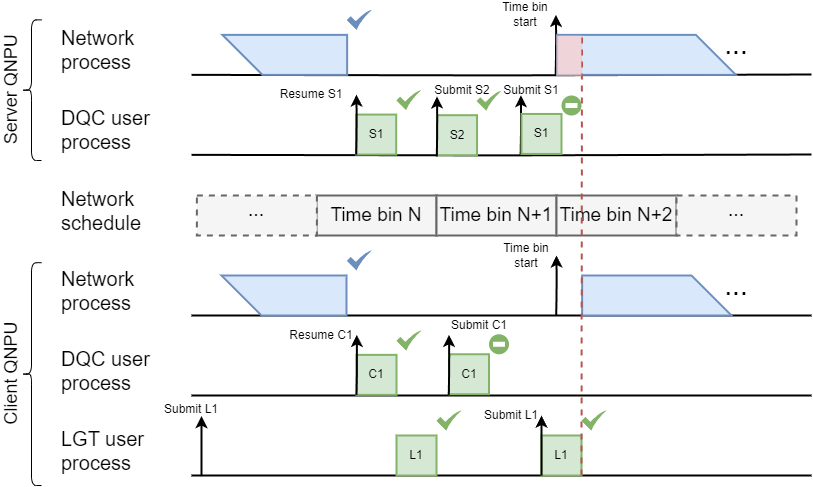}
\caption{Example scheduling pattern of multitasking one \ac{DQC} application (on client and server) and one \ac{LGT} application (on client only), where the server must wait for client to finish its \ac{LGT} user process (red area); the symbol and color coding is the same as in~\cref{fig:multitasking-default}. At the start of time bin $N+1$, the server activates the network process to handle the request that was put by the previous S1 execution. However the client only starts some time later during the time bin, since it first needs to finish executing L1 for the \ac{LGT} user process.}
\label{fig:multitasking-wait-on-client}
\end{figure*}

For the above sequences for user processes, only the internal order is fixed; the time in between steps depends on the \ac{QNPU} scheduler, as well as the time at which the \ac{CNPU} submits subroutines. Furthermore, since there are multiple user processes at the same time (for the server, only for $N > 1$), the above steps happen for each user process $U_i$ and the steps are interleaved. \cref{fig:multitasking-default,fig:multitasking-2-apps,fig:multitasking-wait-on-client} show examples of how these user processes can be interleaved on both client and server \ac{QNPU}.

\subsubsection{DQC and LGT interleaving}

We investigate the degree of interleaving the execution of \ac{DQC} and \ac{LGT}, in particular how many \ac{LGT} subroutines are executed when a \ac{DQC} process is waiting: The client \ac{QNPU} executes both \ac{DQC} and \ac{LGT} user processes. \ac{DQC} user processes are often in the waiting state. This happens when their C1 subroutine is suspended, waiting for the network process to handle their entanglement request. The network process is only activated at the beginning of a time bin, which happens only every 10\,ms, or when a user process finishes executing a subroutine, the latter not occurring very frequently for low number of programs $N$. Furthermore, \ac{DQC} user processes can be in the idle state, namely when they completed execution of C1 for some iteration $i$ of the \ac{DQC} circuit, but are still waiting for the \ac{CNPU} to send C1 for iteration $i+1$. In both these types of `gaps` (waiting or idle), \ac{LGT} subroutines can be executed (each taking $\approx$2.4\,ms). \cref{tab:multitasking_numbers} lists the maximum number of consecutive \ac{LGT} subroutines that were executed in between \ac{DQC} subroutines  for both types of gaps. 

\subsubsection{Subroutine (Quantum block) execution order}

We investigate whether the \ac{QNPU} schedules quantum subroutines in a different order than they arrived from the \ac{CNPU}. As expected, we find that this is the case. Although the \ac{QNPU} handles subroutines from the \ac{CNPU} first-come-first-served, some of these subroutines (in our experiments, precisely the \ac{DQC} subroutines that wait for entanglement) are put into the waiting state. This allows the \ac{QNPU} to schedule other subroutines (in our experiments, we observe \ac{LGT} subroutines being executed), even if they arrived later from the \ac{CNPU} than the waiting \ac{DQC} subroutine. Schematic overviews of such scheduling that we observed are depicted in \cref{sec:multitasking_patterns}.

\subsubsection{User process idle times}

We examine the number of times, and the duration, that a user process is idle waiting for submission of a subroutine from the \ac{CNPU} as a function of $N$: A user process is \textit{idle} when there are currently no subroutines associated with the process pending to be executed. This means that the \ac{QNPU} waits, at least for this user process, until the \ac{CNPU} sends the next subroutine for the user process. \cref{tab:multitasking_numbers} lists the number of times and durations of moments at which all client \ac{QNPU} user processes are idle. This number and their durations decrease for larger values of $N$. This is expected since there are more active processes, and hence more subroutines being sent from the \ac{CNPU} for different processes. In most cases, when finishing a subroutine for user process $U$, there is then another user process $U'$ already waiting with another subroutine to execute.

\subsubsection{Network process start delays}

We examine the scheduling behaviour of the network process in relation to user processes. We expect that due to the fact we use a non-preemptive scheduler, a network process may not be activated at the start of a network time bin, due to a user process still being executed. We investigate the occurrence of such events in our multi-tasking experiment, including the delay with which the network process is started in such a scenario (see \cref{tab:multitasking_numbers}): When a user process submits and entanglement request to the network stack, this request is handled at the earliest when the network process is activated. This happens either at the start of the next network time bin, or when a user process finishes a subroutine. Therefore, there is often some time in between submitting the request and the network process handling it. This waiting time is in most cases bounded by 10\,ms, since that is the length of a time bin, and all time bins are assigned to networking in our experiment. However, in some cases the client may still be executing a \ac{LGT} subroutine when a new time bin starts, delaying the start of the network process until this subroutine has finished. We expect however that in all cases, as soon as such an \ac{LGT} subroutine finishes, the \ac{QNPU} scheduler then immediately schedules the network process, and not another \ac{LGT} subroutine. We found that the maximum difference between time bin start and network process start is 2.59\,ms, which verifies that indeed at most one \ac{LGT} subroutine is sometimes executed during a time bin start (\ac{LGT} subroutine execution duration being $\approx$2.4\,ms.)

We remark that with increasing $N$, the network process is delayed more frequently by a \ac{LGT} subroutine. This is expected due to the fact more subroutines from different user processes await execution. Consequently, with increasing $N$ it also happens more frequently that the client and server do not start execution of the network process in the same time-bin (see below).

\subsubsection{Client waits for server and vice versa}

In order to better understand the concurrent execution of multiple applications (here \ac{DQC} and \ac{LGT}) and corresponding programs, we investigate scenarios and times in which the client waits for the server (or vice versa). 

The client and server open an \ac{ER} socket at the beginning of each \ac{DQC} application. So, during runtime, there are $N$ \ac{ER} sockets opened on the server \ac{QNPU} (one for each \ac{DQC} process) and $N$ \ac{ER} sockets opened on the client \ac{QNPU} (one for each \ac{DQC} process). In each \ac{DQC} application, the client \ac{QNPU} user process for that \ac{DQC} application is the `initiator' (see \cref{sec:design_er_socket}). This means that as soon as the client user process submits a request for entanglement (from within C1), both server and client \ac{QNPU} start their network process to handle it (at the start of the next time bin, and provided the network process should not first handle a request from a user process from another \ac{DQC} application).

It can happen that the client \ac{QNPU} and server \ac{QNPU} do not start their network process at the same time bin. This mostly happens when one of the nodes is still busy executing a user process subroutine when a time bin starts, as explained above. If this happens, the \ac{QNPU} that did already start their network process sends entanglement instructions to their \ac{QDevice}, but this will not result in physical entanglement attempts since the other \ac{QDevice} is not available (leading to a entanglement sync failure, see \cref{sec:qdevice-sync}). \cref{tab:multitasking_numbers} lists the number of times that this happened.

For each of the $N$ \ac{DQC} applications that are running on client and server, and for each execution of the \ac{DQC} circuit in those applications, there is a single entanglement request from the client (in C1) and a single entanglement request from the server (in S1). For each of these request pairs, the client at some point starts the network process and handles this request, and the server at some point starts the network process and handles its corresponding request. For each such pair of requests, the following scenarios can happen:
\begin{enumerate}
    \item Client and server \ac{QNPU} start their network process in the same time-bin (one of them may start a bit later than the start of the time-bin because it needs to complete a quantum subroutine).
    \item The client starts its network process in time-bin $k$ but the server starts it at some time-bin $>k$. This happens when the server still has a qubit in memory when time-bin $k$ starts. Therefore, the server cannot activate its network process yet. A qubit still being in memory happens when the server \ac{QNPU} has executed S1 for some \ac{DQC} process (which produced an entangled qubit) but has not yet executed S2 (in which the qubit is measured and hence freed).
    \item The server starts its network process in time-bin $k$ but the client starts it at time-bin $k+1$. This happens (although rarely) when the client user process puts the entanglement request to the network stack just before the start of $k$. The server will immediately start attempts at $k$, but the client itself is still processing and `misses' $k$; the client then only starts at time-bin $k+1$.
\end{enumerate}

\cref{tab:multitasking_numbers} lists how often the above scenarios happen for each $N$.

\begin{table*}
    \centering
    \begin{tabular}{|r|c|c|c|c|c|}
    \hline
    \textbf{Parameter} & \textbf{N = 1} & \textbf{N = 2} & \textbf{N = 3} & \textbf{N = 4} & \textbf{N = 5} \\ 
    \hline
    average no. \ac{LGT} subroutines in between any \ac{DQC} subroutines & 0.83 & 1.42 & 1.59 & 1.65 & 1.65 \\
    max no. consecutive \ac{LGT} subroutines in between \ac{DQC} subroutines & 3 & 4 & 6 & 7 & 8 \\
    max no. consecutive \ac{LGT} subroutines when a \ac{DQC} is in waiting state & 2 & 3 & 4 & 4 & 4 \\
    \% of times that $\geq 1$ \ac{LGT} subroutines fills time waiting for time bin  & 56 & 81 & 99 & 99 & 100 \\
    no. times that network process is delayed by a \ac{LGT} subroutine & 88/360 & 212/720 & 554/1080 & 940/1440 & 1170/1800 \\
    no. time windows in which all user processes are idle & 399 & 56 & 4 & 1 & 0 \\
    Average length of idle time window (ms) & 10 & 5.9 & 5.9 & 8.3 & --- \\
    Maximum length of idle time window (ms) & 152 & 31 & 15 & 8.3 & --- \\
    \% client and server start network process at same time bin & 95.8 & 58.2 & 42.1 & 42.4 & 38.5 \\
    \% server started network process 1 time bin after client & 0.8 & 37.9 & 50.1 & 50.9 & 53.1 \\
    \% server started network process $> 1$ time bins after client & 0.0 & 3.3 & 7.7 & 6.6 & 8.4 \\
    \% client started network process 1 time bin after server & 3.3 & 0.6 & 0.1 & 0.1 & 0.0 \\
    \hline
    \end{tabular}
    \caption{Overview of values derived from the multitasking experiments in which $N$ \ac{DQC} applications (on client and server) and $N$ \ac{LGT} applications (client only) were executed concurrently, for $N \in \{1, 2, 3, 4, 5\}$.}
    \label{tab:multitasking_numbers}
\end{table*}


\clearpage
\onecolumn
\section{Multitasking scheduling patterns}
\label{sec:multitasking_patterns}

\clearpage

\clearpage

\clearpage
\section{Application source code}
\label{sec:app_source}

\begin{figure*}[htbp]
  \centering
  \begin{minipage}{\textwidth}
\begin{lstlisting}[language=Python, caption={}]
# Bases to measure the final server state in.
# Note: for efficiency reasons, only bases +Y and -Y were
# used for alpha=pi/2, and +Z and -Z for alpha=pi.
MEAS_BASES = ["+X", "+Y", "+Z", "-X", "-Y", "-Z"]

# Server code (no parameters).
class DelegatedComputationServer(Application):
    def _rotate_basis(self, qubit: Qubit, basis: str) -> None:
        right_angle = math.pi / 2
        if basis == "+X":
            qubit.rot_Y(angle=-right_angle)
        elif basis == "+Y":
            qubit.rot_X(angle=right_angle)
        elif basis == "-X":
            qubit.rot_Y(angle=right_angle)
        elif basis == "-Y":
            qubit.rot_X(angle=-right_angle)
        elif basis == "-Z":
            qubit.X()

    def run(self, context: ApplicationContext) -> Dict[str, Any]:
        outcomes = {}
        for basis in MEAS_BASES:
            # Create EPR pair with client
            epr = context.epr_sockets[0].recv_keep()[0]
            # Compile and send subroutine S1.
            context.connection.flush()
            # Wait and receive delta from client.
            delta = context.app_socket.recv_float()
            # Local gates using delta.
            epr.rot_Y(angle=math.pi / 2)
            epr.rot_X(angle=delta)
            epr.rot_X(angle=math.pi)
            # At this point, the server has qubit state |psi>.
            # Measure in particular basis (part of tomography).
            self._rotate_basis(qubit=epr, basis=basis)
            m_s = epr.measure(store_array=False)
            # Compile and send subroutine S2.
            context.connection.flush()
            # Receive and store result (m_s).
            m_s = int(m_s)
            outcomes[basis] = m_s
        return outcomes
\end{lstlisting}
  \end{minipage}
  \caption{\acf{DQC} source code for the server.}
  \label{src:dqc_server}
\end{figure*}

\clearpage

\begin{figure*}[htbp]
  \centering
  \begin{minipage}{\textwidth}
\begin{lstlisting}[language=Python, caption={}]
# Bases to measure the final server state in.
# Note: for efficiency reasons, only bases +Y and -Y were
# used for alpha=pi/2, and +Z and -Z for alpha=pi.
MEAS_BASES = ["+X", "+Y", "+Z", "-X", "-Y", "-Z"]

# Client code, with parameters alpha and theta.
class DelegatedComputationClient(Application):
    def __init__(self, alpha: float, theta: float):
        self._theta = theta
        self._alpha = alpha

    def run(self, context: ApplicationContext) -> Dict[str, Any]:
        outcomes = {}
        for basis in MEAS_BASES:
            # Create EPR pair with server
            epr = context.epr_sockets[0].create_keep()[0]
            # Local gates
            epr.rot_Y(angle=math.pi / 2)
            epr.rot_X(angle=self._theta)
            epr.rot_X(angle=math.pi)
            # Measurement
            m_c = epr.measure(store_array=False)
            # Compile and send subroutine C1
            context.connection.flush()
            # Receive and store result (m_c).
            outcomes[basis] = m_c
            # Compute and send delta.
            delta = self._alpha - self._theta + m_c * math.pi
            context.app_socket.send_float(delta)
        return outcomes
\end{lstlisting}
  \end{minipage}
  \caption{\acf{DQC} source code for the client.}
  \label{src:dqc_client}
\end{figure*}

\clearpage

\begin{figure*}[htbp]
  \centering
  \begin{minipage}{\textwidth}
\begin{lstlisting}[language=Python, caption={}]
# Bases to measure the final server state in.
MEAS_BASES = ["+X", "+Y", "+Z", "-X", "-Y", "-Z"]

# Local tomography code with axis and angle parameters.
class ClientLocalApp(Application):
    def __init__(self, axis: str, angle: float) -> None:
        self._axis = axis
        self._angle = angle

    def run(self, context: ApplicationContext) -> Dict[str, Any]:
        outcomes = {}
        # Loop over all 6 cardinal measurement bases.
        for basis in self.MEAS_BASES:
            # Create and initialize a qubit in the |0> state.
            q = Qubit(context.connection)
            # Rotate it to one of the 6 cardinal states.
            if self._axis == "X":
                q.rot_X(angle=self._angle)
            else:
                q.rot_Y(angle=self._angle)
            # Measure it in the current measurement basis.
            self._rotate_basis(qubit=q, basis=basis)
            outcomes[basis] = q.measure(store_array=False)
        # Compile send the subroutine containing the above instructions.
        context.connection.flush()
        return outcomes
\end{lstlisting}
  \end{minipage}
  \caption{\acf{LGT} source code.}
  \label{src:lgt}
\end{figure*}

\clearpage

\begin{figure*}[htbp]
  \centering
  \begin{minipage}{\textwidth}
\begin{lstlisting}[language=Python, caption={}]
# Instantiate the programs to run.
programs = []
er_socket_ids = {}
for i in range(N):
    dqc_program = create_dqc_client_program()
    lgt_program = create_lgt_program()
    programs.append(dqc_program)
    programs.append(lgt_program)
    er_socket_ids[dqc_program] = i  # assign unique ID for ER socket
    

# Create a thread pool that can be executed by the OS hosting the CNPU.
tpe = ThreadPoolExecutor()

# For each program, submit a piece of code that executes the whole program.
for program in programs:
    runner = program_runner(program, er_socket_ids[progam])
    tpe.submit(runner)  # submit it to the thread pool

# Block until the OS hosting the CNPU has finished all programs.
tpe.wait()


# Code for running a single program.
def program_runner(program, er_socket_id):
    # Create connection with QNPU
    qnpu_connection = connect_qnpu()

    # Use connection to setup processes.
    qnpu_connection.register_program()
    for remote_node in program.remote_nodes:
        er_socket = ERSocket(
            remote_node=remote_node.name,
            er_socket_id=er_socket_id,
            remote_er_socket_id=er_socket_id)
        qnpu_connection.open_er_socket(er_socket)
    
    # classical sockets with other programs do not go through the QNPU
    create_classical_sockets()
    

    # Execute the program code; it can use the connection to send subroutines
    # to the QNPU and receive results.
    run(program, qnpu_connection)
\end{lstlisting}
  \end{minipage}
  \caption{Pseudocode illustrating the \ac{CNPU} runner. It can instantiate multiple programs, like the ones defined in~\cref{src:dqc_server,src:dqc_client,src:lgt}. Each program is submitted for concurrent execution to a thread pool executor which is managed by the host \ac{OS}. Each program independently sets up a connection with the \ac{QNPU}, and executes the program code itself.}
  \label{src:cnpu_runner}
\end{figure*}

\lstdefinelanguage{netqasm}
{
    morekeywords={
        add, add, sub, addm, subm,
        jmp, bez, bnz, beq, bne, blt, bge,
        set, store, load, undef, lea,
        array, qalloc, qfree,
        wait_all, wait_any, wait_single,
        ret_reg, ret_arr,
        meas,
        create_epr, recv_epr,
        init, x, y, z, h, s, k, t, rot_x, rot_y, rot_z, cnot, cphase, cx_dir, cy_dir,
        APPID, NETQASM,
    }
    sensitive=false,
    morecomment=[l]{//},
    morecomment=[s]{/*}{*/},
    morecomment=[s][\color{blue}]{\#\ }{\ },
    morestring=[b]",
}

\clearpage

\begin{figure*}[htbp]
  \centering
  \begin{minipage}{\textwidth}
\begin{lstlisting}[basicstyle=\small\ttfamily, language=netqasm, caption={}]
   array 10 @0
   array 1 @1
   store 0 @1[0]
   recv_epr(2,0) 1 0   // submit request for entanglement 
   set R0 0
LOOP3:
   beq R0 1 LOOP_EXIT3
   set R3 0
   set R4 0
   set R5 0
   set R6 0
LOOP:
   beq R6 10 LOOP_EXIT
   add R3 R3 R0
   add R6 R6 1
   jmp LOOP
LOOP_EXIT:
   add R4 R0 1
   set R6 0
LOOP1:
   beq R6 10 LOOP_EXIT1
   add R5 R5 R4
   add R6 R6 1
   jmp LOOP1
LOOP_EXIT1:
   wait_all @0[R3:R5]  // wait until entangled qubit is ready
   set R3 9            // check which Bell state
   set R4 0
LOOP2:
   beq R4 R0 LOOP_EXIT2
   add R3 R3 10
   add R4 R4 1
   jmp LOOP2
LOOP_EXIT2:
   load R2 @0[R3]      // load Bell state type into R2
   set R1 0
   bne R2 3 IF_EXIT
   rot_z R1 16 4       // correction for Phi-
IF_EXIT:
   bne R2 1 IF_EXIT1
   rot_x R1 16 4       // correction for Psi+
IF_EXIT1:
   bne R2 2 IF_EXIT2
   rot_x R1 16 4
   rot_y R1 8 4
   rot_x R1 16 4
   rot_y R1 24 4       // corrections for Psi-
IF_EXIT2:
   beq R0 0 IF_EXIT3   // no correction for Phi+
IF_EXIT3:
   add R0 R0 1
   jmp LOOP3
LOOP_EXIT3:
   ret_arr @0
   ret_arr @1
\end{lstlisting}
  \end{minipage}
  \caption{\ac{NetQASM} subroutine S1 of the \ac{DQC} application. Compiled by the \ac{DQC} server program code listed in~\cref{src:dqc_server}.}
  \label{src:netqasm_dqc_s1}
\end{figure*}

\clearpage

\begin{figure*}[htbp]
  \centering
  \begin{minipage}{\textwidth}
\begin{lstlisting}[basicstyle=\small\ttfamily, language=netqasm, caption={}]
set Q0 0
rot_y Q0 1 1
set Q0 0
rot_x Q0 1 0
set Q0 0
rot_x Q0 1 0
set Q0 0
meas Q0 M0
qfree Q0
ret_reg M0
\end{lstlisting}
  \end{minipage}
  \caption{\ac{NetQASM} subroutine S2 of the \ac{DQC} application. Compiled by the \ac{DQC} server program code listed in~\cref{src:dqc_server}. The exact gates may differ depending on the iteration of the program loop and the $\delta$ value sent by the client.}
  \label{src:netqasm_dqc_s2}
\end{figure*}

\clearpage

\begin{figure*}[htbp]
  \centering
  \begin{minipage}{\textwidth}
\begin{lstlisting}[basicstyle=\small\ttfamily, language=netqasm, caption={}]
   array 10 @0
   array 1 @1
   store 0 @1[0]
   array 20 @2
   store 0 @2[0]
   store 1 @2[1]
   create_epr(1,0) 1 2 0    // submit request for entanglement
   set R0 0
LOOP2:
   beq R0 1 LOOP_EXIT2
   set R3 0
   set R4 0
   set R5 0
   set R6 0
LOOP:
   beq R6 10 LOOP_EXIT
   add R3 R3 R0
   add R6 R6 1
   jmp LOOP
LOOP_EXIT:
   add R4 R0 1
   set R6 0
LOOP1:
   beq R6 10 LOOP_EXIT1
   add R5 R5 R4
   add R6 R6 1
   jmp LOOP1
LOOP_EXIT1:
   wait_all @0[R3:R5]    // wait until entangled qubit is ready
   add R0 R0 1
   jmp LOOP2
LOOP_EXIT2:              // end of entanglement creation code
   set Q0 0
   rot_y Q0 1 1          // local gate in C1
   set Q0 0
   rot_x Q0 1 0          // local gate in C1
   set Q0 0
   meas Q0 M0            // measurement
   qfree Q0
   ret_arr @0
   ret_arr @1
   ret_arr @2
   ret_reg M0
\end{lstlisting}
  \end{minipage}
  \caption{\ac{NetQASM} subroutine C1 of the \ac{DQC} application. Compiled by the \ac{DQC} client program code listed in~\cref{src:dqc_client}. The exact gates may differ depending on the \ac{DQC} parameters $\alpha$ and $\theta$.}
  \label{src:netqasm_dqc_c1}
\end{figure*}

\clearpage

\begin{figure*}[htbp]
  \centering
  \begin{minipage}{\textwidth}
\begin{lstlisting}[basicstyle=\small\ttfamily, language=netqasm, caption={}]
set Q0 0
qalloc Q0
init Q0
rot_y Q0 3 1
meas Q0 M0
qfree Q0
qalloc Q0
init Q0
rot_x Q0 1 1
meas Q0 M1
qfree Q0
qalloc Q0
init Q0
meas Q0 M2
qfree Q0
qalloc Q0
init Q0
rot_y Q0 1 1
meas Q0 M3
qfree Q0
qalloc Q0
init Q0
rot_x Q0 3 1
meas Q0 M4
qfree Q0
set Q0 0
qalloc Q0
init Q0
rot_x Q0 1 0
meas Q0 M5
qfree Q0
ret_reg M0
ret_reg M1
ret_reg M2
ret_reg M3
ret_reg M4
ret_reg M5
\end{lstlisting}
  \end{minipage}
  \caption{\ac{NetQASM} subroutine L1 of the \ac{LGT} application. Compiled by the \ac{LGT} program code listed in~\cref{src:lgt}. The exact gates may differ depending on the iteration of the program loop.}
  \label{src:netqasm_lgt_l1}
\end{figure*}

\clearpage
\section{Traces}
\label{sec:traces}

In our \ac{NV} experiments, the \ac{CNPU}, \ac{QNPU} and \ac{QDevice}, on both client and server nodes, trace (i.e. record the timestamps of) events happening on their system. The events that are traced on the \ac{CNPU} and \ac{QNPU} are listed in~\cref{tab:host_events,tab:qnpu_events}, respectively. The \ac{NV} \ac{QDevice} separately records messages received (physical instructions from the \ac{QNPU}, see~\cref{tab:qdevice-instructions}) and responses sent back to the \ac{QNPU}(see~\cref{tab:qdevice-return-values}).

\begin{figure*}
\centering
\includegraphics[width=0.8\linewidth]{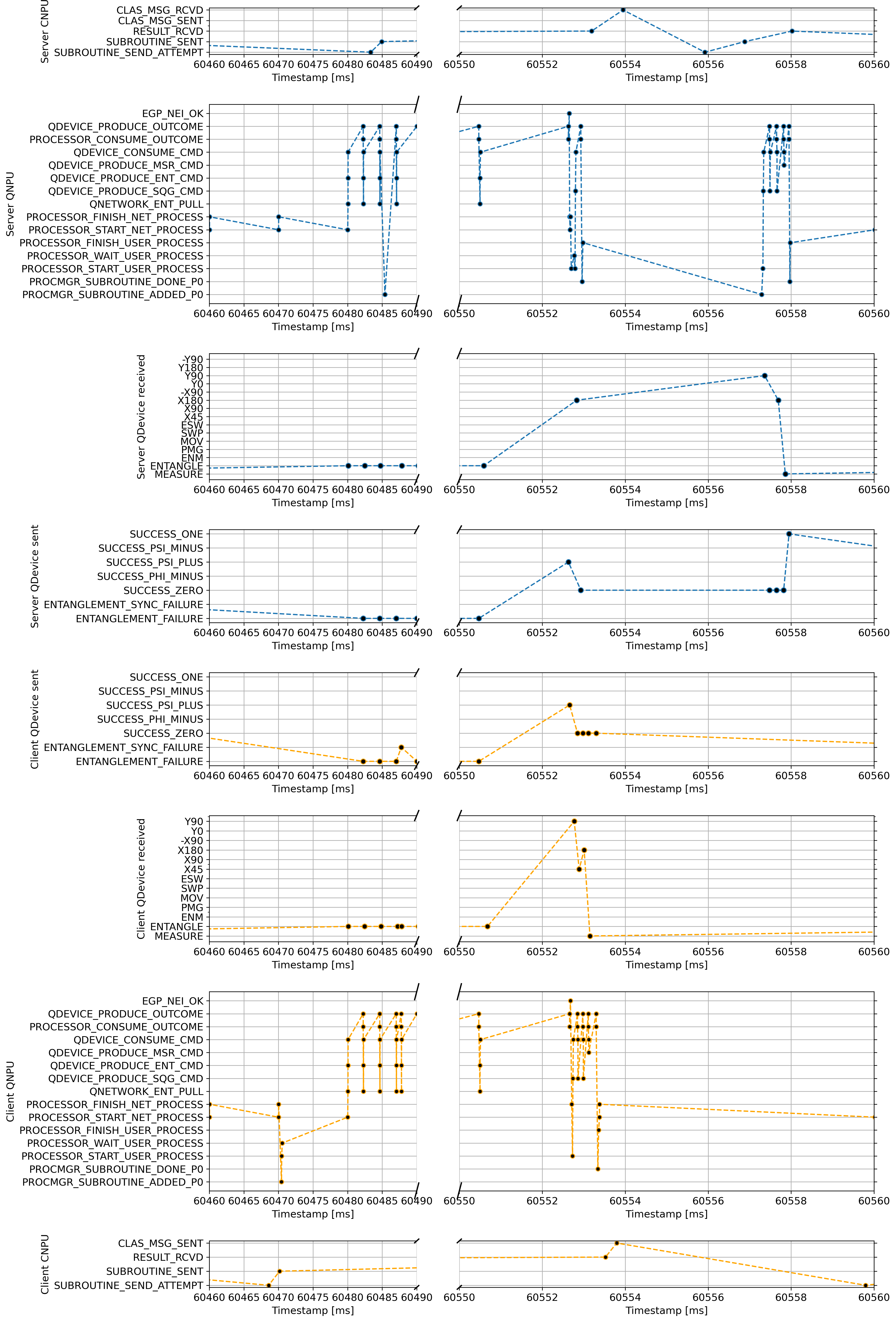}
\caption{Full-stack event trace for one particular execution of the \ac{DQC} circuit. Between timestamps 60490 and 60550 are more entanglement attempts which are cut out for the sake of clarity.}
\label{fig:delcomp-trace-example}
\end{figure*}

\Cref{fig:delcomp-trace-example} shows a full-stack trace slice of a single execution of the \ac{DQC} circuit. This particular sequence of events started at offset 60460\,ms from the start of the experiment. The following events (among others) can be seen:
\begin{itemize}
    \item At $\approx$\,60470\,ms: client \ac{CNPU} sends subroutine C1 to the \ac{QNPU}; it is received slightly after on the \ac{QNPU} (\texttt{PROCMGR\_SUBROUTINE\_ADDED\_P0}).
    \item Slightly after 60470\,ms: the \ac{QNPU} starts the user process containing C1; it hits the entanglement instruction and moves the process to the waiting state (\texttt{PROCESSOR\_WAIT\_USER\_PROCESS}).
    \item At 60480\,ms: the first next time bin starts, starting the network process on both client and server. This results in \texttt{ENTANGLE} commands being sent to the \acp{QDevice} by both client and server.
    \item Between 60480 and 60550\,ms: the two \acp{QDevice} repeatedly attempt entanglement but fail (each \texttt{ENTANGLE} instruction from the \ac{QNPU} starts one batch; each \texttt{ENTANGLEMENT\_FAILURE} return message indicates the batch failed).
    \item Meanwhile at 60485\,ms, the server \ac{CNPU} sends subroutine S1 to the \ac{QNPU}.
    \item At $\approx$\,60552.5\,ms, the \acp{QDevice} succeed in entanglement generation, producing a $\ket{\Psi^+}$ Bell pair.
    \item After this, the client and server finish C1 and S1, respectively. The client sends instructions for local gates ending with a \texttt{MEASURE} physical instruction. The server starts S1, hits the \texttt{recv\_epr} instruction, goes into the waiting state, gets immediately unblocked (since the entangled pair was already created) and sends a bell state correction gate to the \ac{QDevice} (\texttt{X180}).
    \item At $\approx$\,60553.5\,ms, the client \ac{CNPU} receives the result of C1 (\texttt{RESULT\_RCVD}), and sends the classical message $\delta$ to the server \ac{CNPU} (\texttt{CLAS\_MSG\_SENT}).
    \item At $\approx$\,60554\,ms, the server \ac{CNPU} receives $\delta$ (\texttt{CLAS\_MSG\_RCVD}).
    \item At $\approx$\,60557\,ms, the server \ac{CNPU} sends S2 to the \ac{QNPU}. The \ac{QNPU} executes the user process containing S2 which involves sending local quantum instructions to the \ac{QDevice} ending with a measurement.
    \item At $\approx$\,60558\,ms, the \ac{QNPU} sends the result of S2 to the \ac{CNPU}.
\end{itemize}

\clearpage

\begin{table*}[htpb]
    \centering
    \begin{tabular}{|c|l|}
    \hline
    \textbf{Event name} & \textbf{Description} \\ 
    \hline
    \texttt{SUBROUTINE\_SEND\_ATTEMPT} & Try to send subroutine to \ac{QNPU} \\
    \texttt{SUBROUTINE\_SENT} & Subroutine sent to \ac{QNPU} \\ 
    \texttt{RESULT\_RCVD} & Subroutine results received from \ac{QNPU} \\
    \texttt{CLAS\_MSG\_SENT} & Classical message sent to other node \\
    \texttt{CLAS\_MSG\_RCVD} & Classical message received from other node \\
    \hline
    \end{tabular}
    \caption{\ac{CNPU} events that are traced (recorded with their timestamps) during application execution.}
    \label{tab:host_events}
\end{table*}

\begin{table*}[htpb]
    \centering
    \begin{tabular}{|c|l|}
    \hline
    \textbf{Event name} & \textbf{Description} \\ 
    \hline
    \texttt{SCHEDULER\_ARRIVE\_USER\_PROCESS} & A user process goes to the Ready state \\
    \texttt{SCHEDULER\_SCHEDULE\_USER\_PROCESS} & A user process goes to the Running state \\
    \texttt{SCHEDULER\_ARRIVE\_NET\_PROCESS} & Network process goes to the Ready state \\
    \texttt{SCHEDULER\_SCHEDULE\_NET\_PROCESS} & Network process goes to the Running state \\
    \texttt{PROCMGR\_SUBROUTINE\_ADDED\_P<i>} & New subroutine received from \ac{CNPU} for process <i> \\
    \texttt{PROCMGR\_SUBROUTINE\_DONE\_P<i>} & A subroutine for process <i> finished execution \\
    \texttt{PROCESSOR\_START\_USER\_PROCESS} & Processor starts or resumes executing a user process \\
    \texttt{PROCESSOR\_WAIT\_USER\_PROCESS} & Processor suspends a user process and puts it in the Waiting state \\
    \texttt{PROCESSOR\_FINISH\_USER\_PROCESS} & Processor stops executing a user process \\
    \texttt{PROCESSOR\_START\_NET\_PROCESS} & Processor starts or resumes executing the network process \\
    \texttt{PROCESSOR\_FINISH\_NET\_PROCESS} & Processor stops executing the network process \\
    \texttt{QDEVICE\_PRODUCE\_<cmd>\_CMD} & Processor prepares <cmd> command for the \ac{QDevice} \\
    \texttt{QDEVICE\_CONSUME\_CMD} & \ac{QDevice} reads the next command from the \ac{QNPU} \\
    \texttt{QDEVICE\_PRODUCE\_OUTCOME} & \ac{QDevice} sends result to the \ac{QNPU} \\
    \texttt{PROCESSOR\_CONSUME\_OUTCOME} & Processor reads \ac{QDevice} result \\
    \texttt{QNETWORK\_ENT\_PULL} & Network stack pulls instruction from the EGP \\
    \texttt{EGP\_NEI\_OK} & QEGP notifies that EPR pair has been created \\
    \hline
    \end{tabular}
    \caption{\ac{QNPU} events that are traced (recorded with their timestamps) during application execution. <i> can be any number from 0 to 9 (`subroutine added` and `subroutine done` events are not traced for processes with ID 10 or larger). <cmd> can be any physical instruction.}
    \label{tab:qnpu_events}
\end{table*}

\end{appendices}

\end{document}